\documentclass[useAMS,usenatbib,iop,numberedappendix
–]{mn2e} 

\usepackage{siunitx}
\usepackage{lipsum,graphicx,multicol}
\usepackage{amsmath,amsfonts,amssymb}
\usepackage{makeidx}
\usepackage{graphicx}
\usepackage{epstopdf}
\usepackage{multirow}
\usepackage{rotating}
\usepackage{color}
\usepackage{tablefootnote}
\usepackage{lscape, afterpage}
\usepackage{subfigure}
\usepackage{tablefootnote}
\usepackage[T1]{fontenc}
\usepackage{aecompl}
\usepackage{subfig}
\usepackage{float,lscape}
\usepackage{afterpage}
\usepackage{adjustbox}

\definecolor{ored}{rgb}{1.00,0.27,0.00}

\title[Contribution of BGGs to the total baryon content of groups]{Brightest 
group galaxies-II: the relative contribution of BGGs to the total baryon 
content of groups at $ z<1.3 $}
\author[Gozaliasl et al.]{ Ghassem Gozaliasl,$^{1,2,3}$\thanks{E-mail: 
ghassem.gozaliasl@utu.fi},
Alexis Finoguenov $^{2,4}$,
Habib G. Khosroshahi$^{5}$,
\newauthor
Bruno M. B. Henriques $ ^{6} $,
Masayuki Tanaka$^7$,
Olivier Ilbert$^8 $, 
Stijn Wuyts$^{9}$,
\newauthor
Henry J. McCracken $^{10}$, and
Francesco Montanari $^{2,3}$\\
\newline\\
 $^1$Finnish Centre for Astronomy with ESO (FINCA), University of Turku\\ 
 V\"{a}is\"{a}l\"{a}ntie 20, FI-21500 PIIKKI\"{O},Finland\\  
  $^2$Department of Physics, University of Helsinki, P. O. Box 64, FI-00014, 
  Helsinki, Finland\\
  $^3$Helsinki Institute of Physics, P.O. Box 64, FI-00014 University of Helsinki, Finland\\
    $^{4}$Max Planck-Institute for Extraterrestrial Physics, P.O. Box 1312, Giessenbachstr. 1., D-85741 Garching, Germany\\
  $^5$School of Astronomy, Institute for Research in Fundamental Sciences (IPM), Tehran, Iran\\
 $^6$Institute for Astronomy, ETH Zurich, CH-8093 Zurich, Switzerland\\
 $^7$ National Astronomical Observatory of Japan, 2-21-1 Osawa, Mitaka,\\
  Tokyo 181-8588, Japan\\
 $^8$ Aix Marseille Universit\'{e}, CNRS, Laboratoire \v{d} Astrophysique de Marseille, UMR 7326, F-13388 Marseille, France\\ 
 $^{9}$ Department of Physics, University of Bath, Claverton Down,
Bath, BA2 7AY, UK\\
 $^{10}$CNRS, UMR 7095 \& UPMC, Institut \v{d}Astrophysique de Paris, 98bis boulevard Arago, 75014 Paris, France\\
}

\begin{document}

\maketitle


\label{firstpage}
\begin{abstract}
We performed a detailed study of the evolution of the star formation rate (SFR) 
and  
 stellar mass of 
 the brightest group galaxies (BGGs) and their relative contribution to the 
 total 
 baryon 
budget within $R_{200}$ ($f^{BGG}_{b,200}$). The sample comprises 407 BGGs 
selected from X-ray groups ($M_{200}=10^{12.8}-10^{14} \;M_{\odot}$) out to  $z 
\sim
1.3 $ identified in the COSMOS, XMM-LSS, AEGIS fields. We find that BGGs 
constitute two distinct populations of quiescent and star-forming galaxies and 
their mean SFR is $\sim2$ dex higher than the median SFR at $ z<1.3 $. Both the 
mean and the 
median SFRs decline with time by $>2$ dex. The mean (median) 
of stellar mass has grown by $0.3$ dex since $ z=1.3$ to the present day. We 
show
that up to $\sim45\% $ of 
the stellar mass growth in a star-forming BGG can be due to its
star-formation activity. With respect to  $f^{BGG}_{b,200}$, we find it to 
increase with decreasing redshift by 
$\sim0.35$ dex while decreasing 
with halo mass in a redshift dependent manner. We show that the slope of the 
relation between  $f^{BGG}_{b,200}$ and halo mass increases negatively with 
decreasing redshift. This trend is driven by an 
insufficient star-formation in BGGs, compared to the halo growth rate. We 
separately show the BGGs with the 20\% highest $f^{BGG}_{b,200}$ are generally 
non-star-forming galaxies and grow in mass by processes not related to star 
formation
(e.g., dry mergers and tidal striping). We present the $ M_\star-M_h $ and $ 
M_\star/M_h-M_h $ relations  and 
compare them with semi-analytic model predictions and a number of 
results from the 
literature. We quantify the intrinsic scatter in stellar mass of BGGs  at 
fixed 
halo mass ($\sigma_{log M_{\star}}$) and find that $\sigma_{log M_{\star}}$ 
increases from 0.3 dex  at $ z\sim0.2 $ to 0.5 dex at $ z\sim1.0 $ due to the 
bimodal 
distribution of stellar mass.

\end{abstract}

\section{Introduction} \label{Introduction}

The baryon content of the universe and its partitioning between different 
components, e.g., hot/cold gas and stars, is one of the most important 
observations in cosmology. Clusters of galaxies are thought to have baryon 
fractions that approach the cosmic mean, with most of the baryons in the form 
of 
x-ray emitting hot gas and stars, and are particularly important in this context
 \citep{white1991galaxy}. This has been confirmed by previous studies which 
 found that, after including baryons in stars,
 the baryon content in the most massive clusters closely matches that measured 
 from observations of the CMB 
\citep{White93,David95,vikhlinin2006chandra,Allen08,dunkley2009five,Simionescu11,Bulbul16}.
 These 
features  
make clusters important tools to probe cosmological parameters and cosmic 
evolution. Several gravitational (e.g., mergers, tidal stripping) and 
non-gravitational (e.g., outflows from the active galactic nuclei (AGN), 
supernovae explosions) processes act on cluster components and play a major 
role in driving cluster galaxy evolution  
\citep{Evrard97,Mohr99,Roussel00,Lin03,Allen04,McCarthy07,Allen08,Ettori09,
	Giodini09,McGaugh09,Andreon10,Allen11,Simionescu11,Dvorkin15}.
 These processes could account for the deviations reported between the universal 
baryon fraction and that of low mass clusters. The baryon fraction in low mass 
clusters 
or galaxy groups with halo masses $M_h<10^{14}\;M_\odot$ is generally 
smaller than the baryon fraction in massive galaxy clusters (Mathews et al 2005, Sanderson 
et al 2013), possibly due to AGN feedback 
\citep{mccarthy2010case,mccarthy2011}. Admittedly, observations suggest that 
some galaxy groups with a large X-ray to optical luminosity ratio 
($L_{X}/L_{opt}$) such as fossil galaxy groups \citep{khosroshahi2007scaling} 
represent systems with a baryon fraction close to the cosmic mean value, $f_b$ 
= 0.16 \citep{mathews2005baryonically}. 

While studies of the cluster/group baryon fractions have been mostly focused on 
the estimate of the baryons contained in their galaxies and the hot 
intracluster/intragroup gas, understanding the relative contribution of the 
satellite galaxies and the brightest cluster/group galaxies (BCGs/BGGs) is 
highly important for  precise modelling of galaxy formation especially in 
low-mass haloes. 
\cite{Giodini09} have shown that the stellar mass fraction contained in 91 
galaxy groups/clusters at $ 0.1<z<1.0 $ selected from the COSMOS survey scales 
with total mass as  $M_{500}^{-0.37\pm0.04} $ ($ M_{500}$ corresponds to 
the halo mass at the radius at which the over-density is 500 times the 
mean density) and is independent of redshift. \cite{Gonzalez13} have also found 
that the
fraction of the baryons residing in stars and hot-gas are strong functions of 
the total mass and scale as $ f_{star}\propto M_{500}^{-0.45\pm0.04} $  
and $ f_{gas}\propto M_{500}^{0.26\pm0.03}$, 
indicating that the baryons contained in stars become important in low mass 
haloes. Determining the contribution of stars to the total baryon fraction in 
groups, as 
opposed to massive clusters, is also important because baryonic effects (e.g., 
radio feedback) are more significant in groups 
\citep[e.g.,][]{giodini2012galaxy}. The 
primary goal of the present study is to quantify the contribution of central 
galaxies to the total baryonic mass of hosting groups. 

In order to separate the role of different physical mechanisms in  galaxy 
evolution, a number of studies have constrained stellar-to-halo mass (SHM) 
relations and ratios as a function of time using the abundance matching 
technique   \citep[e.g.][]{behroozi2010comprehensive,Moster10,Behroozi10}, the
conditional luminosity function technique proposed by  
\cite{yang2003constraining}, the halo occupation distribution (HOD) formalism 
\citep[e.g.][]{berlind2002halo,kravtsov2004dark,Moster10}, and by  combining 
the HOD, N-body simulations, galaxy clustering, and galaxy-galaxy lensing 
techniques \citep[e.g.,][]{Leauthaud12,Coupon15}. Distinguishing the properties 
of central galaxies from those of satellite galaxies in studies based only on 
the distribution of luminosity or stellar mass is challenging  
\citep[e.g.,][]{George11}. By combining  several observables and techniques 
(e.g. HOD, galaxy-galaxy lensing, galaxy clustering) one can  probe a global  
SHM relation for central galaxies and satellite galaxies 
\citep[e.g.,][]{Leauthaud12,Coupon15}.  
  \cite{Coupon15},   for example, used multi-wavelength data of $ \sim60000  $ 
  galaxies with 
spectroscopic redshifts in the CFHTLenS/VIPERS field to
constrain the 
relationship between central/satellite mass and halo mass, characterising the 
contributions from central and satellite galaxies in the SHM relation. In this 
paper,  we directly identify the BGGs using their precise redshifts and  
estimate stellar masses using the 
broad band  Spectral Energy Distribution (SED) fitting technique 
\citep{Ilbert10}  as used by 
\cite{Coupon15}. We utilise 
the advantages of the X-ray selection of galaxy groups and a wealth of  
multi-wavelength, high signal-to-noise ratio observations such as the 
UltraVISTA 
survey in the COSMOS field \citep{laigle2016cosmos2015} to investigate the SHM 
relation for the central galaxies over 9 
billion years. We aim to quantify the intrinsic (lognormal) scatter in stellar 
mass at fixed redshift in  observations and compare them to the recently 
implemented  
semi-analytic model (SAM) by \cite{Henriques15}. 

 This paper is the second in a series of three studying the evolution of the 
 properties of BGGs. We use a sample of 407 X-ray galaxy groups with 
 halo masses ranging from $\sim10^{12.8} $ to $ 10^{14} M_{\odot} $ at $ 
 0.04<z<1.3  $  selected from the XMM-LSS \citep{Gozaliasl14}, COSMOS 
 \citep{Finoguenov07,George11} and AEGIS \citep{erfanianfar13} fields.
 
  In the first paper in this series \citep{gozaliasl2016brightest}, we  
  presented our data and the sample selection criteria. We studied the distribution of 
  stellar mass ($ M_{\star} $) and (specific) star-formation rate (SFR) of the BGGs 
  and  found  that the stellar mass 
  distribution of the BGGs evolves towards a normal distribution with decreasing 
  redshift. We also showed that the average $ M_{\star} $ of BGGs grows  by a 
  factor of $ \sim2 $ from $ 
  z=1.3 $ to the present day. This $ M_{\star} $ growth slows down at $ z<0.5$ 
  in 
  contrast 
  to the SAM predictions. We also revealed that BGGs are not completely 
  quenched 
  systems, and  about $20\pm3$\% of them with stellar mass of $\sim 10^{10.5} 
  M_{\odot} $ continue  star-formation with rates 
  up to $SFR\sim200\;M_{\odot}yr^{-1}$.  
  
In this paper, we measure the total baryon content of galaxy groups  and 
compute the ratio of the stellar mass of BGGs to the total baryonic mass of 
haloes within $ R_{200}$ as $ f^{BGG}_{b,200} $ and investigate whether this 
ratio changes as a function of redshift and halo mass. We showed that the 
mean value of the SFR of BGGs is considerably higher than the median value of 
SFR and the mean value is influenced by the very high SFRs. Thus, we 
decided 
to investigate the evolution of both mean and median values 
of SFR, $ M_{\star} $, and $f^{BGG}_{b,200} $, individually. Similarly to the 
first  
paper of this series, we  use observations here to probe the predictions by 
four 
SAMs based on the Millennium simulation as presented in \citet[][hereafter 
B06]{Bower06}, \citet[][hereafter DLB07]{deLucia07}, \citet[][hereafter 
G11]{Guo11}, and \citet[][hereafter H15]{Henriques15}.  
 
This paper is organised as follows: we briefly describe our sample in section 
2; Section 3 presents the relation between $f^{BGG}_{b,200} $ and $ M_{h} $ ($ 
M_h  $ corresponds to $ M_{200c}$ or $ M_{200m}$ , where the internal density 
of haloes is 200 times the critical or mean density of the universe). We  
investigate the smoothed distribution  of $f^{BGG}_{b,200}$ in different 
redshift bins. We also examine the 
redshift evolution of the mean (median) value of $ SFR, M_{\star} $ and  
$f^{BGG}_{b,200}$. Section 3 also presents the SHM relation and ratio and 
assigns a lognormal scatter in the stellar mass of BGGs at fixed halo mass. We 
compare our findings with a number of results from the literature. Section 4 
summaries the results and conclusions.

 Unless stated otherwise, we adopt a cosmological model, with
 $(\Omega_{\Lambda}, \Omega_{M}, h) = (0.70, 0.3, 0.71$), where the
 Hubble constant is parametrised as 100 h km s$^{-1}$ Mpc$^{-1}$ and
 quote uncertainties as being on the 68\% confidence level.
  
 \begin{figure}
 	\includegraphics[width=0.495\textwidth]{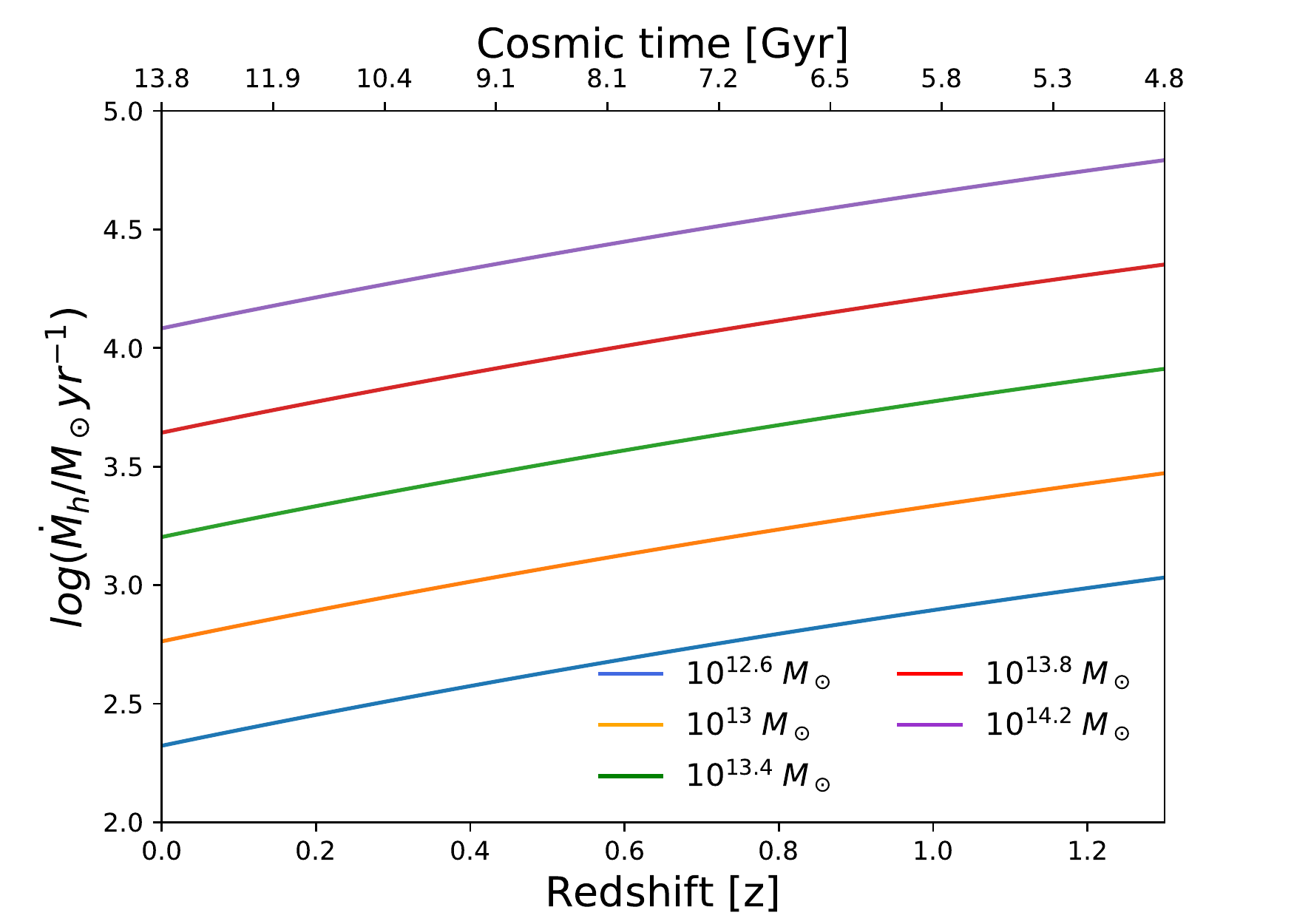}
 	\includegraphics[width=0.495\textwidth]{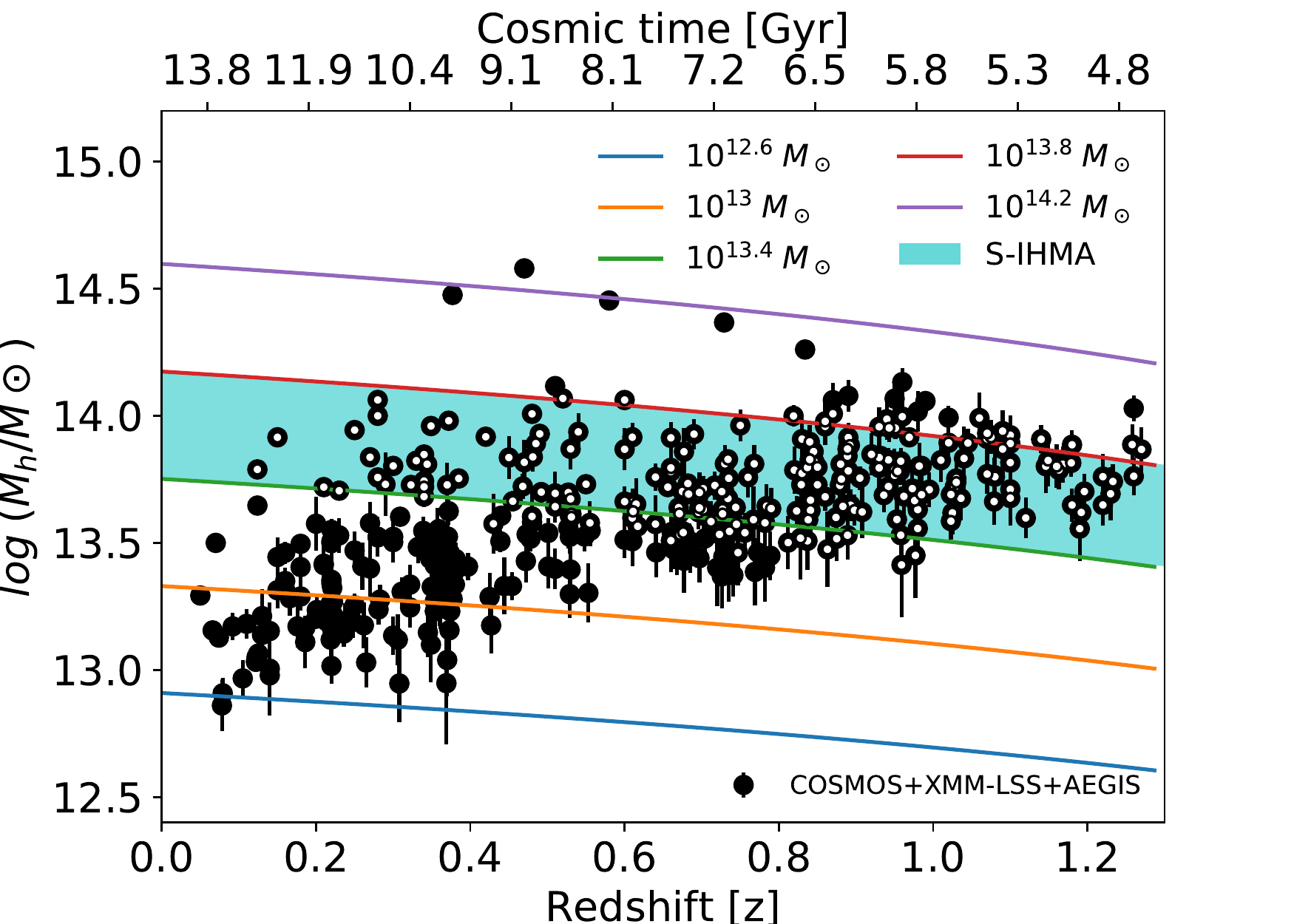}
 	\caption{\textit{(Upper panel)} Mean mass accretion rate of dark matter on 
 		to haloes ($ 
 		\dot{M}_h$) as a function of cosmic time and  redshift from $ z=1.3 $ to 
 		$ 
 		z=0 $ in Millennium 
 		simulations I \& II \citep[following equation 2 in 
 		][]{fakhouri2010merger}. The solid lines show trends for a set of haloes 
 		of given 
 		masses,  $M_h= 10^{12.6}\;  M_\odot, 10^{13}\;  M_\odot, 
 		10^{13.4}\; M_\odot, 10^{13.8}\;  M_\odot, 10^{14.2}\;  
 		M_\odot$.  
 		\textit{(Lower panel)} The halo mass of X-ray galaxy groups selected from 
 		COSMOS, AEGIS, and XMM-LSS fields as a function of cosmic time (z) 
 		(filled 
 		and open circles). Solid lines illustrate the redshift evolution of $ M_h 
 		$ 
 		for a set of typical haloes with a given initial halo mass (as 
 		mentioned 
 		in the upper panel) from $ z=1.3 $ to the present day. In order to 
 		investigate 
 		the 
 		impact of the halo mass growth on the evolution of stellar properties of 
 		galaxies such as stellar mass growth, we define a new sample of galaxy 
 		groups (open circles) which lie in the highlighted area 
 		(S-IHMA).}\label{mhrate}
 \end{figure}
 
\section{Data of BGGs} 
\subsection{Sample definition and BGG selection}\label{sample}
 We use galaxy group catalogues, with $M_{h}\sim5\times10^{12} $ to 
 $10^{14.5}M_{\odot} $ at $0.04<z<1.9$, which have been selected from the 
 COSMOS \citep{Finoguenov07,George11},  XMM--LSS \citep{Gozaliasl14}, and AEGIS 
 \citep{erfanianfar13} fields.  
  To ensure the high quality of the photometric redshift of groups, we constrain our study to the redshift
range of $ 0.04<z<1.3 $ and study groups with halo mass ranging from $ 
M_{h}\simeq7.25\times10^{12} $ to $ 1.04\times10^{14} (M_{\odot})$. As in 
Fig. 1 of paper I \citep{gozaliasl2016brightest}, we define five 
subsamples of galaxy groups considering their halo mass-redshift plane as 
follows: 

(S-I)   $0.04 <$ z $< 0.40$ $ \& $ $12.85 < log(\frac{M_{200}}{M_{\odot}})  \le 13.50 $ 

(S-II) $0.10 <$ z $\leq 0.4$ $ \& $  $13.50 < log(\frac{M_{200}}{M_{\odot}}) \le 14.02 $

(S-III) $0.4 <$ z $\leq 0.70$ $ \& $  $13.50 < log(\frac{M_{200}}{M_{\odot}}) \le 14.02 $

(S-IV) $0.70 <$ z $\leq 1.0$ $ \& $  $13.50 < log(\frac{M_{200}}{M_{\odot}}) \le 14.02 $

(S-V) $1.0 <$ z $\leq 1.3$ $ \& $  $13.50 < log(\frac{M_{200}}{M_{\odot}}) \le 14.02 $  

Four subsamples (S-II to S-V) cover a similar narrow halo mass range, which 
allows us to compare the stellar properties of BGGs within haloes of the same 
masses at different redshifts. The subsample of S-I has a similar redshift 
range to that of S-II but with a different halo mass range, which enables us to 
inspect the impact of halo mass on the properties of BGGs at $ z<0.4 $.

The full details of the sample selection, stellar mass and halo mass 
measurements have been presented in \cite{gozaliasl2016brightest} and 
\cite{Gozaliasl14A}. We estimate the halo mass using the $ L_x-M_h $ relation 
as presented in \cite{leauthaud2009weak}.  We also assume a 0.08 dex extra 
error in the halo mass estimate in our analysis, which corresponds to a 
log-normal scatter of the $ 
L_x-M_h $ relation \citep{allevato2012occupation}. Table \ref{mhmser} presents 
the mean stellar mass and halo mass with corresponding statistical and 
systematic errors for S-I to S-V. The systematic error corresponds to 
uncertanities in the stellar and halo mass measurements. We note that these 
systematic errors are taken from the galaxy and group 
catalogues by  
\cite{Finoguenov07,wuyts2011galaxy,erfanianfar13,Gozaliasl14A,laigle2016cosmos2015}.

According to the cold dark matter (CDM) hierarchical structure formation
paradigm, dark matter haloes grow by accretion of matter and merging with other 
(sub)haloes  \citep{frenk1988formation}. It is well-known that many of the 
observed galaxy properties correlate with the environment such as the known 
positive correlation between the stellar mass of the BCGs/BGGs
and the halo mass of their host dark matter haloes 
\citep{gozaliasl2016brightest}. The slope of this correlation is 
less than unity, implying that the cluster
growth is faster than the BCG growth at the same time. As a 
result, it is important that the effect of halo 
growth is taken into account, when galaxy growth is determined. To do this, we 
use the 
results presented by \cite{fakhouri2010merger} who construct merger trees of 
dark matter haloes and estimate their merger rates and mass growth rates using 
the joint data set of the Millennium I and Millennium-II simulations. We use 
equation 2 by \cite{fakhouri2010merger} and   determine the mean halo mass 
growth 
rates for some typical haloes of given masses ($M_h= 10^{12.6}\;  M_\odot, 
10^{13}\;  M_\odot, 10^{13.4}\; M_\odot, 10^{13.8}\;  M_\odot, 10^{14.2}\;  
M_\odot$) at $ z=1.3 $. As shown 
in Fig. \ref{mhrate}, the mean mass growth rate slowly decreases with cosmic 
time for all halo masses. Considering these mass accretion rates, we determine 
whether  $ M_h $ of these set of haloes grow with cosmic time (z) from $ z=1.3 
$ to the present day (solid lines in the lower panel of Fig. \ref{mhrate}).

Following the method used by \cite{groenewald2016investigating}, we construct 
an evolutionary sequence of galaxy groups at $ z<1.3 $, then select a new 
sample of galaxy groups such that their masses lie in the 
narrow highlighted cyan area and between two halo mass limits (green and red 
lines), which represent the $ M_h $ 
growth for two dark matter haloes with initial masses of $ 10^{13.4} $ and $ 10^{13.8} 
M_\odot $ at $ z=1.3 $, respectively. If a group with $ M_{h}\pm1\sigma $  
falls in the highlighted area, it is also  included in the sample. We analyse this new sample of groups and the associated BGGs at $ 0.2<z<1.3 $ in \S 
\ref{s3.2} and compare the results with those for subsamples  of S-II to S-V, 
where 
the halo mass 
range is the same for all subsamples and halo mass growth is not taken into account. 
Hereafter, we  will refer to the new Sample with Including Halo Mass Assembly 
effect as S-IHMA. 

We select most of the BGGs from groups by cross-matching their spectroscopic 
redshifts with the redshift of groups. BGGs with no spectroscopic observations 
are selected  using their 
photometric redshift 
\citep{wuyts2011galaxy,mccracken2012ultravista,Ilbert13,laigle2016cosmos2015} 
and the colour-magnitude diagram of group members, as described in detail 
in  \cite{Gozaliasl14A}. In this paper, the 
physical properties of BGGs in the COSMOS field are taken from the 
COSMOS2015 catalogue \citep{laigle2016cosmos2015}.  This catalogue contains $ 
1,182,103 $  objects with a high photometric redshift precision of $ 
\sigma_{\Delta z/(1+z_{sec})}=0.007 $ in the 1.5 deg$^2$ UltraVISTA-DR2 region. 
For more detail on the photometric redshift calculation/precision and physical 
parameter measurement of galaxies, we refer the reader to 
\cite{laigle2016cosmos2015}.

The sample selection for this work relies on the detection of the outskirts 
of X-ray galaxy groups. As such, this sample is unbiased towards the 
presence of the cool cores and therefore the properties of the BCGs. The 
difference in the completeness are minimized by selecting a relatively 
narrow range of halo masses for the study. As reported in Tab. \ref{mhmser}, 
the differences ($ \lesssim0.11 $ dex) among  the mean halo mass of haloes for 
S-II 
to S-V 
lie within the total error.  
\begin{table*}
	\caption{The average systematic error (SYSE) and the statistical error on 
		the mean (SEM) of $ log(M_{\star}/M_\odot) $ and $ log(M_h/M_\odot) $ for 
		S-I to 
		S-V.	The error values are given in dex.}
	\begin{tabular}{lllllll}
		\hline
		\hline
		\centering
		Sample &$ \langle log(M_{\star}/M_\odot)\rangle $ & $ dM_{\star}$(SYSE) & 
		$ 
		dM_{\star}$(SEM) &$ \langle log(M_h/M_\odot) \rangle$ & $ dM_h $(SYSE) & 
		$ 
		dM_h$(SEM)\\
		\hline\\
		S-I & 10.84 & 0.12 &  0.08 &13.28& 0.16 & 0.01 \\
		S-II &11.06 & 0.15 &  0.07 &13.68& 0.12 & 0.02 \\
		S-III &11.15 & 0.16 &  0.07 &13.69& 0.15 & 0.01 \\
		S-IV & 11.02 & 0.15 &  0.05 &13.75 &0.15 & 0.01 \\
		S-V &10.89 & 0.19 &  0.05 & 13.79& 0.16 & 0.02 \\
		\hline\hline
	\end{tabular} \label{mhmser}	
\end{table*} 
  \begin{figure}
	\includegraphics[width=0.495\textwidth]{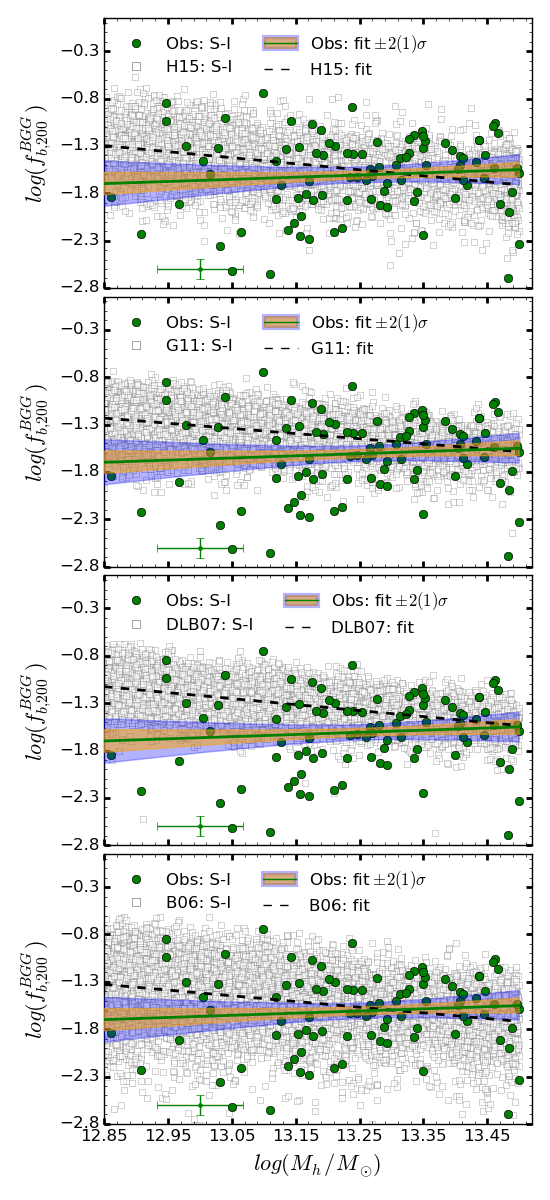} 
	\caption[]{ The relative contribution of the stellar populations of the BGGs to 
		the total baryonic mass of hosting groups, $log(f_{200}^{BGG})$, as a 
		function of halo mass ($log( M_{h}/M_\odot) $). Each row from top to 
		bottom 
		compares the observational data (filled green circles) with the data 
		taken 
		from SAMs of H15, G11, DLB07, and B06 (open gray squares), respectively. 
		The best-fit relation to the data in observations with associated $ 
		\pm1\sigma $ and $ \pm2 \sigma $ confidence intervals are shown as the 
		solid 
		green lines and the highlighted  orange and blue regions. The best-fit 
		relation for the SAMs are plotted as the dashed black lines. In each plot, 
		we 		have shown the scale of the median of observed uncertainties on the $ log 
		(M_h/M_\odot) $ and $log(f_{200}^{BGG})$ estimates with green error bars.}
	\label{fb_m200_1}
\end{figure}

\subsection{Semi-analytic models}         
We compare our results with a number of theoretical studies,
namely with four semi-analytic models (SAMs): \cite{Bower06} (B06);
\cite{deLucia07} (DLB07); \cite{Guo11} (G11); \cite{Henriques15} (H15). All the 
models are based on merger
trees from the Millennium Simulation \citep{springle2005cosmological} which
provides a description of the evolution of 
dark matter structures in a cosmological volume.
While B06, DLB07 and G11 use the simulation in its original
WMAP1 cosmology, H15 scales the merger trees to follow the evolution
of large scale structures expressed for the more recent cosmological 
measurements.
With respect to the treatment of baryonic physics, B06 uses the GALFORM version
of the Durham model,  while DLB07, G11, and H15 follow
the Munich L-Galaxies model.

In \cite{Gozaliasl14A}, we described some important features
of the G11, DLB07, and B06 models. We thus, briefly describe the recent
improvements and modifications in the H15 model here and the reader
is referred to the  \cite{Henriques15} for further
details. With respect to the previous version of L-Galaxies, in addition
to the implementation of a PLANCK cosmology, \cite{Henriques15} uses
the \cite{henriques2013simulations} model for the reincorporation of gas ejected
from SN feedback. The scaling of reincorporation time with virial
mass, instead of virial velocity, suppresses  star-formation
in low mass galaxies at earlier times and results in an excellent match
between theoretical and observed stellar mass functions at least since $z=3$.

In addition, the H15 model assumes that ram-pressure stripping is
only effective in clusters ($ M_{vir}>1.2\times 10^{14}\; M_\odot$) and has a 
cold
gas surface density threshold for  star-formation that is $\sim$ two times
smaller than in earlier models. These two modifications ensure that satellite
galaxies retain more fuel for  star-formation and continue to form
stars for longer. This eases a long standing problem with satellite
galaxies in theoretical models being quenched too quickly and provides
a good match to quenching trends as a function of environment \citep{henriques2016galaxy}.
Finally, H15 modified the AGN radio mode accretion rate in order
to enhance accretion at $z<0.5$ with respect to earlier times
and ensure that galaxies around M* grow significantly down to that redshift,
but are predominantly quenched in the local universe.

\begin{table*}
	\caption[]{The parameters of the best-fit relation, $ log(f_{b,200}^{BGG})= 
	\alpha \times 
	log(M_{200}/M_\odot) + log(\beta)$, obtained by linear least 
	squares (LLS) and Markov chain Monte Carlo (MCMC) 
	methods for both  observations and SAMs. The first column presents the id of 
	subsamples. The second and the third columns list the optimised parameters 
	with 68\% confidence intervals by the MCMC method. The fourth and fifth 
	columns present the parameters with $ \pm1\sigma$ error obtained by LLS. }
	\begin{tabular}{lllll} 
		\hline\hline  
		Subsamples & $ \alpha_{MCMC} $ & $ log(\beta_{MCMC}) $ &$ \alpha_{LLS}$ & $log(\beta_{LLS})$ \\    
		\hline\hline 
		\textbf{S-I}\\  
		Obs&$ +0.209 ^{+ 0.166 }_{- 0.196 }$& $ -4.393 ^{+ 2.598 }_{- 2.203 }$ & $+ 0.228 \pm 0.274 $ & $ -4.627 \pm 3.624 $ \\
		
		H15&$ -0.601 ^{+ 0.022 }_{- 0.022 }$& $ 6.412 ^{+ 0.288 }_{- 0.291 }$& $ -0.6377 \pm 0.0215  $ & $ 6.890 \pm 0.283 $\\
		G11&$ -0.601 ^{+ 0.022 }_{- 0.021 }$& $ 6.413 ^{+ 0.28 }_{- 0.286 }$ & $ -0.544 \pm 0.015  $ & $ 5.762 \pm 0.190 $\\
		DLB07&$ -0.603 ^{+ 0.014 }_{- 0.014 }$& $ 6.622 ^{+ 0.182 }_{- 0.184 }$  & $ -0.632 \pm 0.014  $ & $ 7.00 \pm 0.178 $\\
		B06&$ -0.491 ^{+ 0.021 }_{- 0.021 }$& $ 4.949 ^{+ 0.272 }_{- 0.27 }$& $ -0.601 \pm 0.022  $ & $ 6.402 \pm 0.283 $ \\   
		
		\hline \\
		\textbf{S-II}  \\
		Obs&$ -0.573 ^{+ 0.289 }_{- 0.19 }$& $ 6.077 ^{+ 2.61 }_{- 3.944 }$& $ -0.679\pm 0.338  $ & $ 7.520 \pm4.627 $\\   
		H15&$ -0.564 ^{+ 0.032 }_{- 0.033 }$& $ 5.912 ^{+ 0.446 }_{- 0.433 }$ & $ -0.598 \pm 0.0328 $ & $ 6.389 \pm 0.449 $ \\
		G11&$ -0.562 ^{+ 0.032 }_{- 0.033 }$& $ 5.891 ^{+ 0.452 }_{- 0.436 }$&$ -0.506 \pm 0.0359  $ & $ 5.247 \pm 0.493 $\\
		DLB07&$ -0.557 ^{+ 0.035 }_{- 0.035 }$& $ 6.006 ^{+ 0.481 }_{- 0.485 }$ & $ -0.573 \pm 0.034  $ & $ 6.221 \pm 0.472 $ \\
		B06&$ -0.495 ^{+ 0.057 }_{- 0.055 }$& $ 4.96 ^{+ 0.757 }_{- 0.779 }$& $ -0.542 \pm 0.057  $ & $ 5.610 \pm 0.784 $\\
		\hline \\
		\textbf{S-III} \\
		
		Obs &$ -0.371 ^{+ 0.313 }_{- 0.278 }$& $ 3.361 ^{+ 3.792 }_{- 4.294 }$& $ -0.493 \pm 0.301  $ & $ 5.021 \pm 4.115 $\\
		H15&$ -0.625 ^{+ 0.028 }_{- 0.028 }$& $ 6.731 ^{+ 0.381 }_{- 0.382 }$& $ -0.641 \pm 0.028  $ & $ 6.943 \pm 0.390 $
		\\
		G11&$ -0.626 ^{+ 0.029 }_{- 0.029 }$& $ 6.743 ^{+ 0.398 }_{- 0.396 }$& $ -0.500 \pm 0.020 $ & $ 5.101 \pm 0.269 $
		\\
		DLB07&$ -0.563 ^{+ 0.019 }_{- 0.019 }$& $ 6.007 ^{+ 0.255 }_{- 0.258 }$ & $ -0.574 \pm 0.019  $ & $ 6.153 \pm 0.255 $\\
		B06&$ -0.623 ^{+ 0.032 }_{- 0.032 }$& $ 6.583 ^{+ 0.434 }_{- 0.436 }$& $ -0.668 \pm  0.032 $ & $ 7.204 \pm 0.433 $\\
		\hline \\
		\textbf{S-IV}\\
		Obs &$ -0.05 ^{+ 0.277 }_{- 0.28 }$& $ -1.23 ^{+ 3.836 }_{- 3.799 }$& $ -0.269 \pm 0.301  $ & $ 1.770 \pm 4.255 $\\
		
		H15&$ -0.566 ^{+ 0.031 }_{- 0.032 }$& $ 5.925 ^{+ 0.434 }_{- 0.422 }$& $ -0.585 \pm 0.033  $ & $ 6.186 \pm 0.454 $\\
		G11&$ -0.57 ^{+ 0.032 }_{- 0.033 }$& $ 5.966 ^{+ 0.447 }_{- 0.437 }$& $ -0.590 \pm 0.030  $ & $ 5.164 \pm0.415$\\
		DLB07&$ -0.59 ^{+ 0.026 }_{- 0.026 }$& $ 6.321 ^{+ 0.36 }_{- 0.353 }$& $ -0.600 \pm 0.030  $ & $ 6.434 \pm 0.365 $\\
		B06&$ -0.67 ^{+ 0.048 }_{- 0.049 }$& $ 7.133 ^{+ 0.665 }_{- 0.652 }$& $ -0.709 \pm 0.048  $ & $ 7.669 \pm 0.653 $\\
		\hline \\
		\textbf{S-V}\\
		Obs &$ +0.087 ^{+ 0.282 }_{- 0.403 }$& $ -3.291 ^{+ 5.542 }_{- 3.883 }$ & $+ 0.230 \pm 0.570 $ & $ -5.252 \pm 7.860$\\
		
		H15&$ -0.515 ^{+ 0.037 }_{- 0.038 }$& $ 5.234 ^{+ 0.518 }_{- 0.506 }$&$ -0.521 \pm 0.041 $ & $ 5.318 \pm 0.554$ \\
		
		G11&$ -0.515 ^{+ 0.04 }_{- 0.04 }$& $ 5.239 ^{+ 0.541 }_{- 0.548 }$& $ -0.499 \pm 0.043 $ & $ 4.997 \pm 0.595 $\\
		DLB07&$ -0.598 ^{+ 0.039 }_{- 0.04 }$& $ 6.384 ^{+ 0.54 }_{- 0.54 }$ & $ -0.599 \pm 0.040 $ & $ 6.406 \pm 0.541$\\
		B06&$ -0.537 ^{+ 0.068 }_{- 0.07 }$& $ 5.232 ^{+ 0.951 }_{- 0.939 }$& $ -0.578 \pm 0.072 $ & $ 5.800 \pm 0.978 $\\
		\hline
		\hline
		
	\end{tabular}
	\label{fbm200}
\end{table*}
   \begin{figure*}
   	\includegraphics[width=0.495\textwidth]{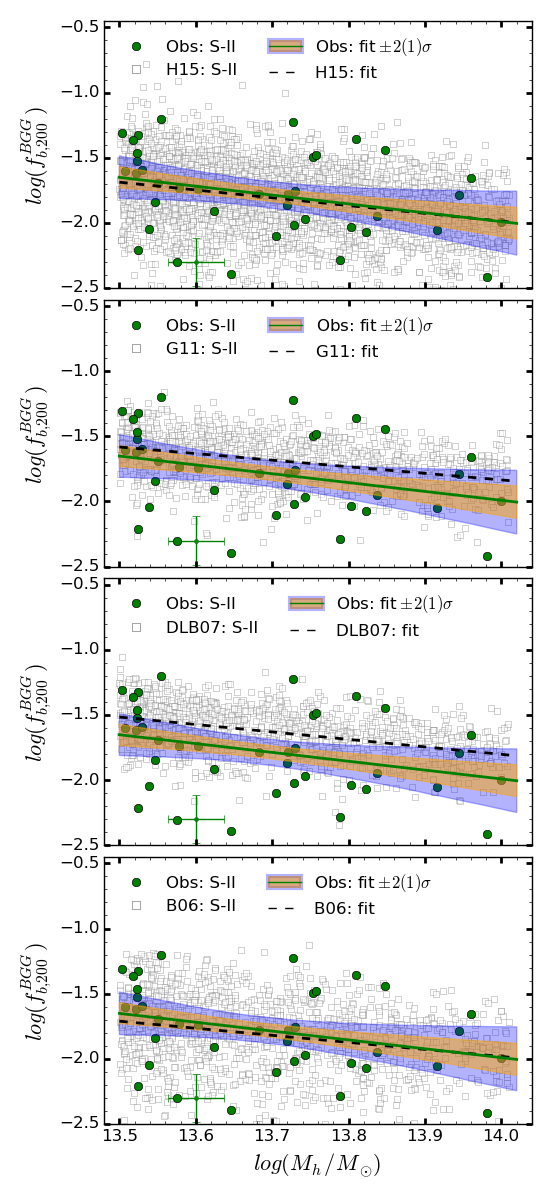}  \includegraphics[width=0.495\textwidth]{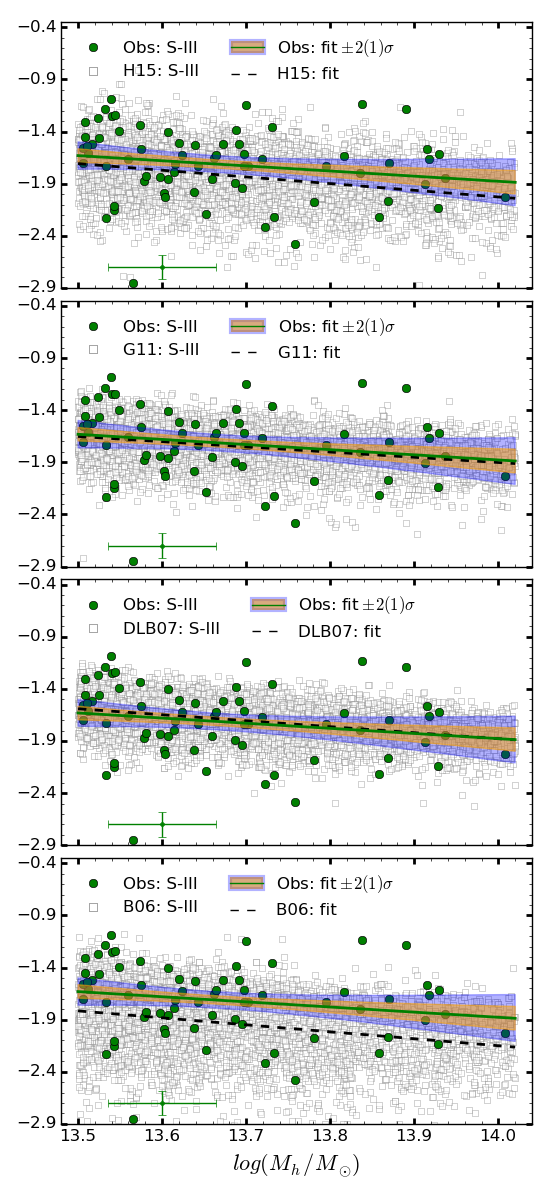} 
   	
   	\caption[]{ The relative contribution of the BGGs to the total baryonic mass 
   	of hosting groups, $log(f_{200}^{BGG})$, as a function of halo mass 
   	($log( M_{h}/M_\odot) $). Same as in Fig. \ref{fb_m200_1} but for S-II 
   	(left column) and S-III (right column), respectively.}
   	\label{fb_m200_23}
   \end{figure*}
   \begin{figure*}
   	\includegraphics[width=0.495\textwidth]{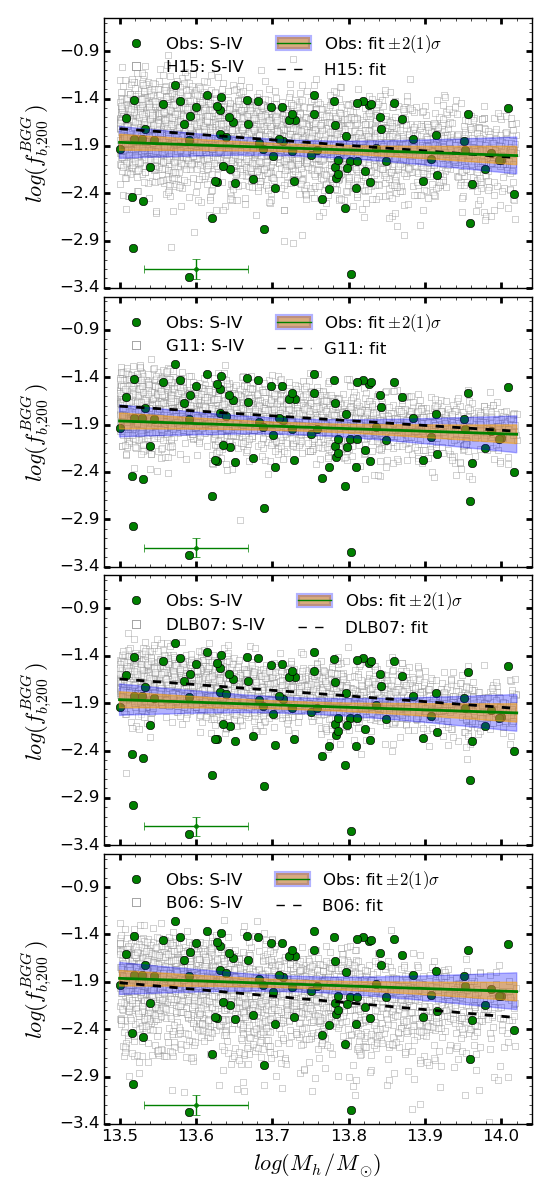}  \includegraphics[width=0.495\textwidth]{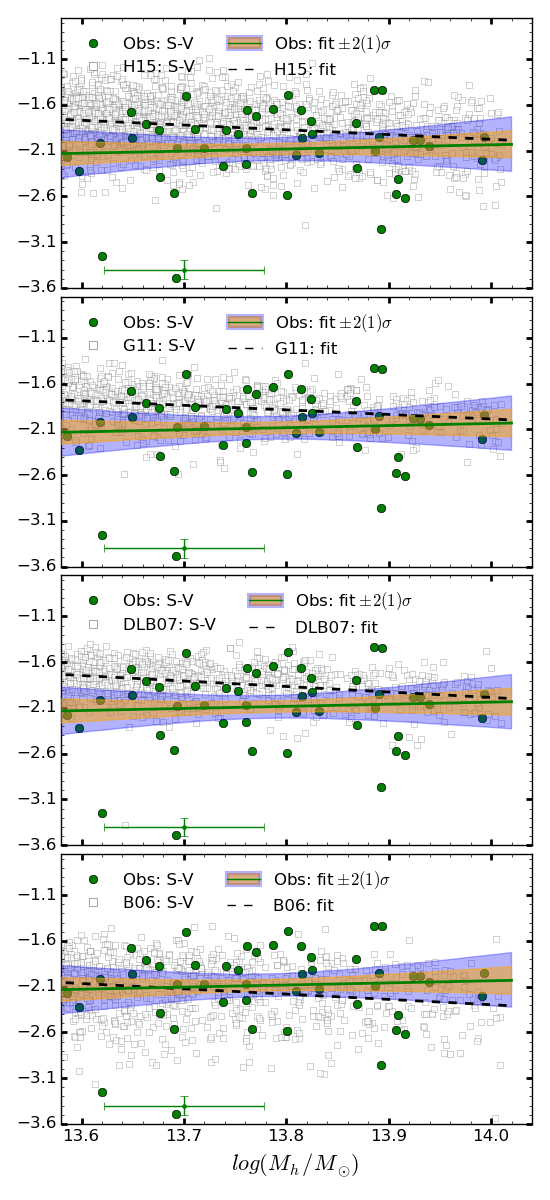} 
   	
   	\caption[]{ The relative contribution of the BGGs to the total baryonic mass 
   	of hosting groups, $log(f_{200}^{BGG})$, as a function of halo mass 
   	($log( M_{h}/M_\odot) $). The same as in Fig. \ref{fb_m200_1} but for 
   	S-IV 
   	(left column) and S-V (right column), respectively.}
   	\label{fb_m200_45}
   \end{figure*}
\section{Results}
\subsection{The halo mass dependence of the stellar baryon fractions contained in the BGGs}\label{s3.1}
Galaxy clusters/groups  are large enough to represent the mean matter distribution of the Universe \citep{White93}, thus the ratio of their total baryonic mass (stars, stellar remnants, and gas; $ M_b $) to their total mass  including dark matter ($ M_{tot}$) is expected to match the ratio of $ \Omega_{b}$ to $\Omega_{m} $ for the universe 
\begin{equation}
f_{b}=\frac{M_{b}}{M_{tot}}=\frac{\Omega_{b}}{\Omega_{m}}.
\label{fb1} 
\end{equation}

We use cosmological parameters from the full-mission Planck satellite observations of temperature and polarisation anisotropies of the cosmic microwave
background radiation  (CMB) and Baryon Acoustic Oscillations (BAO) and estimate the observed baryon fraction of the universe ($
\Omega_{b}h^{-2}=0.02226\pm0.00023 , \Omega_{m}h^{-2}=0.1415\pm 0.0019$, $ f_b=0.1573\pm0.0037$) \citep{Planck15} and determine the total baryonic mass of each galaxy group within $ R_{200}$ as $M_{b,200}=f_{b}M_{200}$. 

In order to quantify the contribution of the BGG stellar component to the total group baryons
in observations and SAMs, we estimate the ratio of the stellar mass of 
BGGs to the total baryonic mass of hosting groups as follows  
\begin{equation} 
 f_{b,200}^{BGG}=\dfrac{M_{\star}^{BGG}}{M_{b,200}}=(\frac{\Omega_{b}}
 {\Omega_{m}})^{-1}(\frac{M^{BGG}_*}{M_{200}}), 
 \label{fb3}
\end{equation}
where $ f_{b,200}^{BGG}$ defines the fraction of the total baryon of a group 
within $ r\sim R_{200}$ contained in stars of the BGG. In \S \ref{s3.1} to 
\ref{s3.3}, we only compare our results with predictions from four SAMs 
\citep{Bower06,deLucia07,Guo11,Henriques15}. The halo mass of groups  ($ M_h $) 
in all these models correspond to  $ M_{200c}$, we thus prefer to also use the 
$  
M_{200c} $ of groups in observations.  However, we convert $  M_{200c} $ to $  
M_{200m} $ in the rest of results presented in \S \ref{SHM} to \ref{shmr}.

In Fig. \ref{fb_m200_1} to \ref{fb_m200_45} , we  focus on the halo mass 
dependency of the $ 
f_{b,200}^{BGG}$. The data associated with the 
observed sample are shown as filled green circles while the data for the SAMs as 
shown with open grey squares. Panels from top to bottom compare observations 
with the SAM 
of H15, G11, DLB07, and B06, respectively.  We approximate the observed and 
predicted data using a power law relation \citep[e.g.,][]{Giodini09} given by
\begin{equation}
f_{b,200}^{BGG}=\beta \times (\dfrac{M_{h}}{M_\odot})^{\alpha},
\end{equation} \label{power_law}
where $ \alpha $ and $ \beta $ present the power low exponent and  constant of 
the $ f_{b,200}^{BGG}-M_{h} $ relationship. We take advantage of the properties 
of logarithms and convert this relation into a linear relationship given by 

\begin{equation}
log\;(f_{b,200}^{BGG})=log\;(\beta) + \alpha\times log\; 
(\dfrac{M_{h}}{{M_\odot}}),
\end{equation} \label{log_law}
we fit this equation to data and quantify the best-fit and optimised 
parameters by the Linear Least Squares approach and the  Markov chain Monte 
Carlo (MCMC) method in Tab. \ref{fbm200}. Since the fitted parameters by two 
methods are comparable in some cases, we decide to report both parameters in 
this table. The first column of Tab. \ref{fbm200} presents the subsample 
ID. The second and third columns present $ \alpha_{MCMC} $ and $ log\; 
(\beta_{MCMC}) $ with corresponding 68\% confidence interval, respectively. The 
$ 4^{th} $ and $ 5^{th} $ columns report $ \alpha_{LSS} $ and $ log\; 
(\beta_{LLS}) $ with $ \pm 1 \sigma $ uncertainties, respectively. 

In Fig. \ref{fb_m200_1} to \ref{fb_m200_45}, the solid green and dashed black lines illustrate the LLS best-fit relations in observations and SAMs, respectively. Our major findings are as follows:
\begin{enumerate}
\item  For S-I (Fig. \ref{fb_m200_1}), we find that $log(f_{b,200}^{BGG})$ 
shows no significant dependence on $ log(M_{h}/M_\odot)$, while 
models show that $log(f_{b,200}^{BGG})$ decreases with increasing $ 
log(M_{h}/M_\odot)$. We note that the best-fit relation to the observational 
data might 
be affected due to insufficient number of the low-mass haloes at $ 
log(M_{h}/M_\odot)<13.15$. Beyond this mass, observations and models become 
consistent. 
	
\item For S-II (Fig. \ref{fb_m200_23}, left), we find that 
 $log(f_{b,200}^{BGG})$ decreases  as a function of increasing $ 
log(M_{h}/M_\odot)$, in a good agreement with all model predictions within $\pm 
2\sigma $  errors. Among the models, H15 and B06 are more successful in predicting the observed 
trend.

\item For S-III (Fig. \ref{fb_m200_23}, right), 
$log(f_{b,200}^{BGG})$ also decreases  as a function of increasing $ 
log(M_{h}/M_\odot)$. The models are consistent with the observed trend within 
$\pm 2\sigma$ errors. It 
appears that B06 underestimates  $log(f_{b,200}^{BGG})$ at a given halo mass. 
	
\item For S-IV (Fig. \ref{fb_m200_45}, left), we find that 
$log(f_{b,200}^{BGG})$   decreases slowly as a function of increasing $ 
log(M_{h}/M_\odot)$, in agreement with models within $\pm 2\sigma$ 
uncertainties.  

\item For S-V (Fig. \ref{fb_m200_45}, right), we find that 
$log(f_{b,200}^{BGG})$  increases slowly as a function of increasing $ 
log(M_{h}/M_\odot)$, which is in contrast with most of the model predictions.	
\end{enumerate}

 In summary,  the observed $log(f_{b,200}^{BGG})$ is found to decrease as a 
 function of 
 increasing $ log(M_{h}/M_\odot)$ a  trend which is mildly redshift dependent.  
 The models predict a similar trend but with no significant
 dependence on redshift. 
 
 At $ z<0.4$, we find that $log(f_{b,200}^{BGG})$ - $ log(M_{h}/M_\odot)$ relation within haloes with $ M_h<10^{13.5}\; M_{\odot}$ show an opposite trend compared to the trend within massive haloes.

SAMs generally reproduce the observed $ f_{b,200}^{BGG}-M_{200}$ relation  of 
BGGs for S-II to S-V within the uncertainty, however, they fail to adequately 
predict this relation for S-I.
 
    \begin{figure*}
	\includegraphics[width=0.88\textwidth]{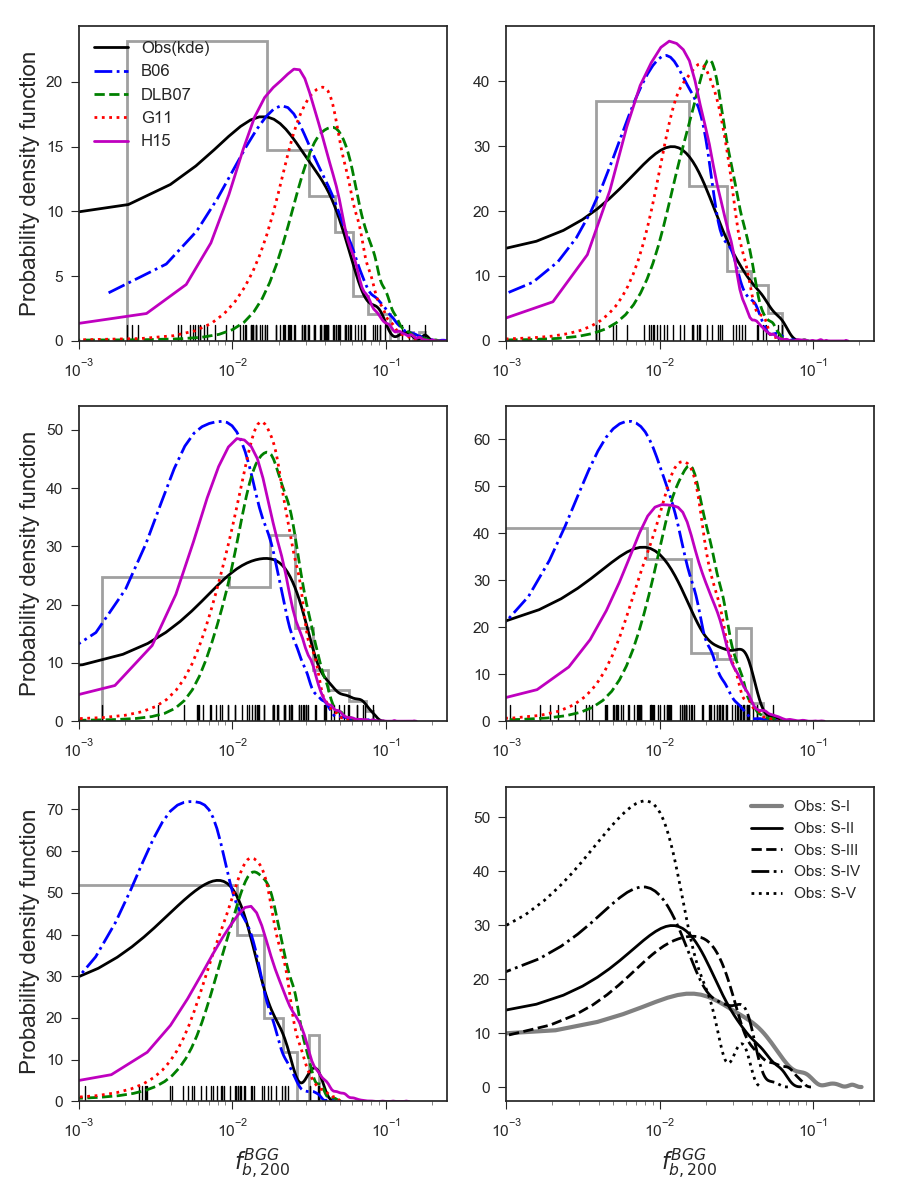}
	\caption[]{ Distribution of $f_{200}^{BGG}$ in observations (solid 
	grey 
	histogram). The smoothed observed distribution (KDE) (solid black line) is 
	compared to those from 	SAMs of G11 (dotted red line), DLB07 (dashed green 
	line), B06 (dashed-dotted 
	line), and H15 (solid magenta line), respectively. The lower right panel 
	compares the observed KDE functions among S-I to S-V and that of the whole 
	sample of 407 BGGs in observations (solid cyan distribution). The large 
	black ticks (rugs) present the observed one-dimensional density along 
	$f_{200}^{BGG}$-axis. }
	\label{fb}
\end{figure*}
\subsection {Distribution of $ f_{b,200}^{BGG}$ }\label{s3.3}

In Fig. \ref{fb}, we  present the distribution of $ f_{b,200}^{BGG}$ 
(grey histogram). To compare the observations to models, we use the Kernel Density 
Estimation (KDE) technique \citep{rosenblatt1956} and determine 
 the smoothed distribution of $f_{b,200}^{BGG}$ in observations (solid black 
line) and the SAMs of H15 (solid magenta line), G11 (dotted red line), DLB07 ( 
dashed green line), and B06 (dash-dotted blue line). The y-axes in this 
distribution display the probability density function and they are normalised 
in 
a way that the area under the curves is unity. The lower right panel present 
the 
smoothed distribution of $ f_{b,200}^{BGG}$ for S-I to S-V and the full sample 
of BGGs in observations. The rest of panels illustrate the results in 
observations and models for S-I to S-V, separately. Our main findings in each 
panel of Fig. 
\ref{fb} are as follows:

\begin{enumerate}
	\item For S-I,  the $ f_{b,200}^{BGG} $ distribution spans between $ 
	\sim $ 
	0.002 and 0.18. All SAMs overestimate  the position of the  centre of the 
	peak (mean value) in the observed 
	distribution. Among them, H15 and B06 predictions are closer to 
	the observations.
	
	\item For S-II,  the  $ f_{b,200}^{BGG} $ distribution extends over 
	0.004 to 
	0.06. All the models overestimate the height of the peak. However, H15 and 
	B06 
	 predict correctly the position of the centre of the observed peak.
	
	\item For S-III, the $f_{b,200}^{BGG} $ distribution  ranges between 
	0.001 
	and 0.08. All models overestimate  the height  
	of the peak in the observed distribution. H15 and B06 underestimate
	the position of the centre of the  peak. It appears that models also 
	underestimate the observed probability distribution function at high $ 
	f_{b,200}^{BGG} $.
	
	\item The $f_{b,200}^{BGG}$ distribution for S-IV spans 
	between 0.0005 and 
	0.055. The height of the peak is overestimated by all the models. Among 
	models, 
	H15 better predicts the observations. 
	
	\item The $f_{b,200}^{BGG}$ distribution for S-V extends from 0.0003 
	to 
	0.037. DLB07, G11, and H15 all over-predict the centre of the peak, while  
	the B06  prediction is in a good agreement with observations.	
	\item The lower right panel of Fig. \ref{fb} compares the 
	smoothed 
distribution of $f_{b,200}^{BGG}$ in observations for S-I to S-V. We find that 
the position of 
the centre of the peak tends to move to lower values with increasing redshift. 
In addition, we observe that the height of the peak also increases with 
increasing redshifts of BGGs. The significant changes in the distribution of 
$f_{b,200}^{BGG}$  occur at $ z\sim0.7 $. This comparison shows that 
$f_{b,200}^{BGG}$ evolves with redshift. As a result, the fraction of BGGs 
that  
contribute strongly to the total baryon budget of hosting groups increases with 
decreasing redshift, suggesting that BGGs may grow considerably in stellar mass 
at $ z<1.3 $.

In addition, we find that the $f_{b,200}^{BGG}$ 
distribution for S-I skews 
more to higher values along the x-axes compared to BGGs within massive groups. 
This 
	indicates that the central galaxies in the low-mass haloes contribute 
	strongly to 
	the total baryonic content of haloes.
\end{enumerate}

 \subsection{Evolution of SFR, $ M_{\star}$, and $ f_{b,200}^{BGG}$ of BGGs 
 }\label{s3.2}
 
 In Fig. \ref{fb}, we find evidence for the redshift evolution of the 
$ 
 f_{b,200}^{BGG}$ distribution. To understand the origin of this evolution, we investigate   the 
 stellar mass, SFR, and $ log(f_{b,200}^{BGG})$ of the BGGs for 
  S-II to S-V  and S-IHMA. We described both data sets in 
 detail in \S \ref{sample}. We determine the mean and median  values of $ 
 log(SFR/M_\odot yr^{-1}) $ (hereafter $ log(SFR)$) , $log(M_{\star}/M_\odot)$ 
 (hereafter $log(M_{\star}$)), and $log(f_{200}^{BGG})$ for S-II to S-V. We 
 also measured these quantities for S-IHMA within 5 redshift bins.  
 
\begin{figure*}

 	 \includegraphics[width=0.495\textwidth]{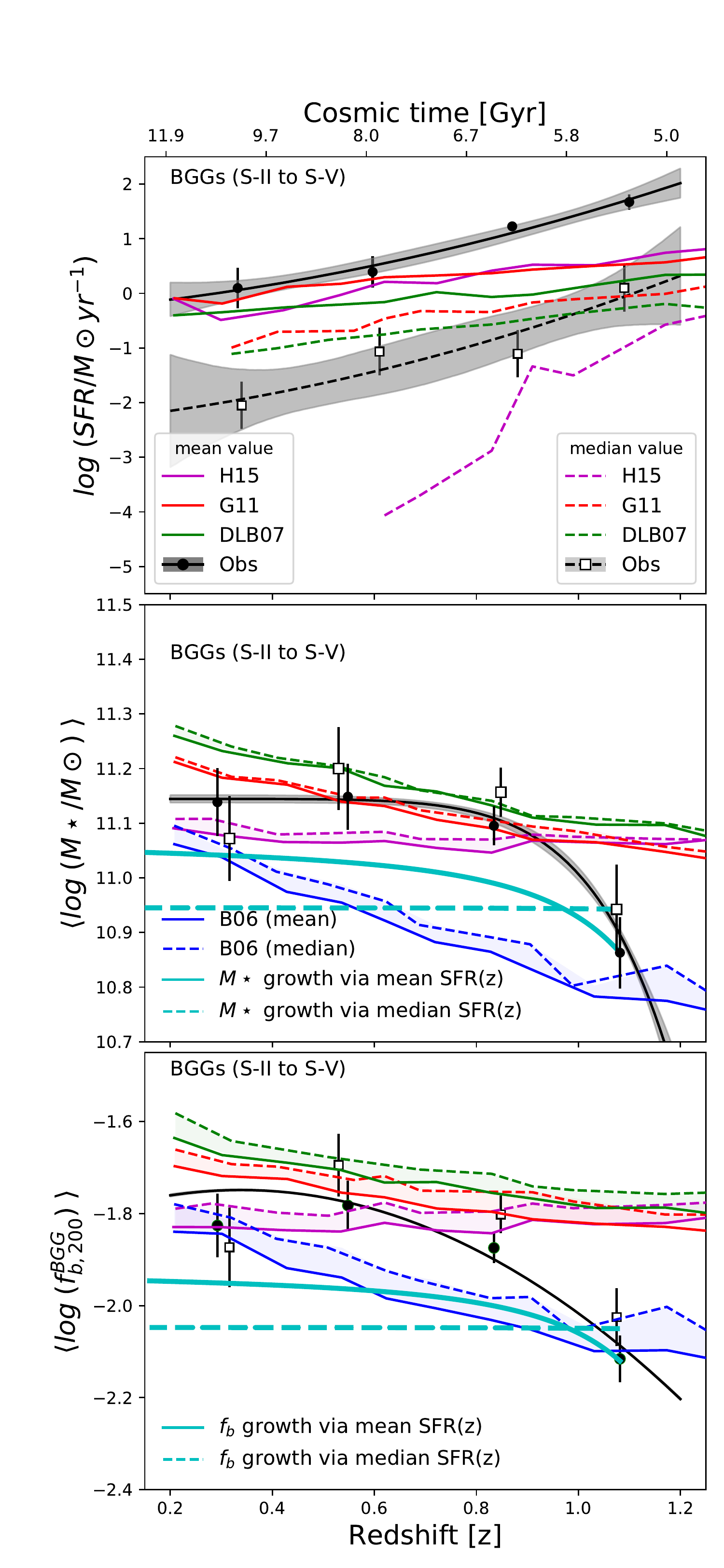}
 	  	 \includegraphics[width=0.495\textwidth]{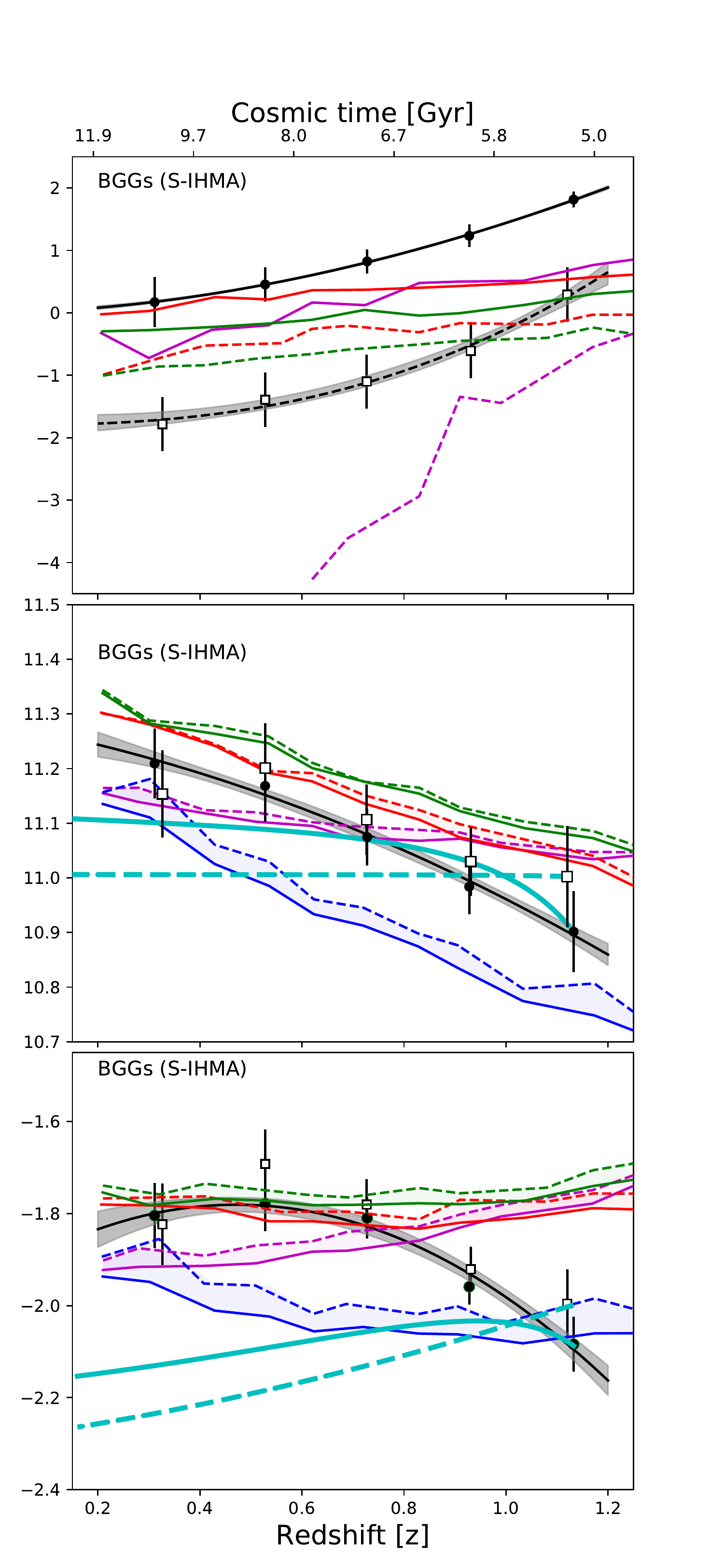}

 	\caption[] {{\textit(Upper panels)}  the mean (median) of $ 
 	log(SFR/M_\odot yr^{-1}) $ as a function of redshift
 	for S-II to S-V (left panel) and  S-IHMA (right panel). The solid (dashed) 
 	black lines show the best-fit function to the mean (median) of $ 
 	log(SFR/M_\odot yr^{-1}) $ versus redshift in observations. The mean 
 	(median) 
 	of $ 
 	log(SFR/M_\odot yr^{-1})-z$ relations in predictions by H15, G11, and DLB07 
 	are shown with solid (dashed) magenta, red, and green lines, respectively.
 	\textit{(Middle panels)} same as in the upper panel but for
 	$log\;(M_{\star}/M\odot)$ as a function of redshift. We highlight the area 
 	between the mean and median trends in model predictions. The cyan solid 
 	(dashed) line shows the $M_{\star}$  growths for typical BGGs 
 	through the mean (median) SFR-z relations with initial stellar masses which 
 	correspond to the mean (median) of stellar mass of BGGs for S-V.  {\textit 
 	(Lower panels)} the average/median value of $ 
 	log(f_{200}^{BGG})$  versus redshift for S-II to S-V (left panel) 
 	and S-IHMA (right panel). Same as in the middle panel the shaded regions 
 	present the model predictions.  The solid (dashed) cyan 
 	line shows the  growth of $log(f_{200}^{BGG})$ through  star-formation for 
 	two typical BGGs 
 	with initial baryon fractions which correspond to the mean (median) value 
 	of $log( f_{200}^{BGG})$ for S-V. Stars in these two typical BGGs are formed 
 	in time with mean (median)
 	rate which we approximate  in the upper panel.}
 	\label{fb_sfr_ms}
 \end{figure*}

  We note that the halo mass growth is taken into account in the sample 
  selection for S-IHMA, while this effect is not considered in the BGG 
  selection for S-II to S-V.  The left and right columns of Fig. 
  \ref{fb_sfr_ms} 
 show  results for S-II to S-V and S-IHMA, respectively.

To gain further knowledge on the relative contribution of the BGG 
stars to the total baryon content of groups, we repeat the same computation for 
20\% of the BGGs with the highest $ f_{200}^{BGG}$ in each redshift bin for 
both S-II to S-V and S-IHMA, as presented in Fig. \ref{fb_sfr_ms20}.

The upper, middle and bottom panels of Fig. \ref{fb_sfr_ms} and Fig. 
\ref{fb_sfr_ms20} show the mean and median values of $ 
log(SFR/M_\odot yr^{-1}) $ (hereafter $ log(SFR)$) , $log(M_{\star}/M_\odot)$ 
(hereafter $log(M_{\star}$)), and $log(f_{200}^{BGG})$ 
versus redshift. We describe these relationships in \S 
\ref{sfr-z} to \S \ref{fb-z}, respectively.
 \subsubsection{The SFR evolution of BGGs} 
 \label{sfr-z}
 
 In \cite{gozaliasl2016brightest}, we have shown that $ 
\sim25\% $ and $ \sim60\% $ 
of the BGGs lie on the main sequence of the star forming galaxies at $ 
0.4<z<0.5 
$ and $ 1<z<1.3 $, respectively.  The SFRs of these galaxies can reach up to 
$ \sim $ 100 $ \times $ solar 
mass per year, in a good agreement with a similar study on 90 BCGs selected 
from the X-ray galaxy clusters in the 2500-square-degree South Pole Telescope 
survey  
by 
\cite{mcdonald2016star}. They have also found that the SFR of 34\% of the BCGs 
exceeds $ 10 M_\odot yr^{-1} $ at $ 0.25 <z<1.25 $ and this fraction rises up 
to $ 92^{+6}_{-31}$\% at $ z>1$. 

In this paper, we study the redshift evolution of the 
mean (median)  $log (SFR)$ of  the BGG to find out what fraction of the $ 
M_{\star} $  
and  $ f^{BGG}_{b,200}$  growth at  $z<1.3$ could be driven by  
the star formation.  
 
The upper panel in Fig. \ref{fb_sfr_ms} illustrates mean (median) 
 $log(SFR)$ as a function of redshift for BGG within S-II to 
 S-V (left) and BGG within S-IHMA (right). The filled/open black points show 
the  data in the observations. The solid (dashed) green, red, and magenta lines 
indicate 
 the mean (median) $log(SFR)-z$  relations in the SAMs by DLB07, G11, and H15, 
 respectively. We find that the BGG SFRs in both observations and models follow 
 a bimodal distribution, implying that the mean SFR deviates significantly 
 from the median SFR. In addition, this indicates that BGGs consist of two 
 distinct populations of quiescent and star forming galaxies.

The data in observations shows a rapid evolution at $ z>0.7$. This 
causes the observed trend over $ 0.1<z<1.3$ to deviate from a linear 
form. Thus, 
 we use two non-linear relations such as
 
 \begin{equation}
f(z)=\alpha\times z^{\beta} + \gamma,
 \end{equation}
   or the quadratic polynomial 
 \begin{equation}
f(z)=a+b\times z+ c\times z^2 ,
 \end{equation} 
to reproduce the trend in the observations. We fit both relation to the data 
and select the best one. We also note that 
these best-fit relations may not be valid beyond the redshift range of our BGG 
 sample ($0.1<z<1.3$). The highlighted grey area represents $ \pm1 \sigma $ 
 confidence intervals.
 
Just as in Fig.\ref{fb_sfr_ms}, we measure the redshift evolution of 
 mean (median) $ log (SFR) $  for 20\% of BGGs with the highest $ 
 f_{b,200}^{BGG} 
 $ for S-II to S-V and S-IHMA in the upper panel of Fig. \ref{fb_sfr_ms20}. 
 Tab.  \ref{sfr_tb} presents the best-fit $log(SFR)-z $ relations.
  
We summarise the observed and predicted $ log(SFR)-z$ relations as 
follows:
\begin{enumerate}
\item The mean (median) $ log(SFR)$ of BGGs for both S-II to S-V and 
S-IHMA 
decrease considerably with cosmic time by $\sim2 $ dex since $z=1.3$. A 
significant decline occurs at $ z>0.7$. Including the halo mass growth in  
sample selection has no 
considerable impact on the evolution of star formation in BGGs at $ z<1.3$. 

The mean $ log(SFR)$ of 20\% of BGGs with the highest  $ 
f_{b,200}^{BGG}$ for 
both S-II to S-V and S-IHMA also
decrease in time by $\sim2 $ and $ 4 $ dex since $z=1.3$ to the present day, 
respectively. It appears
that accounting for the halo mass growth in sample selection causes the   
SFR of BGGs 
with the highest $ f_{b,200}^{BGG}$ to decline more efficiently.

 The median $log(SFR)$ of BGGs (20\%) with the highest  $ 
 f_{b,200}^{BGG}$ show no significant changes with redshift within the
 uncertainties at $ 0.2<z<1.2$. 
 The trend for S-IHMA shows a decrease by $\sim$2 dex from z=1.2 to z=0.7 
 followed by an increase of 
 about 1 dex at low redshifts.

\item All the SAMs consistently predict that mean (median) $ log(SFR) $ 
decreases  with  redshift by around $0.5-1$ dex. At $z>0.6$, they underestimate the 
 observed mean  $ log(SFR)$ up to $\sim1.5$ dex. At lower redshift, their 
 predictions are closer to the observations. The halo mass growth  has no 
 considerable effect on the $log(SFR)-z $ relation in model predictions. Except 
 for H15, all the models make a similar prediction for the redshift evolution 
 of the BGG SFR. 
 In the H15 model, the median $ log(SFR) $ begins to drop significantly at 
 $z=0.6$. The trends of 
the mean (median) $log(SFR)-z $ relations in models for the BGGs (20\%) with 
the highest 
$ f_{b,200}^{BGG}$ do not deviate significantly to those of the full BGG 
sample. 

\item  In agreement with the models, mean $log(SFR)$ is higher than the 
median $log(SFR)$ at a given redshift by at least $\sim1$ dex. This finding 
shows that 
BGGs generally consist of two distinct types of galaxies: dead/quiescent objects
with no  star-formation activity and  main-sequence/star-forming galaxies. 
Thus, the mean SFR of BGGs is influenced by outliers having either very low or 
very high 
values of SFRs. It suggests that the mean value of the BGG SFR could not 
be a typical indicator of star formation activity for all BGGs. 
 
\end{enumerate}

\begin{table*}
\caption{We summarise the relation of mean (median) value of $ log(SFR/M\odot 
yr^{-1}) $, $log(M\star/M_\odot$), and  
$log(f_{b,200}^{BGG}$) with redshift(z) in observations (as illustrated in Fig. 
\ref{fb_sfr_ms} and Fig. \ref{fb_sfr_ms20}) using Eq. 
(5):$f(z)=\alpha\times z^{\beta} + \gamma$ or 
Eq.(6):$ 
f(z)=a+b\times z+ c\times z^2$. We have reported the best-fit relations for 
the full sample of BGGs and 20\% of BGGs with the highest f$_{b,200}^{BGG}$.  
Since the mean and median values of $log(M\star/M_\odot$), and  
$log(f_{b,200}^{BGG})$ are consistent within errors at fixed redshift, we just 
present the relation 
between the mean value of these quantities and redshift. The first column 
presents the relation and 
subsample. The columns from 2 to 7 list the  best-fit parameters.} 
\centering
\begin{tabular}{llllllll}
\hline\hline \\
Relation (sample)& a&b&c& $ \alpha $ & $\beta$ & $\gamma$\\
\hline \\
\textbf{All BGG sample}\\ 
Mean log(SFR)-z (S-II to S-V)  & -2.38 $\pm$ 2.11 & 0.9 $\pm$ 6.6 & 1.12 $\pm$ 
4.58& -&-&-\\
Median log(SFR)-z (S-II to S-V) & -0.31 $\pm$ 0.63 & 0.81 $\pm$ 1.98 & 0.95 
$\pm$ 1.37&-&-&-\\
Mean log(SFR)-z (S-IHMA)&- & -& -& 2.56$\pm $0.56&1.54$\pm $ 0.19&-1.8 $ \pm $ 
0.16\\
Median log(SFR)-z (S-IHMA)&- & -& 
-&1.83$\pm$0.1&1.43$\pm$0.04&0.01$\pm$0.04\\	\\

\textbf{20\% of BGGs with high $ f_{b,200}^{BGG} $}\\
Mean log(SFR)-z (S-II to S-V) & $-1.74 \pm 1.7$&$ 2.96 \pm 5.04$&$ -0.81 \pm 
3.26$&-&-&-\\ 
Mean log(SFR)-z (S-IHMA)&$-1.23\pm0.52$&$0.43\pm1.64$&$1.96\pm1.15$&-&-&-\\
			\hline   \hline  \\
\textbf{All BGG sample}\\ 		
Mean log(M$ \star $)-z (S-II to S-V)&-&-&-&7.55 $\pm$ 1.02 & -0.14 $\pm$ 0.02 & 
11.14 $\pm$ 0.01\\			
Mean log(M$\star$)-z 
(S-IHMA)&-&-&-&$1.54\pm0.19$&$-0.31\pm0.02$&$11.27\pm0.02$\\\\	
\textbf{20\% of BGGs with high $ f_{b,200}^{BGG} $}\\ 
Mean log(M$ \star $)-z (S-II to 
S-V)& 11.63 $ \pm $ 0.18& -0.37 $ \pm $ 0.54& 0.19 $ \pm $ 0.35&-&-&-\\  
Mean log(M$\star$)-z 
(S-IHMA) &$11.52\pm0.23$&$0.59\pm0.71$&$-0.78\pm0.5$&-&- & -\\
		\hline   \hline \\
\textbf{All BGG sample}\\  
Mean log(f$_{b,200}^{BGG}$)-z (S-II to S-V)& -1.82 $\pm$ 0.01 & 0.41 $\pm$ 0.02 
& -0.61 $\pm$ 0.01&-&-&-\\      
Mean log(f$_{b,200}^{BGG}$)-z (S-IHMA)&$-1.94\pm0.08$&$0.68\pm 
0.24$&$-0.72\pm0.16$&-&-&-\\ \\
\textbf{20\% of BGGs with the highest $ f_{b,200}^{BGG} 
$}\\ 
Mean log(f$_{b,200}^{BGG}$)-z (S-II to S-V)& -1.35 $ \pm $ 0.14& 0.08 $ \pm $ 
0.41& -0.21$  \pm $ 0.27&-&-&-\\        
Mean log(f$_{b,200}^{BGG}$)-z 
(S-IHMA)&$-1.68\pm0.15$&$1.48\pm0.47$&$-1.38\pm0.33$&-&-&-\\
\hline \hline 
		\end{tabular}
	\label{sfr_tb}
\end{table*}

\subsubsection{Stellar mass evolution of the BGGs } 
\label{ms-z}
The hierarchical structure formation theory predicts that the most
massive 
 galaxies (e.g., BCGs/BGGs) in the universe should form in the late epochs. 
 In a semi-analytic modelling of galaxy formation based on 
 the Millennium simulation \citep{springle2005cosmological}, 
 \cite{deLucia07} found that the BCG mass is assembled relatively late. They 
 obtain
 50\% of their final mass at z < 0.5 by galactic merging. Using a
 deep near-IR data of BCGs, \citep[e.g.,][]{collins2009,Stott10}
showed that BCGs exhibit little growth in stellar mass at $ z<1$. 
\cite{Lidman2012} 
 studied the  correlation between the stellar mass of BCGs and the halo mass of their 
 hosting X-ray clusters and found that the stellar mass of BCGs grows 
 by a factor of 1.8 since $ z=0.9 $ to $ z=0.2 $. Such contradictory 
 findings indicate that the detail of the stellar mass growth of BCGs/BGG still 
 remains an unresolved issue in galaxy formation.
  
\begin{figure*}
	
	\includegraphics[width=0.495\textwidth]{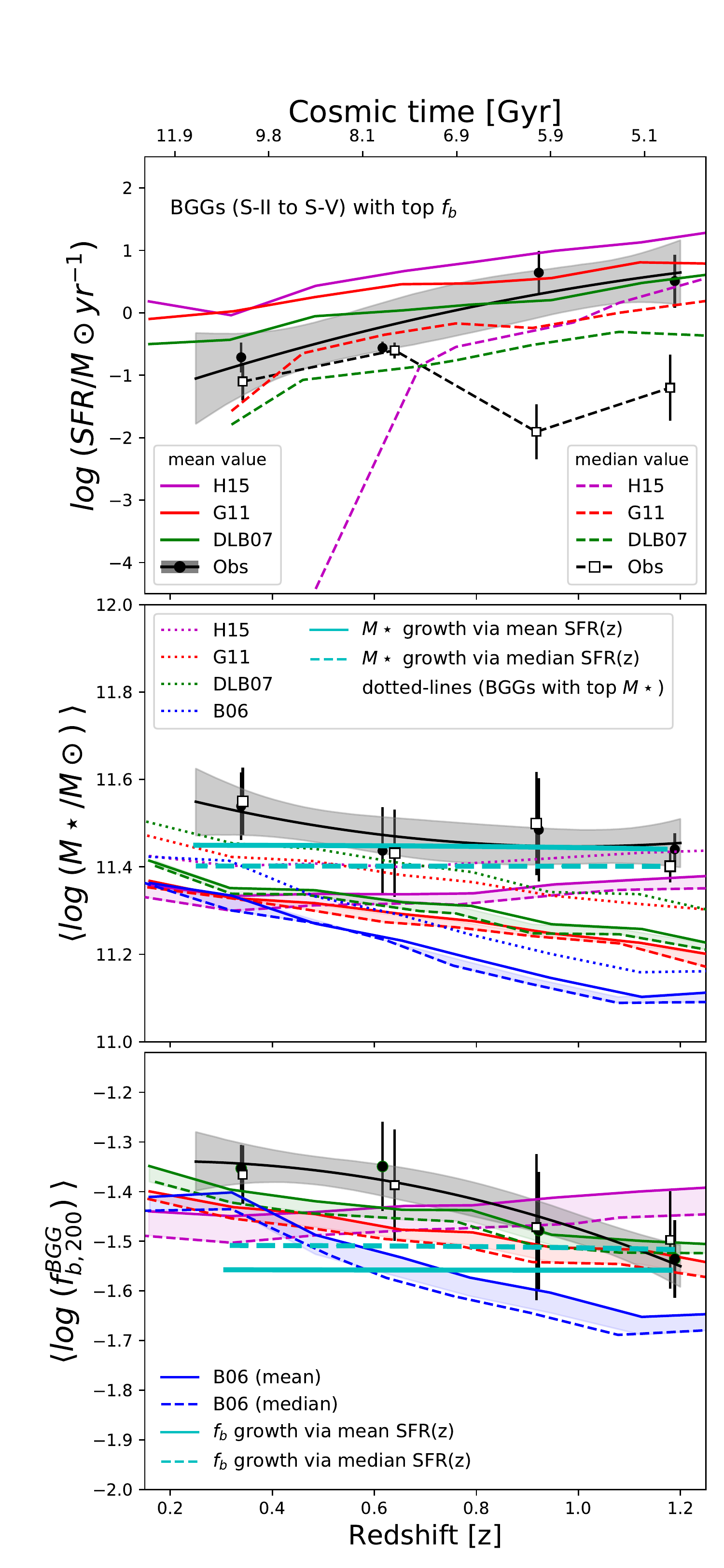}
	\includegraphics[width=0.495\textwidth]{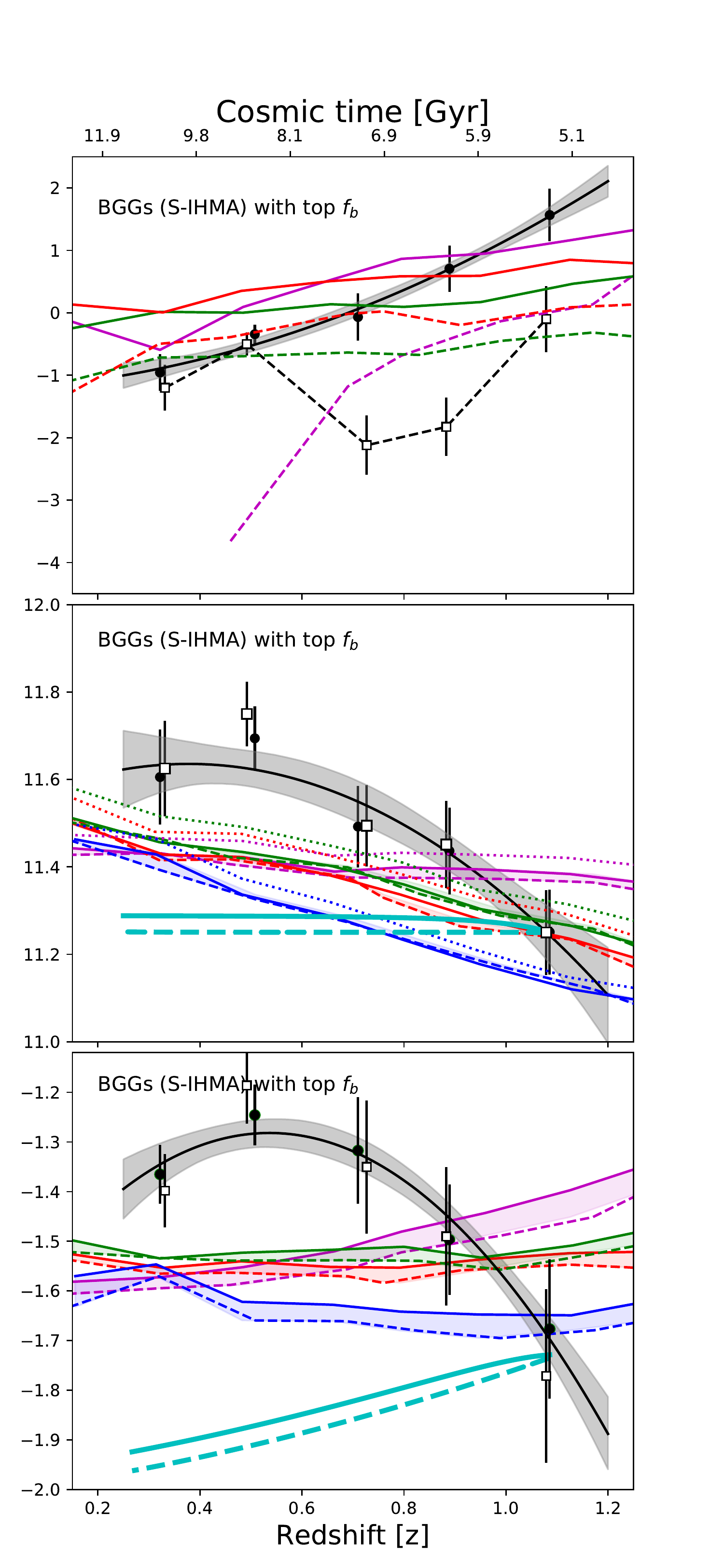}

	\caption[] {Same as in Fig. \ref{fb_sfr_ms} but for the 20\% of the BGGs 
	with 
	the 
	highest $ f_{200}^{BGG}$. In the middle left panel, the dotted magenta, red, 
	green, and blue lines represent the mean stellar mass of 20\% 
	of the BGG with highest $ M_{\star} $ in the H15, G11, DLB07, and B06 models as 
	a 
	function of redshift.}
	\label{fb_sfr_ms20}
\end{figure*}

In the middle panel of Fig. \ref{fb_sfr_ms}, we determine the 
mean (median) value 
of $log(M_\star)$ as a function of redshift for S-II to S-V (left panel) and 
S-IHMA (right panel). We  do
   the same computation for the BGGs (20\%) with the highest $ f_{b,200}^{BGG} 
   $, as shown in the middle panel of Fig. \ref{fb_sfr_ms20}.

 In the middle panel of Fig. \ref{fb_sfr_ms}, the observed trend 
 indicates that 
 the mean (median) stellar mass of BGGs changes with time in a 
 non-linear fashion (filled/open black points). We fit both non-linear 
equations of Eq. (5) and (6) to the relation between the mean stellar mass and 
redshift in observations and choose the best-fit, as presented in Tab. 
\ref{sfr_tb}. The highlighted gray area illustrates $ 
 \pm1 \sigma $ confidence intervals from the fit.
 
The observed mean $  log(SFR)-z $ relation shows that some of BGGs 
 have 
 relatively high-SFRs in particular at $ z>0.7$. To quantify a possible 
 contribution of star-formation in the stellar mass growth of BGGs, we 
 assume a typical galaxy with an initial mass  which equals the mean (median) 
 stellar mass of BGGs within S-V. We then allow this galaxy to form  
 stars at a rate permitted by the mean (median) $log(SFR)-z$
relations, as  approximated in the upper panel of Tab. \ref{sfr_tb}. The 
 solid (dashed) cyan lines represents the $ M_{\star}-z$ relation for this 
 typical galaxy.  
 
 We summarise our findings in the middle panels 
of Fig. \ref{fb_sfr_ms20} and Fig. \ref{fb_sfr_ms} as follows:
 \begin{enumerate}
 	\item We find that the mean and 
 	median 
 	stellar mass of BGGs at a given redshift for observations and model 
 	predictions  are the same and differences between two quantities lie within 
 	the respective errors.
 	
 	\item The mean (median) stellar mass of BGGs for S-II to S-V 
 	and S-IHMA  grows as a function of decreasing redshift from z=1.13 
 	(mean (median)  	redshift of BGGs for S-V) to z=0.31 (mean/median 
 	redshift of 
 	BGGs for S-II) by $ \sim0.31/0.32$ dex. While the net growth of the BGG 
 	stellar mass for S-II to S-V  and S-IHMA are similar, however, their trends 
 	with redshift differ. The  stellar mass of BGGs for S-IHMA  changes more 
 	linearly with redshift  compared to that of S-II to S-V.  We note that the 
 	significant growth of stellar mass of 
 	BGGs for S-II to S-V  occurs  at $ z>0.7$.

 	\item For 20\% of BGGs with the highest $ f_{b,200}^{BGG}$ 
 	for S-II to S-V, mean (median) $log(M_\star)$  grows
 	with cosmic time by 0.11 dex, while this growth for S-IHMA is 0.44 dex. This 
 	growth is 
 	$\sim 0.12$ dex  higher than that seen for the full BGG sample,  
 	indicating that accounting for the halo growth in the sample selection  
 	reveals that  BGGs with the highest $ 
 	f_{b,200}^{BGG}$ assemble more stellar mass by $\sim0.3$ dex, compared to 
 	mass-limited samples (S-II to S-V). 
 	  
 	\item In contrast to observations, the mean (median) $ M_{\star}$ of 
 	BGGs for 
 	S-II to 
 	S-V and S-IHMA in the G11 and DLB07 SAMs
 grow	slowly  in time by about 0.15 dex and 0.25 dex, respectively. The H15 
 model 
 	shows no 
   trend between  $ M_{\star}$ with redshift.  It is also evident that  the 
   Munich SAMs (G11, DLB07, and H15) overestimate the stellar mass of BGGs at $ 
   z>0.7 $.  The net growth of $ M_{\star}$ 
 	for S-II to S-V and S-IHM  in the 
 	B06 model is much closer to the observed trend compared to the other models. 
 	However, this model predicts a linear evolution for the BGG mass and 
 	underestimates $ M_{\star}$ at intermediate  redshifts.
 	 	
 We find that models underestimate the stellar mass for 20\% of BGGs 
 with 
 the highest $ 
  f_{b,200}^{BGG}$ for S-II to S-V at a given redshift. For S-IHMA, their 
  predictions are consistent with observations at z=1, but they underpredict  
  the observed stellar mass at lower redshifts. In the 
  middle panel of Fig. \ref{fb_sfr_ms20}, we have also shown that the evolution 
  of 
  mean stellar mass 
   for 20\% of BGGs with the highest $ M_{\star}$ in the SAM predictions by
  H15, 
  G11, DLB07, and B06 ( dotted magenta, red, green, and blue lines). All the 
  models 
  make predictions which lie under the observed stellar mass.            
 
    \item The Munich SAMs overestimate the BGG mass for the full sample 
    of BGGs 
    for S-V at $ z>0.8 $. On the other hand, they also underestimate the mean 
    SFR of BGGs for S-V. These findings suggest that the 
    star-formation quenching processes in models are more efficient than 
    they  needed to be in the  early epochs and the BGGs should have possibly 
    grown by 
    non-star forming mechanisms such as  dry merger or the tidal striping of 
    stars 
    from satellite galaxies.
    
     \item We find that  the mean $log(M_\star)$ of a typical BGG 
     which  
     obeys the mean $SFR-z $ relation can increase via star-formation by $ 
     \sim0.17$ dex since $ z\sim1.2 $, which may explain about 50\% of the mean 
     mass growth of a typical star forming BGG at $ 0.8<z<1.2 $. At this 
   period, the stellar mass growth for a 
     typical BGG which follows from the median $ SFR-z $ relation is negligible
     (see solid and dashed cyan lines). The mean (median) stellar 
     mass growth of a typical BGG with the highest $ f_{b,200}^{BGG}$ which follows 
     from the mean (median) $ SFR-z $ 
      relation for  20\% of BGGs with the highest $ 
     f_{b,200}^{BGG}$ is also not 
     significant, indicating that BGGs with the highest $ f_{b,200}^{BGG}$ 
     grow mainly in stellar mass through dry merging and tidal stripping.
        
     \item In observations, BGGs with the highest $ f_{b,200}^{BGG} $ 
     are 
     generally more massive than the BGGs with low $ f_{b,200}^{BGG}$ 
     at a fixed redshift by at least 0.25 dex. 
      	 \end{enumerate}

 \subsubsection{The evolution of $f_{200}^{BGG}$} \label{fb-z}  
  Similarly to computation for the stellar mass of BGGs, we 
  determined the 
  redshift evolution of the mean (median) value of $  log(f_{b,200}^{BGG})$ 
   of the full sample of BGGs and  20\% of BGGs with the highest $ 
  f_{b,200}^{BGG}$ for S-II to S-V (left panel) and S-IHMA (right panel) in the
  lower panel of Fig. \ref{fb_sfr_ms} and  Fig. \ref{fb_sfr_ms20}, 
  respectively. 
  
 We quantify what fraction of the growth of $ f^{BGG}_{b,200}$ 
 may be driven by star formation in BGGs (solid and dashed cyan lines). We also use Eq. 
  (5) and (6) and summarise the observed trend.  The best-fit parameters 
  presented in the lower 
  panel of Tab. \ref{sfr_tb}. We describe the observed and 
  predicted trends as follows:
 \begin{enumerate}
 	\item We find that $log(f_{b,200}^{BGG})$ of the entire sample of 
 	BGGs for S-II to S-V and S-IHMA increase as a function of decreasing 
 	redshift by 0.35 and 0.30 dex since $ z\sim1.13 $ (mean/media redshift of 
 	BGGs for S-V) to $ z\sim0.31 $ (the mean/median redshift of BGGs for 
 	S-II). 
 	These growths for 20\% of the BGGs with the highest $ f_{b,200}^{BGG}$ are 0.22 
 	and 
 	0.41 dex, respectively. As a result, the $log(f_{b,200}^{BGG}) $ growth of 
 	20\% of BGGs with the highest $ f_{b,200}^{BGG}$ for S-II to S-V is lower 
 	than that of 
 	the full sample of BGGs by 0.13 dex. However, this growth for 20\% of BGGs 
 	with the highest $ f_{b,200}^{BGG}$ for S-IHMA is 0.1 dex higher than that 
 	of the full sample of BGGs.

  \item It is evident that the observed growth in $ f_{b,200}^{BGG}$ 
  slows down 
  at $ z<0.5 $ possibly due to the star-formation quenching in the BGGs and small 
  evolution of the BGG mass.   
  
  \item For S-II to S-V, the  DLB07 and G11 models predict that  the $ 
  f_{b,200}^{BGG}$ increases slightly with cosmic time. H15 predicts no 
   $ log(f^{BGG}_{b,200})$  growth.  The Munich models 
  overestimate the  mean (median) of $ log(f_{b,200}^{BGG})$ at $ z>0.8 $. This 
  can be explained by overprediction of the BGG mass at $ z>0.8 $. Within the
  models, B06 predictions of the $ 
  f_{b,200}^{BGG}$ growth is closer to the observations. However, B06 
  underestimates $log(f^{BGG}_{b,200})$  at intermediate redshifts. In 
  addition, the Munich models show no significant growth of $ f_{b,200}^{BGG}$ 
  for 20\% BGGs with the highest $ f_{b,200}^{BGG}$. At $ z=0.5$, these models 
  underestimate observations.
  
  For S-IHMA, models estimate a flat trend between 
  $log(f^{BGG}_{b,200})$ and 
  redshift for both the 
  full sample of BGGs and 20\% of BGGs with the highest $ f_{b,200}^{BGG}$. 
  Within model predictions, B06 predicts  
  $log(f^{BGG}_{b,200})$ to grow a little with redshift. In addition, we 
  observe that accounting for the growth of halo mass in the comparison leads 
  to a 0.1 dex larger increase in  $log(f^{BGG}_{b,200})$.
  
 \item We find that $ \overline{f_{b,200}^{BGG}} $ of a typical BGG  
 which 
 grows in stellar mass via  star-formation (as summarized in Tab. 
 \ref{sfr_tb}) can increase with cosmic time by 0.18 dex (solid cyan line in 
 the lower left panel of Fig. \ref{fb_sfr_ms}) when assuming the halo mass 
 of hosting halo remains constant  at $ log(M_{200}/M_\odot)=13.8$ with 
 redshift. In contrast, once we assume this halo to grow in mass according to 
 the rates as illustrated in Fig. 1, the $ log(f_{b,200}^{BGG}) $ growth via  
 star-formation in the BGG (solid cyan line in the lower right panel of Fig. 
 \ref{fb_sfr_ms}) for S-IHMA becomes less effective at $ z\gtrsim0.8$. At 
 $ z\lesssim0.8 
 $, the halo mass growth rate becomes faster than the SFR.
  As a result,
 the growth of $ log(f_{b,200}^{BGG}) $ at low redshifts can be 
 explained by BGG growth mass through non-star forming process (e.g., dry 
 merger and tidal striping).  The $ log(f_{b,200}^{BGG}) 
 $ growth due to the  star-formation is negligible for both the full BGG 
 sample
 and 20\% of the BGGs with the highest  $  f_{b,200}^{BGG} $. 

 \end{enumerate} 
  \begin{figure*}
 	\includegraphics[width=0.75\linewidth, height=12cm,keepaspectratio]{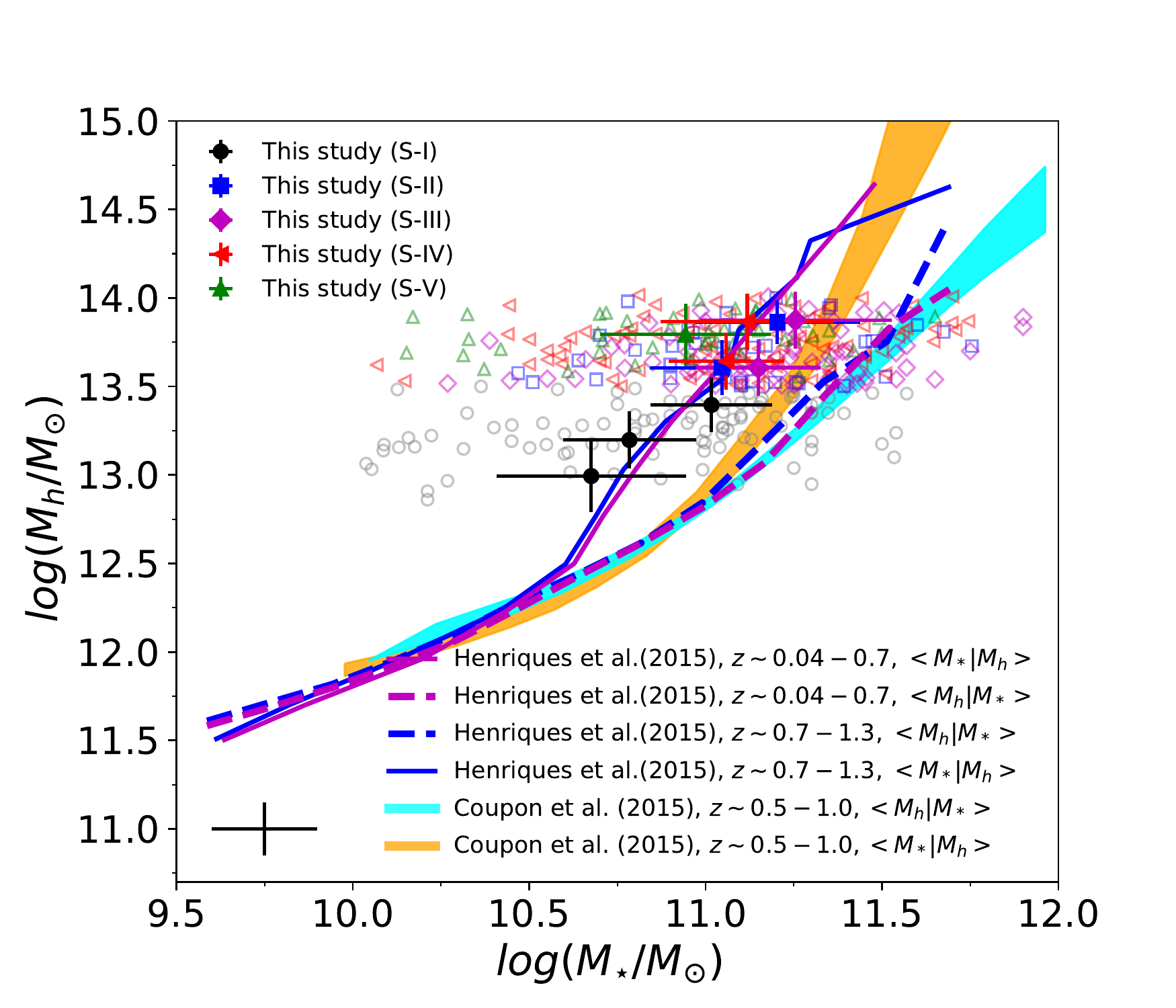}
 	\caption[]{ The $ M_{\star}-M_h $ relation for BGGs in observations, the H15 
 	model, and the result from \cite{Coupon15}. The open symbols show the 
 	observed data for S-I to S-V in different color and symbols. The filled data 
 	points  represent the average $ M_{\star}$ of BGGs in our data at fixed halo 
 	mass 
 	($\langle M_{\star}|M_h\rangle$). The solid magenta and blue lines show 
 	$\langle 
 	M_{\star}|M_h\rangle$ relations  in the SAM presented by H15 at $ 
 	0.04<z<0.7$ and 
 	$ 0.7<z<1.3$ . Similarly, $\langle M_h|M_{\star}\rangle$ relations in this 
 	model 
 	are plotted with dashed blue and magenta lines. The relations from 
 	\cite{Coupon15} results are plotted as orange ($\langle 
 	M_{\star}|M_h\rangle$) and 
 	cyan ($\langle M_h|M_{\star}\rangle$) area. }
 	\label{ms_mh1}
 \end{figure*}
\begin{figure*}
	\includegraphics[width=0.8\textwidth]{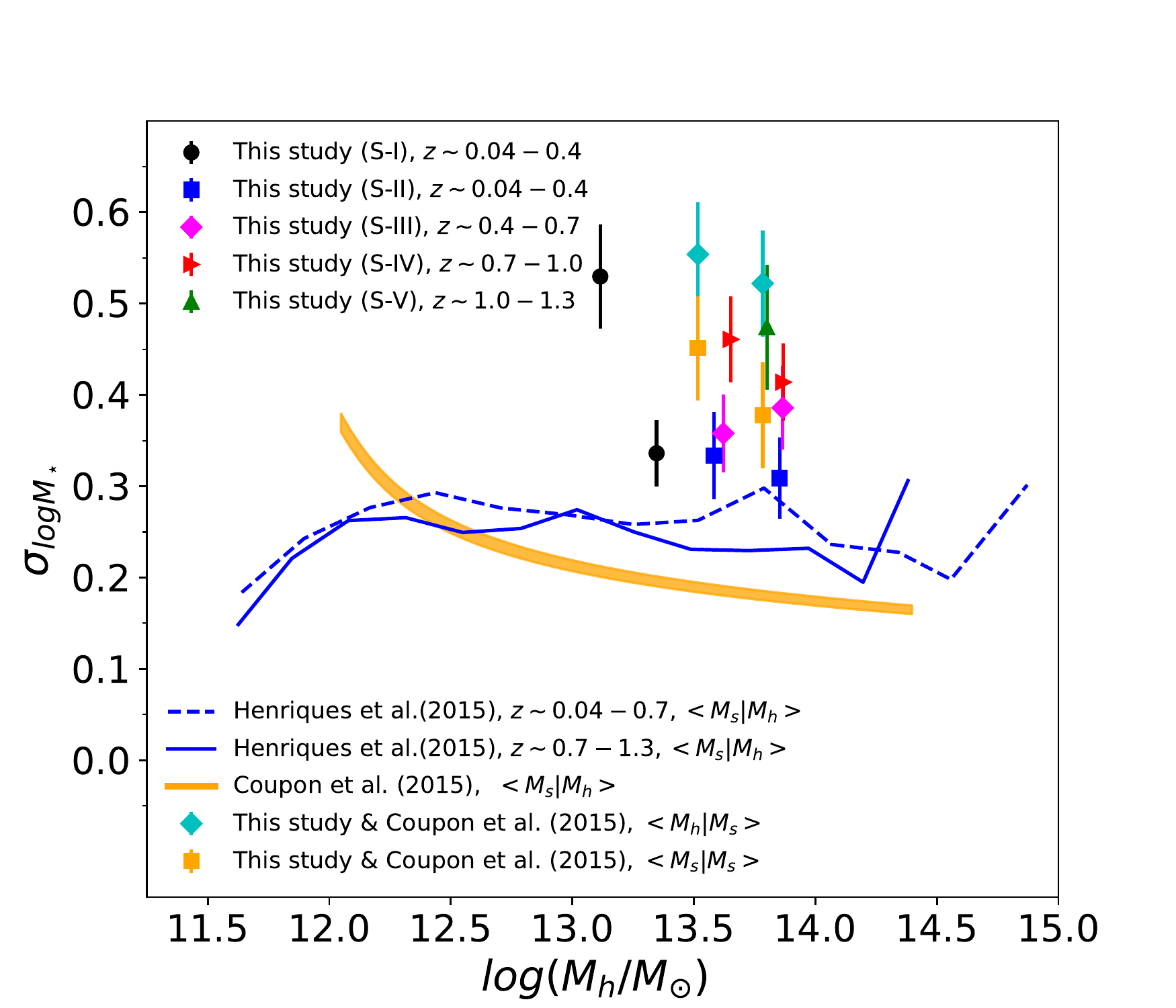}
	\caption[]{ The scatter in the stellar mass of central galaxies ($ \sigma 
	_{log\;M_{\star}}$) as a function of halo mass at a fixed halo mass ($ 
	\langle 
	M_{\star}|M_h \rangle $) in the H15 model at $ 0.04\leq z\leq0.7  $ (dashed 
	blue 
	line) and $ 0.7<z\leq1.3$ (solid blue line). Scatter in $ M_{\star} $ at a 
	given 
	halo mass in observations are plotted as the filled black (S-I), blue 
	(S-II), magenta (S-III), red (S-IV), and green (S-V) circles, respectively.  
	The orange area presents $ \sigma _{log\;M_{\star}}$ as a function of halo 
	mass 
	from \cite{Coupon15} results. The cyan diamonds and orange squares represent 
	the scatter between the mean stellar mass of our BGGs and that from $ 
	\langle M_h|M_{\star}h \rangle $ and $ \langle M_{\star}|M_h \rangle $ 
	relations 
	presented by \cite{Coupon15} .} \label{sigma}
\end{figure*}	` 
\subsection { $ M_{*}-M_{h}  $ relation and the lognormal scatter in $ 
M_{\star} $ at fixed $ M_h $ of the  BGGs}  \label{SHM}

It is well-known that the stellar mass of central galaxies  is correlated with 
the halo mass and the stellar mass-halo mass relation is a 
fundamental relationship which is used to link the evolution of galaxies to that of the host haloes. 
Constraining this relation offers a powerful tool for identifying the role of 
different physical mechanisms over the evolutionary history of galaxies 
\citep{Yang12,Behroozi10,Behroozi13,Moster13,Coupon15}. Every successful galaxy 
formation model is expected to make a reasonable prediction of how galaxies 
grow in mass along with their hosting (sub-)haloes. 

 The $ M_{\star}-M_h $ relation can be parametrised in one of the following 
 ways:  
 representing the average stellar mass of BGGs at fixed halo mass or 
 determining the mean halo mass at fixed stellar mass. Hereafter, we will refer 
 to these formalisms as $ \langle M_{\star}|M_h \rangle $ and $ \langle 
 M_h|M_{\star} 
 \rangle$, respectively \citep[e.g.,][]{Coupon15}.
 
 Figure \ref{ms_mh1} shows the  $ M_{\star}- 
 M_{h}$ relation ($ M_h $ 
 corresponds to $ 
 M_{200m}$, where the mean internal density of the halo is  200 times the mean 
 density of the universe). 
 The halo mass of groups in observations ranges from $10^{12.8}\;M_\odot $ to $ 
 10^{14}\;M_\odot$. Our data also span a wide dynamic range of stellar mass: $ 
 10\lesssim log(M_{\star}/M_\odot) \lesssim 12 $.  Error bars  include both the 
 systematic and statistical errors. We have shown the median error bar on $ 
 M_{\star} 
 $ and $ M_h $ with a black line. In Tab. 1, we have also reported the mean 
 systematic and statistical errors on stellar and halo masses for S-I to S-V. 
The typical total error on the stellar mass and halo mass in our data 
correspond to $ \sim0.22 $ and $ \sim0.16$ dex, respectively. 

In Fig. \ref{ms_mh1}, we have plotted the data for S-I to S-V  with open black, 
blue, magenta, red, and green symbols, respectively. We have also shown the 
mean stellar mass of BGGs in each subsample with filled symbols. The data show 
that 
the stellar mass of BGGs increases as a function 
of increasing halo mass and more massive haloes host generally more massive 
central galaxies. 

Within four SAMs (G11,DLB07,B06,H15) probed in this paper, we selected the most 
recently implemented model of H15 and measured both $\langle 
M_{\star}|M_h\rangle$ and 
$\langle M_h|M_{\star}\rangle$ relations over a large  range of $M_{h}$ from $ 
10^{11.5} ;M_\odot$ to $ 10^{14.75}\;M_\odot$ at two redshift ranges $ 
0.04<z<0.7$ and $ 0.7<z<1.3$. As shown in Fig. \ref{ms_mh1}, we find that both 
$\langle M_h|M_{\star}\rangle$ and  $\langle M_{\star}|M_h\rangle$ show no 
significant 
dependence on redshift.  Both relations also reveal that $ M_{\star}$, similarly, 
increases with increasing $ M_h$, however, at $log(M_h/M_\odot)>12.25$, the 
slope of  the $\langle M_{\star}|M_h\rangle$ relation becomes steeper than  the 
slope 
of the $\langle M_h|M_{\star}\rangle$ relation and the two trends are separated 
from 
each 
other at higher halo messes. The differences in $ M_{\star} $  between $\langle 
M_h|M_{\star}\rangle$ and  $\langle M_{\star}|M_h\rangle$ relations at a given 
$ M_h $ 
appears to increase as a function of increasing $ M_h$ by up to 0.5 dex. This 
means that different type of averaging of halo/stellar  mass can highly 
influence the $ M_{\star}-M_h $  relation due to the scatter in $ M_{\star}$. 
We find a 
good agreement between the $\langle M_{\star}|M_h\rangle$ relation from the H15 
model 
(solid blue and magenta lines) with our data of BGGs (filled symbols with error 
bars). 

Figure \ref{ms_mh1} also  illustrates the best-fit relation of  $\langle 
M_{\star}|M_h\rangle$ 
(orange area) and  $\langle M_h|M_{\star}\rangle$ (cyan area) with associated  
68\% 
confidence limits for central galaxies  at $ 0.5<z<1.0 $ by \cite{Coupon15}.  
At $log(M_h/M_\odot)<12.75$ and $log(M_{\star}/M_\odot)<11$, the two relations  
show 
similar trends, however, when  halo and stellar mass  are increased the slope 
of  
the $\langle M_{\star}|M_h\rangle$  relation becomes steeper  than the slope of 
the 
$\langle M_h|M_{\star}\rangle$ relation and the two trends diverge at $ 
M_h=10^{13} 
;M_\odot$. The $\langle M_{\star}|M_h\rangle$ relation by \cite{Coupon15} 
agrees with 
our observed data within errors. However, the $\langle M_{\star}|M_h\rangle$ 
relation 
from the H15 model better predicts  the observed data compared to the 
relations  of
\cite{Coupon15}. 
The $\langle M_h|M_{\star}\rangle$  relation by \cite{Coupon15} shows a good 
agreement with that from the H15 model. There is a $ \sim0.4 $ dex differences 
between $ M_{\star} $ at a given halo mass between the measurements by 
\cite{Coupon15} and 
H15, which could have arisen from the different methods and quality of data for 
estimating stellar 
mass/halo mass estimates and populating haloes by galaxies (such as the 
difference 
in the HOD model). We note that the stellar masses in our study are measured by 
using 
the Lephare code \citep{Ilbert10} based on the 
broad band SED fitting technique.  \cite{Coupon15} also use this method. 
However, we utilize a wealth of  
multi-wavelength, high signal-to-noise ratio observations such as UltraVISTA 
survey in the COSMOS field \citep{laigle2016cosmos2015} in which the depth of 
image is higher than data for SPIDER by about $ \sim 3 $ (AB magnitude) in $ 
K_s $ band. Thus, the stellar mass and photometric redshift of galaxies in our 
sample have been estimated with high accuracy. In addition,  using different 
stellar population models can also bias the stellar mass estimation.   

 H15 estimated the stellar mass using the \cite{maraston2005evolutionary} model 
 as 
 the default stellar population model with a \cite{chabrier2003galactic}  
 initial mass function (IMF). They have argued that the measurement by 
 \cite{chabrier2007}  IMF also gives similar results for all properties of 
 galaxies. However, the measurement with the older model of \cite{Bruzual2003} 
 shows some differences in properties of galaxies. \cite{Coupon15} computed the 
 stellar masses following a procedure reproduced by \cite{Arnouts2013} and use 
 a library of SED templates based on the 
 stellar population synthesis (SPS) code from \cite{Bruzual2003} with the 
 \cite{chabrier2003galactic} IMF. Adopting a different choice of SPS models and 
 IMF can lead to large systematic errors in stellar mass 
 measurements of up to 0.2 dex \citep{Behroozi10,Coupon15}. In addition, 
 various 
 choice of dust extinction laws can also bias the stellar mass estimates as 
 well. \cite{ilbert2010galaxy} have estimated a 0.14 dex difference between 
 stellar masses measured with the \cite{calzetti2000dust} attenuation law and 
 the \cite{charlot2000simple} dust model. In addition, the assumed cosmology 
 and photometric calibration can also cause further uncertainties in the 
 stellar mass estimates, which are larger than the statistical errors 
 \citep{Coupon15}.
 
Observations indicate that galaxy groups can be very diverse in their BGG 
properties. For example, at fixed group/cluster mass, the $ M_{\star} $ of the 
central  galaxies in fossils are larger compared to normal BGGs 
\citep[e.g.,][]{harrison2012xmm}. Figure \ref{ms_mh1} shows that  
$M_{\star}$ at a given $M_h$  largely scatters due to the fact that 
environments have a
strong impact on galaxy evolution and every BGG may experience a variety of 
environmental effects and evolutionary histories.  

Models based on the sub-halo abundance matching (SHAM) technique, such as  
\cite{Moster13},  assume a lognormal scatter in the stellar mass of (central) 
galaxies 
at fixed halo mass  as $ \sigma_{log\; M_{\star}}\sim 0.18$. 
In Fig. \ref{sigma}, we investigate  whether the scatter in  the stellar mass 
of our central galaxies ($\sigma _ {log\;M_{\star}}$) changes as a function of 
halo 
mass. We compute  $\sigma _ {log\;M_{\star}}$  for two samples of BGGs selected 
from 
the  H15 model at  $ 0.04\leq z\leq0.7  $ and $ 0.7<z\leq1.3$, as shown with 
the dashed and solid blue lines, respectively.  $\sigma _ {log\;M_{\star}}$ 
corresponds to the standard deviation of stellar mass at a given halo mass. We 
find that at the halo masses  studied here, $\sigma _ {log\;M_{\star}}$ 
exhibits no 
trend 
with $ M_h$ and redshift and  it remains constant  around $\sigma _ 
{log\;M_{\star}}\sim 0.25$. In addition, we  also measure  $\sigma _ 
{log\;M_{\star}}$  at 
fixed halo mass for BGGs in observations.  For S-I to S-IV, we determine 
$\sigma _ {log\;M_{\star}}$ in two halo mass bins. For S-I, the stellar mass 
of  
BGGs  spread over a range characterised by $\sigma _ {log\;M_{\star}}= 0.5$.

For S-II, we measure a constant scatter for both halo mass bins with $\sigma _ 
{log\;M_{\star}}\sim0.3$, which is  in a 
good 
agreement with the prediction of the H15 SAM. We estimate a 
redshift-dependent  scatter for S-III, S-IV as $\sigma _ 
{log\;M_{\star}}\sim0.35$, 
$\sigma _ {log\;M_{\star}}\sim0.4$, and  $\sigma _ {log\;M_{\star}}\sim0.45$, 
respectively. 
It is  clearly seen  that the scatter in stellar mass 
of BGGs increases with increasing redshift by $\sim0.15$ dex between $ z\sim0.1 
$ and $ z\sim1.3$.  
In Tab.1, we compare the systematic and statistical errors of the stellar and 
halo masses, the systematic/statistical errors among different sub-samples (S-I 
to S-V) and we see a suggestion that the redshift evolution of $\sigma _ 
{log\;M_{\star}}$ 
might not be due to uncertainties associated to stellar/halo mass measurements. 
At high 
redshifts ($ z\sim1 $), 
haloes host a variety of BGG populations in terms of structure, stellar age, 
and  star-formation activities, thus the high scatter in the stellar mass of
BGGs at high-z is feasible compared to BGGs at low redshifts, where the  
majority of them are quiescent elliptical galaxies.  In 
\cite{gozaliasl2016brightest}, we find that the stellar mass distribution of 
BGGs
deviates from a normal/Gaussian distribution with increasing redshift. We find 
evidence for the presence of a second peak in the stellar mass distribution at 
lower masses. We suggest that the large scatter in the stellar mass of BGGs in 
observation is due to their bimodal stellar mass distribution.

The orange area presents the scatter in  stellar mass at fixed halo mass from 
the
\cite{Coupon15} results. \cite{Coupon15} have presented a parametrised function 
for $ \sigma_{log\; M_{\star}} $ as a function of $ M_{\star} $ (see function 9 
in 
\cite{Coupon15}). Using this function and the $M_{\star}-M_h$ relation, we 
estimate $ 
\sigma_{log M_{\star}} $ as a function of $ M_h$. $ \sigma_{log\;M_{\star}} $ 
increases with decreasing $ M_h$, however, at $log(M_h/M_\odot)>12.5$, $ 
\sigma_{log M_{\star}} $  shows no trend with halo mass and remains constant 
around $ 
\sigma_{log\;M_{\star}}=0.2$.  \cite{Coupon15} report a medium mass  ($ 
M_{\star}\sim10^{10}\;M_\odot $) scatter of $\sigma_{log\;M_{\star}}=0.35$ and 
a high-mass 
($ M_{\star}\sim10^{11}\;M_\odot $) scatter of $ \sigma_{log\;M_{\star}}\simeq 
0.2$.

In Fig. \ref{sigma}, to quantify $\sigma_{log\;M_{\star}}$  between  the $ 
\langle 
M_{\star}|M_h\rangle $  or $ \langle M_h|M_{\star}\rangle $ relations from 
\cite{Coupon15} 
results and  our data at  $0.5<z<1.0$, we follow the procedure that has been 
presented by \cite{leauthaud2009weak} and measure $\sigma_{log\;M_{\star}}$  
using 
the following equation:
\begin{equation}
\sigma_{log\;M_{\star}}=\sqrt{\dfrac{\Delta M_{\star}}{(\gamma-1)}}
\end{equation} \label{sigsm}
where $ \Delta M_{\star}$ is the difference of the $ log\;M_{\star}$  between 
our 
measurement and  that from the $ \langle M_{\star}|M_h\rangle $  or $ \langle 
M_h|M_{\star}\rangle $ relations by \cite{Coupon15}. $ \gamma-1 $ is the slope 
of 
halo mass function ($ dn/d\;log\;M_h \propto M^{-(\gamma-1)}_h $). For more 
details, the reader is referred to \cite{leauthaud2009weak}.

The filled cyan diamonds and  filled orange squares in Fig. \ref{sigma} 
illustrate $ \sigma_{log\;M_{\star}}$  between the mean stellar mass of our 
BGGs and 
those from  $ \langle M_h|M_{\star}\rangle $ and  $ \langle 
M_{\star}|M_h\rangle $  
relations by \cite{Coupon15}, respectively. We find that these dispersions 
agree with our measurements within errors. 

 In summary, we conclude that the observed intrinsic lognormal scatter in the 
 stellar mass of BGGs spans a wide range  from $\sigma_{log\; 
 M_{\star}}\sim$ 
 0.25 to 0.5 at $ z<1.3 $, this holds true over a large range of halo 
 masses. Our measurement is in remarkable agreement with a recent study by 
 \cite{Chiu2016b} who estimated the $ M_{\star}-M_h $  scaling relation for 46 
 X-ray  
 groups detected in the XMM-Newton-Blanco Cosmology Survey (XMM-BCS) with a 
 halo 
 mass range of $ 2 \times10^{13} M_\odot \leq M_{500} \leq 
 2.5\times10^{14}M_\odot  $(median mass $ 8 \times 10^{13} M_\odot $) at 
 redshift $ 0.1 \leq z \leq1.02 $ (median redshift 0.47). They found an 
 intrinsic scatter of $ \sigma_{log M_{\star}|M_{500}}=0.36^{+0.07}_{-0.06}$.  
 These 
 scatters that are measured from the observational data are  higher than that 
 assumed by theoretical models based on the sub-halo abundance matching (SHAM) 
 technique. 
 \begin{figure*}
	\includegraphics[width=0.7\textwidth]{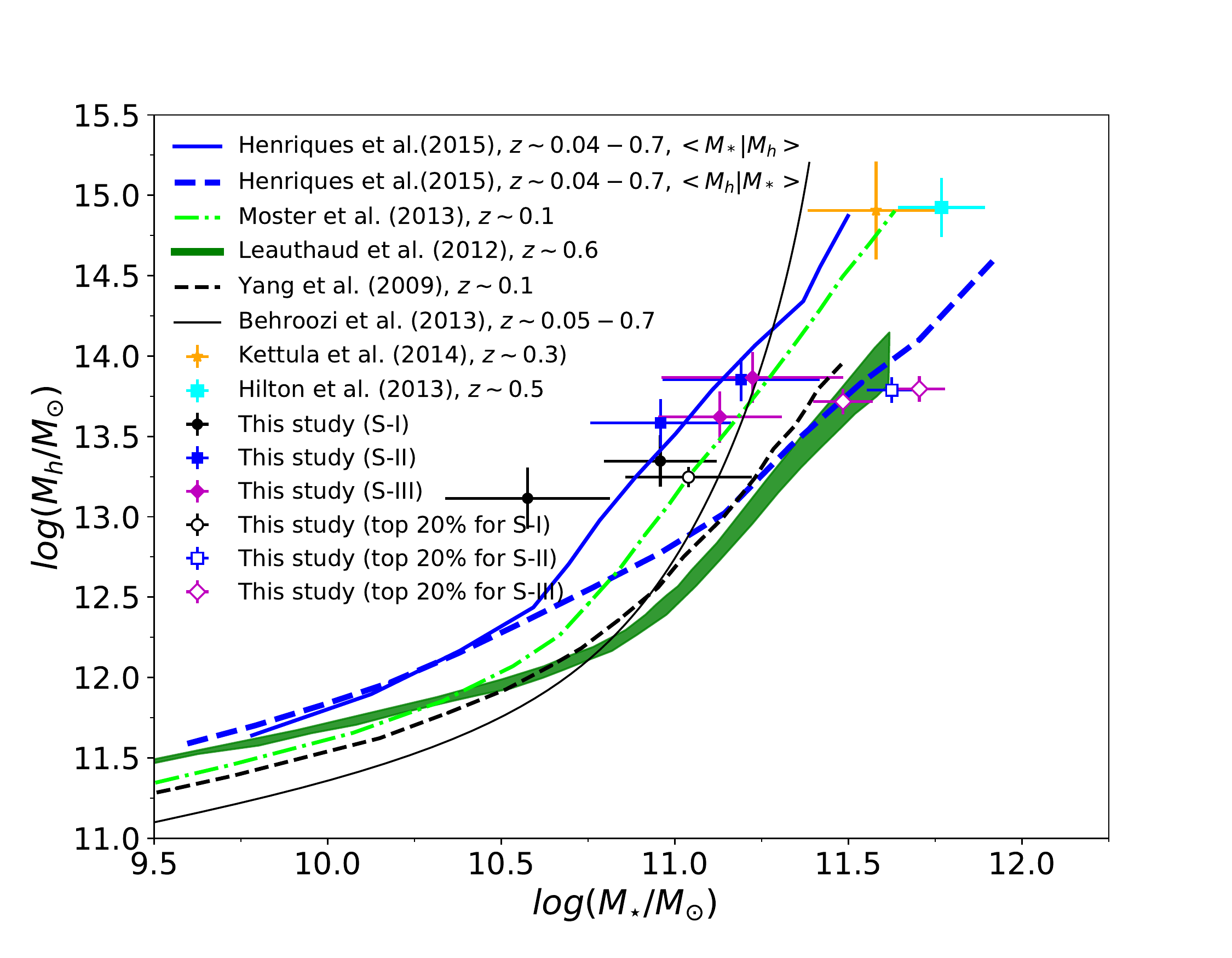}
	\includegraphics[width=0.7\textwidth]{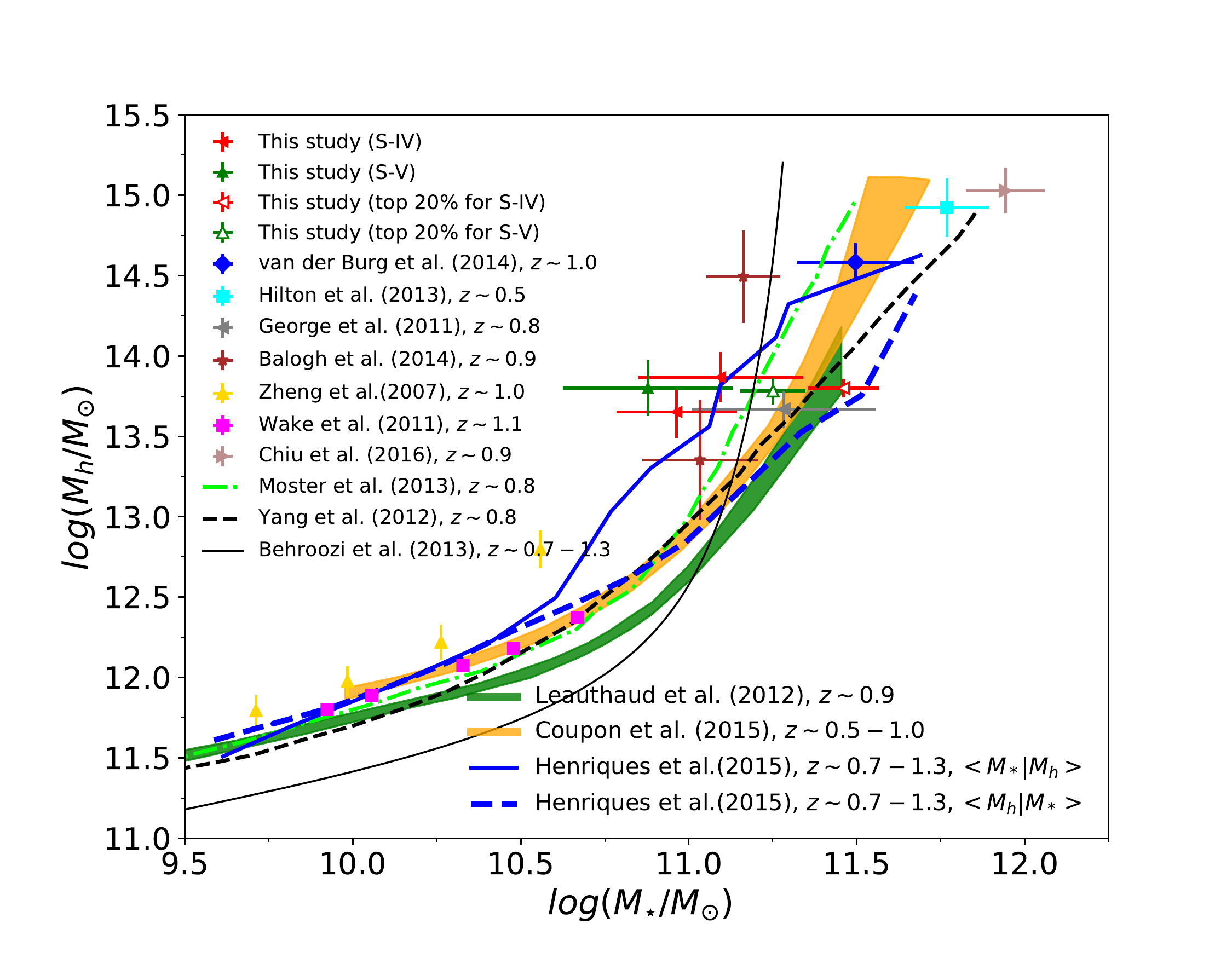}
	\caption[]{The $ M_{\star}-M_h $ relation for BGGs/central galaxies compared 
	with 
	a number of results from the literature at similar redshift ranges: $ z<0.7 
	$ (upper panel) and $ 0.7<z<1.3 $ (lower panel). The results presented here 
	represent the average stellar mass at fixed halo mass, $ \langle 
	M_{\star}|M_h 
	\rangle $, but plotted $ M_h $ as function of $ M_{\star} $ to ease the 
	comparison 
	with the literature. The dashed blue line shows  the $\langle M_h|M_{\star} 
	\rangle $ relation which represents the average halo mass at fixed stellar 
	mass for central galaxies in the H15 model. Following \cite{Coupon15}, the 
	stellar mass of central galaxies in the $ M_{\star}-M_h $ relation by 
	\cite{leauthaud2009weak} (green area) is shifted by  0.2 dex to lower values 
	to  compare directly with other results.  }
	\label{ms_mh}
\end{figure*}

\subsection{Comparison with the literature} \label{com_let}
\subsubsection{$ \langle M_{\star}|M_h \rangle $ }\label{shmr}

In Fig \ref{ms_mh}, we compare the $ M_{\star}-M_h $ relationship for the BGGs with 
a 
number of results from the literature. This relation describes the mean stellar 
mass at fixed halo mass ($\langle M_{\star}|M_h\rangle$), however, to 
facilitate the 
comparison with the literature, we plot $M_h $ as a function of $ M_{\star} $. 
We 
have 
also taken data from Fig. 10 as presented in \cite{Coupon15} and separately 
reproduce this figure here at two redshift ranges, $ z<0.7 $ (upper panel) and 
$ 0.7<z<1.3 $ (lower panel), respectively. 

In the upper panel of Fig. \ref{ms_mh},  we show  $ M_{h}$ as a function of 
$M_{*}$ for S-I, S-II, and S-III, as plotted with filled black, blue and 
magenta symbols with error bars, respectively.  Error bars include both 
1$\sigma$ error on the mean value and the error assigned to the stellar/halo 
mass estimations. We have similarly shown the  $ M_{h}$ versus $M_{*}$ for the 
20\% of BGGs with the highest $ f_{b,200}^{BGG} $ for S-I to S-III with open 
black, blue and magenta symbols. 

 The lower panel of Fig. \ref{ms_mh} presents  $ M_{\star}-M_h $ for S-IV 
 (filled red 
 triangles) and S-V (green triangles). We illustrate  $ M_{h}$ versus $M_{*}$  
 for the 20\% of BGGs with highest $ f_{b,200}^{BGG} $ for S-IV and S-V with 
 open red and green symbols.  The average stellar mass of BGGs increases with 
 halo mass with a minor dependence on redshift. In both panels of Fig. 
 \ref{ms_mh}, we find that $ M_{\star} $ of the 20\% of BGGs with highest $ 
 f_{b,200}^{BGG} $ for each subsample are more massive than that of the entire 
 population of BGGs in the same subsample at fixed halo mass. 
 
In both panels of Fig. \ref{ms_mh}, we show  the $ M_{h}-M_{*}$ relationship 
from \cite{Behroozi13} (solid black line), who constrain it by populating the  
dark matter haloes from N-body simulation employing the SHAM method and using 
the 
observed stellar mass function. They take into account  a number of systematic 
uncertainties such as the errors arising  from the choice of  cosmology, 
stellar population synthesis, and IMF, which  may affect stellar mass 
measurements. 
While this study seems to be consistent with our data at intermediate halo 
masses, it fails to reproduce the observed results from other studies at high 
halo masses. 

Similarly to  Fig. \ref{ms_mh1}, the best-fit $M_{\star}-M_h$ relation 
for the 
central galaxies  with associated  $ 1\sigma $ confidence limits from the 
\cite{Coupon15} result is shown as an orange area. This relation is consistent 
with our observed data within the observational errors. 

Figure \ref{ms_mh} also shows the results from \cite{Leauthaud12} (green area) 
at 
two redshift ranges $ z\sim 0.9 $ and $ z\sim0.6 $, respectively. 
\cite{Leauthaud12}  measured the total stellar mass fraction as a function of 
halo mass at $ z=0.2-1.0 $ using  the HOD model. They also derived the total 
stellar 
mass fraction at group scales using the X-ray galaxy groups catalogue in the 
COSMOS field.  \cite{Coupon15} found a $ \sim0.2  $ dex systematic discrepancy 
between their stellar mass-halo mass scaling relation and that of 
\cite{Leauthaud12} at fixed halo mass. This discrepancy is attributed to the 
differences in the modelling of the HOD, stellar mass measurements and  
separate choices for the dust extinction law. \cite{Coupon15}  estimated $ 
M_{\star} $ 
using the SED fitting method similar to the approach presented in 
\cite{Ilbert10} and our study. While \cite{Leauthaud12} measured  $ M_{\star} $ 
with 
the method described in \cite{bundy2006mass}. \cite{Coupon15} compared the 
stellar mass estimates applying the same IMF and the set of SPS models  
as  \cite{bundy2006mass} and \cite{Ilbert10} and determined an offset of 
$\sim0.2 $ dex. Thus they shift the stellar mass-halo mass relation of
\cite{Leauthaud12} to lower values by 0.2 dex along x-axis ($ M_{\star} $). 
Since we 
have taken the data from  \cite{Coupon15}, we applied this shift to the
relation by \cite{Leauthaud12} to lower the stellar mass by 0.2 dex.

\cite{Moster13}, shown as the  dot-dashed lime line in Fig. \ref{ms_mh}, used 
statistical and  SHAM approaches and provided a redshift-dependent 
parametrisation for the  $M_{\star}-M_h$ relationship of central galaxies. We 
find a 
good agreement between our data and that from \cite{Moster13}.

Figure \ref{ms_mh} shows the average stellar mass of BCGs  versus the average 
halo mass for some  massive clusters  at $ z\sim 0.3 $ in the CFHTLens field  
by \cite{Kettula15} (orange triangle in upper panel). Similarly,  Fig. 
\ref{ms_mh} shows the average stellar mass of BCGs  versus the average halo 
mass for clusters at $ 0.27<z<1.07 $ detected using the Atacama Cosmology 
Telescope, and the Sunyaev-Ze\v{l}dovich (SZ) effect by \cite{Hilton13} (cyan 
square in both upper and lower panel).  The mean stellar mass versus mean halo 
mass from \cite{Chiu16}  results  is shown with a single pink triangle. The 
halo mass range of these studies are above the halo mass range of our groups. 
 
We show as a single blue diamond the average $ M_h $ versus mean $ M_{\star} $ 
from 
\cite{vanderburg2014} in the GCLASS/SpARCS cluster sample at $ z\sim 1$, 
indicating a remarkable agreement with the results of 
\cite{Moster13,Coupon15,Henriques15}. 
 
The dashed black line in the upper panel of Fig. \ref{ms_mh} presents  the 
$M_{\star}-M_h$ relation from  the results of \cite{Yang09}, they use
the SDSS DR4 catalogue of galaxy groups \citep{yang2007galaxy} and  determine 
halo mass through an iteratively calculated group luminosity-mass relation. The 
stellar masses of galaxies  were estimated using the relation between $ M/L $ 
and $ g-r $ colour assuming a WMAP3 cosmology \citep{bell2003estimating}.  
\cite{Yang09} is not fully consistent with our data.

The dashed black line in the lower panel of Fig. \ref{ms_mh} presents the 
$M_{\star}-M_h$ relation  at $ z\sim0.8 $  by \cite{Yang12}. This model also 
overestimates the stellar mass of our BGGs by $ 0.2$ dex.

The  magenta squares with error bars in the lower panel of Fig. \ref{ms_mh} illustrate the results from  the NEWFIRM Medium-Band Survey at $z\sim1.1$ by  \cite{Wake11}. There is a good agreement between the \cite{Wake11}  results  and those from  different studies \citep[e.g.][]{Moster13,Behroozi13,Coupon15}.  

The yellow triangles with error bars present results for the DEEP2, a deep 
spectroscopic survey with high-density $ z\sim1  $ galaxies  from the HOD 
modelling  by \cite{Zheng07}, based on real-space clustering and number density 
measurements.   

The grey triangle  shows the mean stellar mass of BGGs and the  halo mass of 
their host groups at $ 0.5<z<1 $  in COSMOS field  by \cite{George11}. The halo 
mass of groups has been matched to  an X-ray detected sample of groups with 
the  $ L_X-M_h $  relation calibrated weak lensing.  The stellar mass has been 
estimated with an identical method to that of \cite{Leauthaud12}. Our result 
agree with this study within errors. 

The brown triangles represent results from \cite{Balogh14} with halo mass 
estimates made using the kinematics of satellites for 11 groups/clusters at $ 
0.8<z<1 $ within the COSMOS field. 

In the same way as the calculations done in Fig. \ref{ms_mh1}, we show  the  
$\langle 
M_{\star}|M_h\rangle$ (solid blue line) and $\langle M_h|M_{\star}\rangle$ 
(dashed blue 
line) relations from the SAM of H15 at $ z<0.7 $ and $ 0.7\leq z<1.3 $ in Fig. 
\ref{ms_mh}. The $\langle M_{\star}|M_h\rangle$ relation matches remarkably 
well with 
observations in both aforementioned redshift ranges.

 \subsection{$ M_{\star}/M_h -M_h$ relation}\label{shmr}
 As discussed before, the stellar to halo mass ratio ($ M_{*}/M_{h}$, 
 SHMR) is one of the key parameter, which provides a powerful insight into the 
 connection between galaxies and their haloes and a useful clue for comprising 
 the integrated efficiency of the past stellar mass assembly (i.e.,  
 star-formation and mergers) 
 \citep[e.g.][]{Leauthaud12,Coupon15,hudson2015cfhtlens,harikane2016evolution}. 
 The studies of  low-redshift galaxy systems find a SHMR with a peak at a dark 
 matter halo mass of $ M_h\sim 10^{12}\; M_\odot $, independent of redshift 
 \citep[e.g.,][]{Coupon15,hudson2015cfhtlens,harikane2016evolution}. However, 
 \cite{Leauthaud12} claim a redshift evolution of SHMRs from  $ z\sim1 $ to the 
 present day.
 
\begin{figure*}
  	\includegraphics[width=0.8\textwidth]{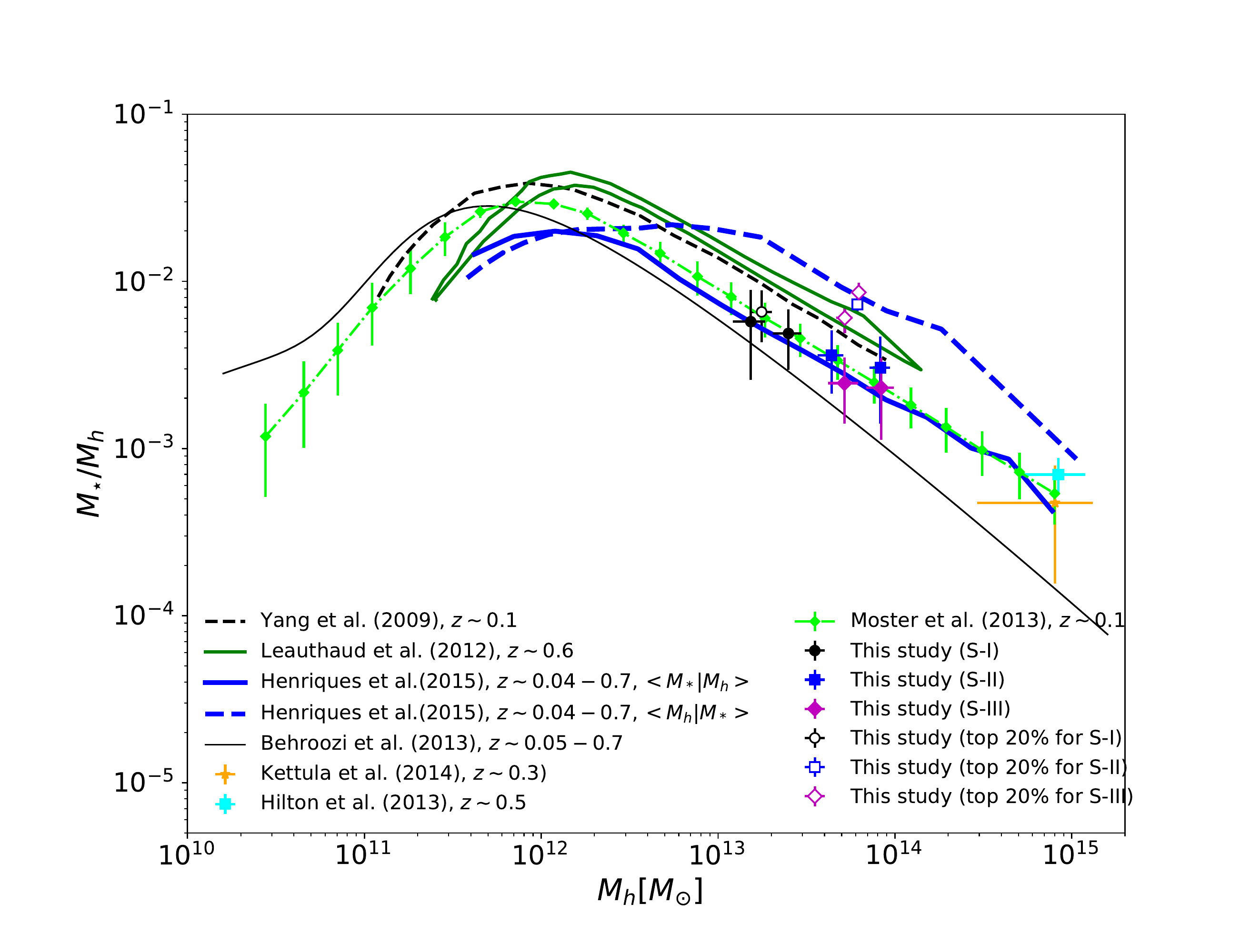}
   	\includegraphics[width=0.8\textwidth]{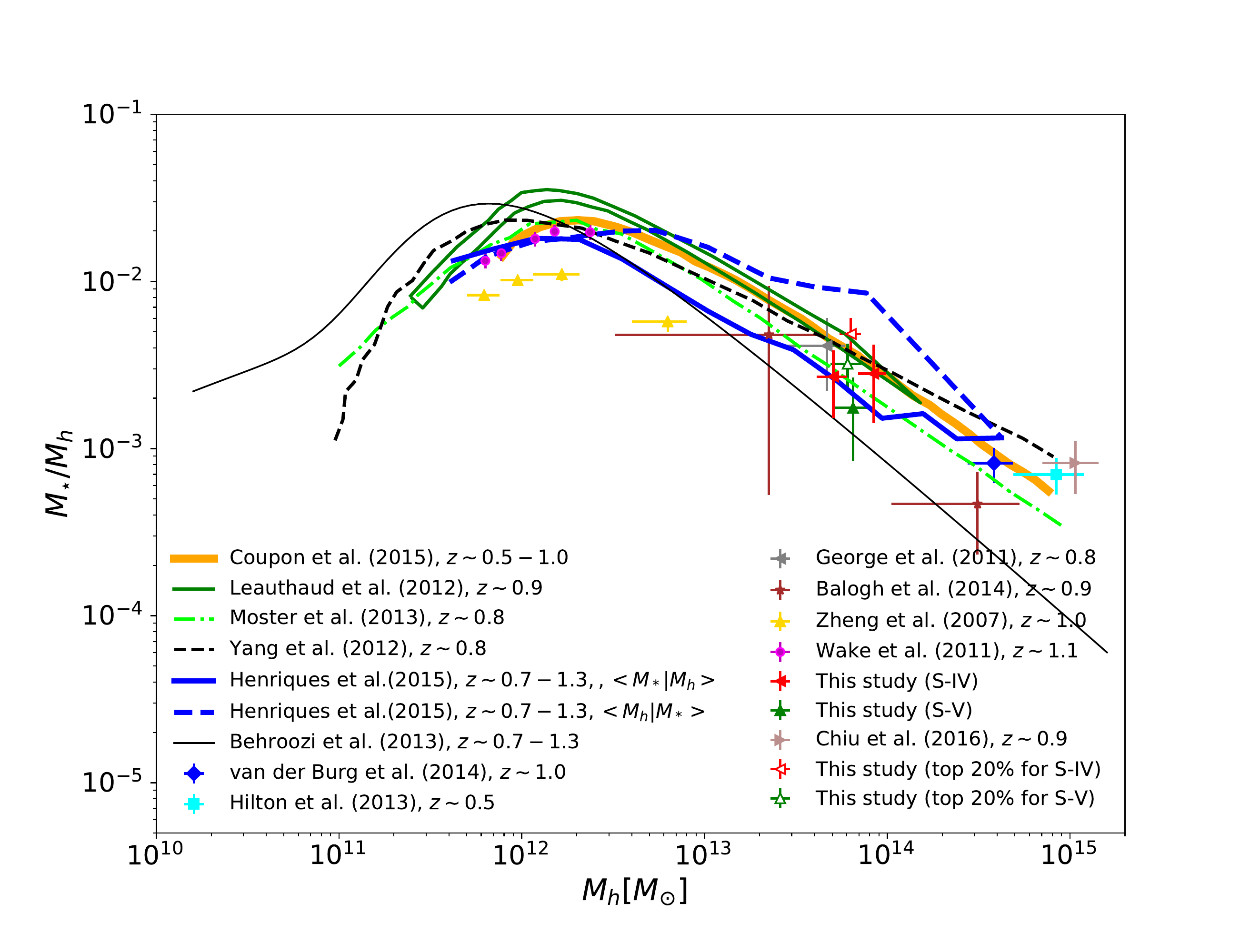}	
  
  	\caption[]{ Same data set as used in Fig. \ref{ms_mh} but for the stellar to 
  	halo mass ratio ($ M_{*}/M_{h}$ or SHMR) of BGGs/central galaxies as a 
  	function of halo mass at $ z<0.7 $ (upper panel) and $ 0.7<z<1.3 $ (lower 
  	panel). }  	\label{fm}
  \end{figure*}

 Figure  \ref{fm} presents the  $ M_{*}/M_{h}$ as a function of $M_h$ for 
 central 
 galaxies at $ 0.04<z<0.7$ (upper panel) and $ 0.7<z<1.3$ (lower panel). Just 
 like in Fig. \ref{ms_mh}, Fig. \ref{fm} also compares results from the 
 literature. For further information on the data and studies, we refer the  
 reader to \S \ref{shmr}. We use the same symbols and line styles for the data 
 set used in Fig. \ref{ms_mh}.  
 
We determine the $ M_{\star}/M_h-M_h$ relation using the H15 model at fixed 
halo and 
stellar masses at $ 0.04<z<0.7$ and $ 0.7<z<1.3 $, as shown in the upper and 
lower 
panels of Fig. \ref{fm}, respectively. 
At $ 0.04<z<0.7 $,  $ M_{\star}/M_h$ at fixed halo mass (blue solid line) peaks 
at $ 
log(M_{h}/M_\odot)=12.08\pm 0.13 $ with an amplitude of $ 0.0200\pm 0.0001$. 
While $ M_{\star}/M_h$ at fixed $ M_{\star} $ (blue dashed line) reaches its 
maximum value 
of  $ 0.0218\pm 0.0003 $
 at $ log(M_{\star}/M_\odot)=10.93\pm 0.01$, where the average halo mass 
 corresponds 
 to $log(M_{h}/M_\odot)= 12.73\pm0.01$. In addition, the $ M_{\star}/M_h-M_h$ 
 relation 
 at fixed stellar mass illustrates a flat peak at the halo mass range  $ 
 log(M_{h}/M_\odot)=12-13 $. At $ 0.7<z<1.3 $,  the $ M_{\star}/M_h$ 
 relation at fixed halo mass (blue solid line) peaks at $ 
 log(M_{h}/M_\odot)=12.09\pm 0.13 $ with an amplitude of $ 0.0181\pm0.0002 $. 
 While this relation at fixed stellar mass (blue dashed line) reaches its 
 maximum value of $ 0.0202\pm 0.0005 $ at $ log(M_{\star}/M_\odot)=10.92\pm 
 0.09$, 
 where host groups have an average halo mass of $log(M_{h}/M_\odot)= 
 12.73\pm0.01$. We find that the $ M_{\star}/M_h-M_h$ relation in the H15 model 
 shows 
 no considerable redshift evolution over $ z<1.3$.

 In the upper panel of Fig. \ref{fm}, our data for S-I to S-III show a good 
 agreement  with most of the $ M_{\star}/M_h-M_h$ relations presented by 
 \cite{Yang09,Behroozi13,Moster13,Henriques15}. Other observational data from 
 \cite{Hilton13,Kettula15} are also consistent with these theoretical studies 
 at high halo masses. The position and height of the peaks agree with the 
 studies by \cite{Yang09,Behroozi13,Moster13}. The majority of these studies 
 estimate that $ M_{\star}/M_h$ reaches its maximum value at around 
 $log(M_h/M_\odot)\sim 11.78-12.3$ with an amplitude of $ 
 \sim0.01-0.02$. \cite{Leauthaud12} measured the highest peak within studies.

 In the lower panel of Fig. \ref{fm}, we find that our data for S-IV and S-V are also in  general agreement with previous studies notably \cite{Behroozi13,Moster13,Coupon15,Henriques15}. There is also a good agreement between other observational results from \cite{Zheng07,Wake11,Hilton13,Balogh14,Burg14,Chiu16} and theoretical models.
 
We illustrate $ M_{\star}/M_h$ versus $M_h$ for  20\% of the  BGGs with the 
highest $ 
f_{b,200}^{BGG} $ for S-I to S-III (upper panel) and for S-IV and S-V (lower 
panel) with open symbols. In both panels of Fig \ref{fm}, we find that $ 
M_{\star}/M_h $ for 20\% of the BGGs with the highest $ f_{b,200}^{BGG} $ for 
each 
subsample are higher than that for the full BGG sample at fixed halo 
mass. 
    
We observe that the scatter in $ M_{\star}/M_h$ increases towards high halo 
masses. 
At $ M_{h}=10^{13.8}\;M_\odot$, we measure a scatter of $ 0.003\pm0.002 $ and 
$ 0.002\pm0.001 $ in $ M_{\star}/M_h$ in observations at $ 0.04<z<0.7 $ and  $ 
0.7<z<1.3$, respectively. This scatter corresponds to the standard deviation of 
$ M_{\star}/M_h$ at fixed halo mass. 
 
 
\section{Summary and conclusions}

We study the relative contribution of stellar mass of the brightest group 
galaxies to the total baryonic mass of hosting groups ($f^{BGG}_{b,200} $) and 
its evolution over $0.04<z<1.3$,  using a unique sample of BGGs selected from 
407 X-ray galaxy groups in the COSMOS, XMM-LSS and
AEGIS fields. We also study the $ M_\star-M_h$ and $ M_\star/M_h-M_h$ 
relations, and quantify the intrinsic (lognormal) 
scatter in the 
stellar mass at a fixed halo mass. We define five 
subsamples of BGGs (S-I to S-V) such that four out of five have similar halo 
mass range. This definition allows us to compare the properties of the BGGs within  
haloes of similar masses at different redshifts. In addition, we also selected 
a 
new sample of the BGGs accounting for the $ M_h$ growth of dark matter 
haloes at $ z<1.3$ (S-IHMA). We compare results between these two samples to 
quantify 
the impact of the halo mass assembly on the properties of BGGs and their growth 
over the past 9 billion years. We interpret  our results  with  predictions 
from four SAMs (H15, G11, DLB07, and B06) based on the Millennium simulation 
and with a number of results from the literature. We summarize our findings as  
follows:

\begin{enumerate}
\item In agreement with the SAM predictions, the relative contribution of  
the BGGs to 
the total baryon content of haloes, $ log(f^{BGG}_{b,200}) $
decreases with increasing $ M_{200}$, at $ z<1.0 $. While $ 
log(f^{BGG}_{b,200}) $  shows no trend with halo mass for the lowest/highest-z  
BGGs for S-I and S-V. Using the linear least square and MCMC techniques, we 
quantified the $ f^{BGG}_{b,200}-M_{200}$ relations (see Tab. \ref{fbm200}) and 
found that the observed relation evolves mildly with redshift. The 
slope of the relation becomes steeper with decreasing redshift while the 
zero-point  ($ log \beta_{MCMC} $) increases significantly. 

\item We obtain an smoothed distribution of the $f^{BGG}_{b,200}$ by
applying a non-parametric method to estimate the KDE function. We found that 
the observed KDE function evolves with 
redshift and tends be skewn at lower values by increasing redshift. In 
addition, the smoothed $f^{BGG}_{b,200}$ distribution of the BGGs within $ z<0.4 $ 
and low mass group (S-I) skews to higher values compared to 
that of the BGGs within the same redshift and massive groups (S-II), indicating
 a high contribution of BGGs to baryons in  low-mass haloes. The semi-analytic 
models probed here 
are not able to reconstruct the observed $f^{BGG}_{b,200}$ distribution, although 
H15 and B06 provide, relatively, better predictions.

\item We measured the  mean and median SFRs of the BGGs as a function of 
redshift 
and found that these quantities decline by $ \sim 2$ dex between z=1.2 to 
z=0.2.  
As the hierarchical structure formation scenario implies, haloes grow 
with cosmic time and their abundance with masses  $ \geq 
10^{12} M_{\odot} $ increases \citep{mo2002abundance}. In this halo 
mass regime, the
influences of environments on galaxy properties within virial radius, heating 
by virial shocks of 
 infalling gas, energy feedback from an AGN, and gravitational heating of a gas 
disk by clumpy mass infall, all are efficiently increased  
\citep[e.g.,][]{lewis20022df,wetzel2012galaxy}. As a result, this makes 
 it 
 likely that the strong evolution of  abundance of such massive haloes play a 
 key role 
 in  quenching star formation in cluster central galaxies.

In agreement with observations, SAMs also predict that the observed 
mean SFR 
decreases with decreasing redshift by about 1 dex between z=1.2 to z=0.2. 
However, they underestimate the mean SFR at a given redshift at $z>0.5$. The 
differences in the mean SFR of BGGs between observations and the models reach 
up to 
1.5 dex at $ z=1$. It may be that the BGGs in these models have ended their 
star-forming activities early.

We found that the median SFR in the G11 and DLB07 models decreases 
slowly with 
cosmic time by $ \sim0.5 $ dex from z=1.2 and this trend overcomes observations 
at $ z<0.8$. The median SFR predicted by H15 model decreases sharply by $\geq4$ 
dex and this model underestimates significantly SFRs in particular at low 
redshifts. 

In agreement with the models, we find no considerable differences among 
the mean (median) SFR  of the full BGG sample between S-II to S-V and S-IHMA. 
This indicates that accounting for the growth of halo mass in  the comparison 
leads to no significant changes in the SFR evolution.

The mean SFR of 20\% of the BGGs with the highest  $f^{BGG}_{b,200}$ 
for 
both S-II 
to S-V and S-IHMA decrease  as a function of decreasing redshift by $ \sim2 $ 
dex and $ \sim3  $
dex, respectively. The trend  for S-II to S-V is slower than that of S-IHMA at 
$z>0.6$. Models predict better evolution the mean (median) SFR of the 20\% of 
BGGs for 
S-II to S-V and S-IHMA than SFR of the full BGG sample.

We found that the BGG SFR exhibits a bimodal distribution. As a result, 
the mean 
SFR is $ \sim 2  $ dex higher than the median  SFR at fixed redshift. Such a  
large gap indicates that the mean SFR of BGGs is 
influenced by outliers (very high and low SFRs). We suggest that 
BGGs 
consist of two distinct populations of star-forming and passive galaxies.

\item We  determined that the  stellar mass of BGGs for both S-II to S-V and 
S-IHMA grows  as a function of decreasing redshift by at least 0.3 dex. While 
the significant growth for  S-II to S-V occurs at $ z>0.7 $ and  slows down at 
low redshifts, the stellar mass for S-IHMA  linearly increases with redshift 
from 
z=1.2 to the present day. 

The mean (median) stellar mass of the 20\% of BGGs with 
the highest  $f^{BGG}_{b,200}$ for S-II to S-V and S-IHMA increases by 0.11 dex 
and 0.44 dex over $ 0.31<z<1.13$. The stellar mass growth for S-IHMA is 
substantially higher than that of S-II to S-V by 0.33 dex, illustrating that 
the halo mass growth results in BGGs with the highest $f^{BGG}_{b,200}$ to grow 
further. We found that 20\% of the BGGs  with the highest $f^{BGG}_{b,200}$ are 
generally 
more massive galaxies compared to BGGs with low  $f^{BGG}_{b,200}$ by at least 
0.2 dex at a given redshift. 

 We show that 45\% of the growth of stellar mass for a typical 
 star forming BGG which follow the mean $SFR-z $ relation (as presented in Tab. 
 3)  could be driven by
 star-formation since $ z<1.3$. However, this growth for a typical BGG which follows the 
 median $SFR-z $ relation is not significant.

 Munich models are found to underestimate the stellar mass growth of 
 BGGs at $ 
 z<1.3$. They also overestimate the stellar mass of BGGs at $ 
z>1.0$. This can be explained  to be due to overestimating the number density 
of 
massive galaxies  and the number of major mergers that BGGs may have 
experienced at 
high-z. For the 20\% BGGs with highest $ f_{b,200}^{BGG}$, models underestimate
the BGGs mass, particularly at intermediate redshifts. We quantified the growth 
of the mean $ M_{\star} $  for the 20\% BGGs with the highest $ M_{\star}$, 
with redshift, in 
models and showed that their trends approach the observed trend, however, their 
predictions are still below observations by $ 1\sigma$. 

\item We investigated the mean (median) $log(f_{b,200}^{BGG})-z$ relation 
of the BGGs  
and found that 
$ log(f_{b,200}^{BGG}) $ increases with decreasing redshift from $ z=1.13 $ to 
$ 
z=0.31 $ by $ 0.33-0.35$ dex. Munich models (H15, G11, and 
DLB07) predict no significant growth for $ log(f_{b,200}^{BGG}) $ with redshift 
and overpredict  the $ f_{b,200}^{BGG} $ at $ z>0.8 $. Since models   
under-predict the SFR at $ z>0.8 $, we conclude that these models possibly 
overestimate either the number of early major mergers or early SFRs of the 
BGGs. 

The $ log(f_{b,200}^{BGG}) $ for the 20\% BGGs with the highest $ 
f_{b,200}^{BGG} $ for S-II to S-V and S-IHMA increases with cosmic time by $ 
0.22$  and $ 0.41 $ dex, indicating that the halo mass growth  
is  efficient  in the $ log(f_{b,200}^{BGG})$ growth. SAMs match data for 
S-II to S-V  better than those of S-IHMA. 

We quantified what fraction of the $ f_{b,200}^{BGG} $ growth for a 
typical 
galaxy may have been driven by star formation in the BGGs. We assume a typical 
galaxy with an initial mass which corresponds to the mean (median) stellar mass 
of BGGs for S-V. We allowed this galaxy to form stars 
according to the mean (median)  SFR-z trends as 
summarised in Tab. \ref{sfr_tb}. We also measured the halo mass growth for the 
hosting group with initial mass which is equal to the mean (median) halo mass 
of 
groups for S-V.  When the halo mass of a typical halo remains constant with 
redshift, $f_{b,200}^{BGG}$ of  the typical star forming BGG  can increase 
via  star formation  by 0.18 dex over $ z<1.3$. In contrast, once we allow the 
halo to grow in mass with redshift, the $f_{b,200}^{BGG}$ 
growth via  star formation  becomes negligible at $ z<0.8 $, indicating that 
the halo 
mass growth rate is more efficient than the star formation rate.  

In addition,  $f_{b,200}^{BGG} $ which follows the 
median SFR-z relation 
indicates no growth through star-formation for both the entire sample of the 
BGGs and  
the 20\% BGGs with the highest $f_{b,200}^{BGG}$.
\item We found no strong correlation between the $f^{BGG}_{b,200}$ and  
the evolutionary state of the groups, probed by the luminosity gap in the 
r-band magnitude between the BGG and the second brightest satellite ($ r\simeq 
0.052 $). 
\item The observed $ M_{\star}/M_h-M_h$ and $M_{\star}-M_h$ 
relations are generally consistent with the SAM predictions. However, there are some 
discrepancies at low and high halo masses which can depend on the statistical 
methods used 
for the comparison. As an example, we obtained both $\langle 
M_{\star}|M_h\rangle$ and $\langle 
M_h|M_{\star}\rangle$ relations for the H15 model and found that the slope of 
the  
$\langle M_{\star}|M_h\rangle$ relation  becomes steeper than that of $\langle 
M_h|M_{\star}\rangle$ relation at $M_h > 10^{12.5}\; M_\odot$. 
\item By comparing a number of $ M_{\star}-M_h$ relations from the
literature, we 
showed that the scatter in stellar mass of BGGs  
($\sigma_{log\;M_{\star}}$) at a given halo mass is one of the key issues which 
leads to inconsistencies between results. 

Using the H15 model, we find that $M_{\star}-M_h$, and 
$\sigma_{log\;M_{\star}}-M_h$ relations show no evolution since $ z\sim1$ and 
$\sigma_{log\;M_{\star}}$ also shows no trend with $ M_h $ at halo mass 
$M_h=10^{12} $ to $ 10^{14.5} M_\odot$. The scatter in the stellar mass of BGGs 
is found to remain constant at around $\sigma _ {log\;M_{\star}} \sim0.25$. 

The scatter in $M_{\star}$ of BGGs in observations is relatively higher 
than the
model predictions and $\sigma_{log\;M_{\star}}$  increases when
increasing redshift from $\sigma _ {log\;M_{\star}} \sim0.3$ at $ z\sim0.2 $ to 
$\sigma _ {log\;M_{\star}}\sim 0.5$ at $ z\sim1.0$. Our findings are in
remarkable 
agreement with the recent research by \cite{Chiu2016b} who estimated 
the $ M_{\star}-M_h $  scaling relation for 46 X-ray  groups detected in the 
XMM-Newton-Blanco Cosmology Survey (XMM-BCS) with a median halo mass of $ 8 
\times 10^{13} M_\odot $ at a median redshift of $z=0.47$. This study measured  
an intrinsic scatter in stellar mass of BGGs by $ \sigma_{log 
M_{\star}|M_{500}}=0.36^{+0.07}_{-0.06}$.  In 
\cite{gozaliasl2016brightest}, we have found 
evidence for the presence of a second peak in the $ M_{\star} $ distribution 
of BGGs 
around $M_{\star}\sim10^{10.5} \; M_\odot$  and showed that the shape of $ 
M_{\star} $ 
distribution deviates from a normal distribution when increasing redshift. As a 
result, the BGGs that make the second 
peak are found to be generally young and star 
forming systems, implying that the presence of high scatter in the $ 
M_{\star}-M_h$ 
relation is a feasible effect due to the presence of BGGs with a variety of 
properties e.g., star formation and stellar age.

\item  The $ M_{\star}/M_h-M_h$ relation which we measure using the 
H15 
model 
 peaks at $ log(M_{h}/M_\odot)=12.08\pm 0.13 $ with an amplitude 
of $ 0.0200\pm 0.0001$ at $0.04<z<0.7 $ and $ 0.0181\pm0.0002 $ at $ 
0.7<z<1.3$, which is roughly independent of redshift. 

At $ M_{h}=10^{13.8}\;M_\odot$, $ M_{*}/M_{h}$ in observations 
scatters by $ 0.003\pm0.002 $ with no dependence on redshift. 

\item  We point out that the different choices of the statistical 
analysis, data fitting techniques, and the ravaging methods, altogether, can 
cause bias results. It is advantageous to distinguish the results which are 
derived for the mean and median values of the BGG properties. Classification of 
BGGs based on 
their SFR/stellar age can obviously provide more insight onto  some of the 
hidden 
properties of BGGs.

\item We have compared observed trends with SAMs based on the Millennium 
Simulation and explored that the models are generally able to predict 
observational results. However, in order to advance our understanding of galaxy 
formation further, a grid of model predictions fully sampling the parameter 
space of amplitude and redshift evolution of the relevant physical processes 
such as star formation and AGN feedback is necessary.
\end{enumerate}

\section{Acknowledgements}
This work has been supported by the grant from the Finnish Academy of Science 
to the University of Helsinki and the Euclid project, decision numbers 266918 
and 1295113. The first author wishes to thank Donald Smart and Christopher 
Haines for the useful 
comments 
and the School of Astronomy in Institute for Research in Fundamental Sciences 
for its support. We used the data of the Millennium simulation and the web 
application providing on-line access to them were constructed as the activities 
of the German Astrophysics Virtual Observatory.
\bibliographystyle{mn2e}
\bibliography{cite}

\begin{thebibliography}{}

\bibitem[\protect\citeauthoryear{Ade, Aghanim, Arnaud, Ashdown, Aumont,
  Baccigalupi, Banday, Barreiro, Bartlett, Bartolo et~al.,}{Ade
  et~al.}{2015}]{Planck15}
Ade P.,  Aghanim N.,  Arnaud M.,  Ashdown M.,  Aumont J.,  Baccigalupi C.,
  Banday A.,  Barreiro R.,  Bartlett J.,  Bartolo N.,    et~al., 2015, arXiv
  preprint arXiv:1502.01591

\bibitem[\protect\citeauthoryear{Allen, Rapetti, Schmidt, Ebeling, Morris \&
  Fabian}{Allen et~al.}{2008}]{Allen08}
Allen S.,  Rapetti D.,  Schmidt R.,  Ebeling H.,  Morris R.,    Fabian A.,
  2008, \mnras, 383, 879

\bibitem[\protect\citeauthoryear{Allen, Schmidt, Ebeling, Fabian \&
  Van~Speybroeck}{Allen et~al.}{2004}]{Allen04}
Allen S.,  Schmidt R.,  Ebeling H.,  Fabian A.,    Van~Speybroeck L.,  2004,
  \mnras, 353, 457

\bibitem[\protect\citeauthoryear{Allen, Evrard \& Mantz}{Allen
  et~al.}{2011}]{Allen11}
Allen S.~W.,  Evrard A.~E.,    Mantz A.~B.,  2011, arXiv preprint
  arXiv:1103.4829

\bibitem[\protect\citeauthoryear{Allevato, Finoguenov, Hasinger, Miyaji,
  Cappelluti, Salvato, Zamorani, Gilli, George, Tanaka et~al.,}{Allevato
  et~al.}{2012}]{allevato2012occupation}
Allevato V.,  Finoguenov A.,  Hasinger G.,  Miyaji T.,  Cappelluti N.,  Salvato
  M.,  Zamorani G.,  Gilli R.,  George M.,  Tanaka M.,    et~al., 2012, The
  Astrophysical Journal, 758, 47

\bibitem[\protect\citeauthoryear{Andreon}{Andreon}{2010}]{Andreon10}
Andreon S.,  2010, \mnras, 407, 263

\bibitem[\protect\citeauthoryear{{Arnouts}, {Le Floc'h}, {Chevallard},
  {Johnson}, {Ilbert}, {Treyer}, {Aussel}, {Capak}, {Sanders}, {Scoville},
  {McCracken}, {Milliard}, {Pozzetti} \& {Salvato}}{{Arnouts}
  et~al.}{2013}]{Arnouts2013}
{Arnouts} S.,  {Le Floc'h} E.,  {Chevallard} J.,  {Johnson} B.~D.,  {Ilbert}
  O.,  {Treyer} M.,  {Aussel} H.,  {Capak} P.,  {Sanders} D.~B.,  {Scoville}
  N.,  {McCracken} H.~J.,  {Milliard} B.,  {Pozzetti} L.,    {Salvato} M.,
  2013, \aap, 558, A67

\bibitem[\protect\citeauthoryear{Balogh, McGee, Mok, Wilman, Finoguenov, Bower,
  Mulchaey, Parker \& Tanaka}{Balogh et~al.}{2014}]{Balogh14}
Balogh M.~L.,  McGee S.~L.,  Mok A.,  Wilman D.~J.,  Finoguenov A.,  Bower
  R.~G.,  Mulchaey J.~S.,  Parker L.~C.,    Tanaka M.,  2014, \mnras, 443, 2679

\bibitem[\protect\citeauthoryear{Behroozi, Conroy \& Wechsler}{Behroozi
  et~al.}{2010a}]{behroozi2010comprehensive}
Behroozi P.~S.,  Conroy C.,    Wechsler R.~H.,  2010a, The Astrophysical
  Journal, 717, 379

\bibitem[\protect\citeauthoryear{Behroozi, Conroy \& Wechsler}{Behroozi
  et~al.}{2010b}]{Behroozi10}
Behroozi P.~S.,  Conroy C.,    Wechsler R.~H.,  2010b, \apj, 717, 379

\bibitem[\protect\citeauthoryear{Behroozi, Wechsler \& Conroy}{Behroozi
  et~al.}{2013}]{Behroozi13}
Behroozi P.~S.,  Wechsler R.~H.,    Conroy C.,  2013, \apj, 770, 57

\bibitem[\protect\citeauthoryear{Bell}{Bell}{2003}]{bell2003estimating}
Bell E.~F.,  2003, The Astrophysical Journal, 586, 794

\bibitem[\protect\citeauthoryear{Berlind \& Weinberg}{Berlind \&
  Weinberg}{2002}]{berlind2002halo}
Berlind A.~A.,  Weinberg D.~H.,  2002, The Astrophysical Journal, 575, 587

\bibitem[\protect\citeauthoryear{Bower, Benson, Malbon, Helly, Frenk, Baugh,
  Cole \& Lacey}{Bower et~al.}{2006}]{Bower06}
Bower R.,  Benson A.,  Malbon R.,  Helly J.,  Frenk C.,  Baugh C.,  Cole S.,
  Lacey C.~G.,  2006, \mnras, 370, 645

\bibitem[\protect\citeauthoryear{{Bruzual} \& {Charlot}}{{Bruzual} \&
  {Charlot}}{2003}]{Bruzual2003}
{Bruzual} G.,  {Charlot} S.,  2003, \mnras, 344, 1000

\bibitem[\protect\citeauthoryear{Bulbul, Randall, Bayliss, Miller,
  Andrade-Santos, Johnson, Bautz, Blanton, Forman, Jones et~al.,}{Bulbul
  et~al.}{2016}]{Bulbul16}
Bulbul E.,  Randall S.~W.,  Bayliss M.,  Miller E.,  Andrade-Santos F.,
  Johnson R.,  Bautz M.,  Blanton E.~L.,  Forman W.~R.,  Jones C.,    et~al.,
  2016, \apj, 818, 131

\bibitem[\protect\citeauthoryear{Bundy, Ellis, Conselice, Taylor, Cooper,
  Willmer, Weiner, Coil, Noeske \& Eisenhardt}{Bundy
  et~al.}{2006}]{bundy2006mass}
Bundy K.,  Ellis R.~S.,  Conselice C.~J.,  Taylor J.~E.,  Cooper M.~C.,
  Willmer C.~N.,  Weiner B.~J.,  Coil A.~L.,  Noeske K.~G.,    Eisenhardt
  P.~R.,  2006, The Astrophysical Journal, 651, 120

\bibitem[\protect\citeauthoryear{Calzetti, Armus, Bohlin, Kinney, Koornneef \&
  Storchi-Bergmann}{Calzetti et~al.}{2000}]{calzetti2000dust}
Calzetti D.,  Armus L.,  Bohlin R.~C.,  Kinney A.~L.,  Koornneef J.,
  Storchi-Bergmann T.,  2000, The Astrophysical Journal, 533, 682

\bibitem[\protect\citeauthoryear{Chabrier}{Chabrier}{2003}]{chabrier2003galactic}
Chabrier G.,  2003, Publications of the Astronomical Society of the Pacific,
  115, 763

\bibitem[\protect\citeauthoryear{{Charlot} \& {Bruzual}}{{Charlot} \&
  {Bruzual}}{2007}]{chabrier2007}
{Charlot} S.,  {Bruzual} G.,  2007, provided to the community but not published

\bibitem[\protect\citeauthoryear{Charlot \& Fall}{Charlot \&
  Fall}{2000}]{charlot2000simple}
Charlot S.,  Fall S.~M.,  2000, The Astrophysical Journal, 539, 718

\bibitem[\protect\citeauthoryear{Chiu, Mohr, McDonald, Bocquet, Ashby, Bayliss,
  Benson, Bleem, Brodwin, Desai et~al.,}{Chiu et~al.}{2016}]{Chiu16}
Chiu I.,  Mohr J.,  McDonald M.,  Bocquet S.,  Ashby M.,  Bayliss M.,  Benson
  B.,  Bleem L.,  Brodwin M.,  Desai S.,    et~al., 2016, \mnras, 455, 258

\bibitem[\protect\citeauthoryear{Chiu, Saro, Mohr, Desai, Bocquet, Capasso,
  Gangkofner, Gupta \& Liu}{Chiu et~al.}{2016}]{Chiu2016b}
Chiu I.,  Saro A.,  Mohr J.,  Desai S.,  Bocquet S.,  Capasso R.,  Gangkofner
  C.,  Gupta N.,    Liu J.,  2016, Monthly Notices of the Royal Astronomical
  Society, 458, 379

\bibitem[\protect\citeauthoryear{{Collins}, {Stott}, {Hilton}, {Kay},
  {Stanford}, {Davidson}, {Hosmer}, {Hoyle}, {Liddle}, {Lloyd-Davies}, {Mann},
  {Mehrtens}, {Miller}, {Nichol}, {Romer}, {Sahl{\'e}n}, {Viana} \&
  {West}}{{Collins} et~al.}{2009}]{collins2009}
{Collins} C.~A.,  {Stott} J.~P.,  {Hilton} M.,  {Kay} S.~T.,  {Stanford} S.~A.,
   {Davidson} M.,  {Hosmer} M.,  {Hoyle} B.,  {Liddle} A.,  {Lloyd-Davies} E.,
  {Mann} R.~G.,  {Mehrtens} N.,  {Miller} C.~J.,  {Nichol} R.~C.,  {Romer}
  A.~K.,  {Sahl{\'e}n} M.,  {Viana} P.~T.~P.,    {West} M.~J.,  2009, \nat,
  458, 603

\bibitem[\protect\citeauthoryear{Coupon, Arnouts, van Waerbeke, Moutard,
  Ilbert, van Uitert, Erben, Garilli, Guzzo, Heymans et~al.,}{Coupon
  et~al.}{2015}]{Coupon15}
Coupon J.,  Arnouts S.,  van Waerbeke L.,  Moutard T.,  Ilbert O.,  van Uitert
  E.,  Erben T.,  Garilli B.,  Guzzo L.,  Heymans C.,    et~al., 2015, \mnras,
  449, 1352

\bibitem[\protect\citeauthoryear{David, Jones \& Forman}{David
  et~al.}{1995}]{David95}
David L.~P.,  Jones C.,    Forman W.,  1995, \apj, 445, 578

\bibitem[\protect\citeauthoryear{De~Lucia \& Blaizot}{De~Lucia \&
  Blaizot}{2007}]{deLucia07}
De~Lucia G.,  Blaizot J.,  2007, \mnras, 375, 2

\bibitem[\protect\citeauthoryear{Dunkley, Komatsu, Nolta, Spergel, Larson,
  Hinshaw, Page, Bennett, Gold, Jarosik et~al.,}{Dunkley
  et~al.}{2009}]{dunkley2009five}
Dunkley J.,  Komatsu E.,  Nolta M.,  Spergel D.,  Larson D.,  Hinshaw G.,  Page
  L.,  Bennett C.,  Gold B.,  Jarosik N.,    et~al., 2009, The Astrophysical
  Journal Supplement Series, 180, 306

\bibitem[\protect\citeauthoryear{Dvorkin \& Rephaeli}{Dvorkin \&
  Rephaeli}{2015}]{Dvorkin15}
Dvorkin I.,  Rephaeli Y.,  2015, \mnras, 450, 896

\bibitem[\protect\citeauthoryear{Erfanianfar, Finoguenov, Tanaka, Lerchster,
  Nandra, Laird, Connelly, Bielby, Mirkazemi, Faber et~al.,}{Erfanianfar
  et~al.}{2013}]{erfanianfar13}
Erfanianfar G.,  Finoguenov A.,  Tanaka M.,  Lerchster M.,  Nandra K.,  Laird
  E.,  Connelly J.,  Bielby R.,  Mirkazemi M.,  Faber S.,    et~al., 2013,
  \apj, 765, 117

\bibitem[\protect\citeauthoryear{Ettori, Morandi, Tozzi, Balestra, Borgani,
  Rosati, Lovisari \& Terenziani}{Ettori et~al.}{2009}]{Ettori09}
Ettori S.,  Morandi A.,  Tozzi P.,  Balestra I.,  Borgani S.,  Rosati P.,
  Lovisari L.,    Terenziani F.,  2009, \aap, 501, 61

\bibitem[\protect\citeauthoryear{Evrard}{Evrard}{1997}]{Evrard97}
Evrard A.~E.,  1997, \mnras, 292, 289

\bibitem[\protect\citeauthoryear{Fakhouri, Ma \& Boylan-Kolchin}{Fakhouri
  et~al.}{2010}]{fakhouri2010merger}
Fakhouri O.,  Ma C.-P.,    Boylan-Kolchin M.,  2010, Monthly Notices of the
  Royal Astronomical Society, 406, 2267

\bibitem[\protect\citeauthoryear{Finoguenov, Guzzo, Hasinger, Scoville, Aussel,
  B{\"o}hringer, Brusa, Capak, Cappelluti, Comastri et~al.,}{Finoguenov
  et~al.}{2007}]{Finoguenov07}
Finoguenov A.,  Guzzo L.,  Hasinger G.,  Scoville N.,  Aussel H.,
  B{\"o}hringer H.,  Brusa M.,  Capak P.,  Cappelluti N.,  Comastri A.,
  et~al., 2007, \apjs, 172, 182

\bibitem[\protect\citeauthoryear{Frenk, White, Davis \& Efstathiou}{Frenk
  et~al.}{1988}]{frenk1988formation}
Frenk C.~S.,  White S.~D.,  Davis M.,    Efstathiou G.,  1988, The
  Astrophysical Journal, 327, 507

\bibitem[\protect\citeauthoryear{George, Leauthaud, Bundy, Finoguenov, Tinker,
  Lin, Mei, Kneib, Aussel, Behroozi et~al.,}{George et~al.}{2011}]{George11}
George M.~R.,  Leauthaud A.,  Bundy K.,  Finoguenov A.,  Tinker J.,  Lin Y.-T.,
   Mei S.,  Kneib J.-P.,  Aussel H.,  Behroozi P.~S.,    et~al., 2011, \apj,
  742, 125

\bibitem[\protect\citeauthoryear{Giodini, Finoguenov, Pierini, Zamorani,
  Ilbert, Lilly, Peng, Scoville \& Tanaka}{Giodini
  et~al.}{2012}]{giodini2012galaxy}
Giodini S.,  Finoguenov A.,  Pierini D.,  Zamorani G.,  Ilbert O.,  Lilly S.,
  Peng Y.,  Scoville N.,    Tanaka M.,  2012, Astronomy \& Astrophysics, 538,
  A104

\bibitem[\protect\citeauthoryear{Giodini, Pierini, Finoguenov, Pratt,
  Boehringer, Leauthaud, Guzzo, Aussel, Bolzonella, Capak et~al.,}{Giodini
  et~al.}{2009}]{Giodini09}
Giodini S.,  Pierini D.,  Finoguenov A.,  Pratt G.,  Boehringer H.,  Leauthaud
  A.,  Guzzo L.,  Aussel H.,  Bolzonella M.,  Capak P.,    et~al., 2009, \apj,
  703, 982

\bibitem[\protect\citeauthoryear{Gonzalez, Sivanandam, Zabludoff \&
  Zaritsky}{Gonzalez et~al.}{2013}]{Gonzalez13}
Gonzalez A.~H.,  Sivanandam S.,  Zabludoff A.~I.,    Zaritsky D.,  2013, \apj,
  778, 14

\bibitem[\protect\citeauthoryear{Gozaliasl, Finoguenov, Khosroshahi, Mirkazemi,
  Salvato, Jassur, Erfanianfar, Popesso, Tanaka, Lerchster et~al.,}{Gozaliasl
  et~al.}{2014}]{Gozaliasl14}
Gozaliasl G.,  Finoguenov A.,  Khosroshahi H.,  Mirkazemi M.,  Salvato M.,
  Jassur D.,  Erfanianfar G.,  Popesso P.,  Tanaka M.,  Lerchster M.,
  et~al., 2014, \aap, 566, A140

\bibitem[\protect\citeauthoryear{Gozaliasl, Finoguenov, Khosroshahi, Mirkazemi,
  Erfanianfar \& Tanaka}{Gozaliasl et~al.}{2016}]{gozaliasl2016brightest}
Gozaliasl G.,  Finoguenov A.,  Khosroshahi H.~G.,  Mirkazemi M.,  Erfanianfar
  G.,    Tanaka M.,  2016, Monthly Notices of the Royal Astronomical Society,
  458, 2762

\bibitem[\protect\citeauthoryear{Gozaliasl, Khosroshahi, Dariush, Finoguenov,
  Jassur \& Molaeinezhad}{Gozaliasl et~al.}{2014}]{Gozaliasl14A}
Gozaliasl G.,  Khosroshahi H.,  Dariush A.,  Finoguenov A.,  Jassur D.,
  Molaeinezhad A.,  2014, \aap, 571, A49

\bibitem[\protect\citeauthoryear{Groenewald}{Groenewald}{2016}]{groenewald2016investigating}
Groenewald D.~N.,  2016, PhD thesis, North-West University (South Africa),
  Potchefstroom Campus

\bibitem[\protect\citeauthoryear{Guo, White, Boylan-Kolchin, De~Lucia,
  Kauffmann, Lemson, Li, Springel \& Weinmann}{Guo et~al.}{2011}]{Guo11}
Guo Q.,  White S.,  Boylan-Kolchin M.,  De~Lucia G.,  Kauffmann G.,  Lemson G.,
   Li C.,  Springel V.,    Weinmann S.,  2011, \mnras, 413, 101

\bibitem[\protect\citeauthoryear{Harikane, Ouchi, Ono, More, Saito, Lin,
  Coupon, Shimasaku, Shibuya, Price et~al.,}{Harikane
  et~al.}{2016}]{harikane2016evolution}
Harikane Y.,  Ouchi M.,  Ono Y.,  More S.,  Saito S.,  Lin Y.-T.,  Coupon J.,
  Shimasaku K.,  Shibuya T.,  Price P.~A.,    et~al., 2016, The Astrophysical
  Journal, 821, 123

\bibitem[\protect\citeauthoryear{Harrison, Miller, Richards, Lloyd-Davies,
  Hoyle, Romer, Mehrtens, Hilton, Stott, Capozzi et~al.,}{Harrison
  et~al.}{2012}]{harrison2012xmm}
Harrison C.~D.,  Miller C.~J.,  Richards J.~W.,  Lloyd-Davies E.,  Hoyle B.,
  Romer A.~K.,  Mehrtens N.,  Hilton M.,  Stott J.~P.,  Capozzi D.,    et~al.,
  2012, The Astrophysical Journal, 752, 12

\bibitem[\protect\citeauthoryear{Henriques, White, Thomas, Angulo, Guo, Lemson
  \& Wang}{Henriques et~al.}{2016}]{henriques2016galaxy}
Henriques B.,  White S.~D.,  Thomas P.~A.,  Angulo R.~E.,  Guo Q.,  Lemson G.,
    Wang W.,  2016, arXiv preprint arXiv:1611.02286

\bibitem[\protect\citeauthoryear{Henriques, White, Thomas, Angulo, Guo, Lemson,
  Springel \& Overzier}{Henriques et~al.}{2015}]{Henriques15}
Henriques B.~M.,  White S.~D.,  Thomas P.~A.,  Angulo R.,  Guo Q.,  Lemson G.,
  Springel V.,    Overzier R.,  2015, \mnras, 451, 2663

\bibitem[\protect\citeauthoryear{Henriques, White, Thomas, Angulo, Guo, Lemson
  \& Springel}{Henriques et~al.}{2013}]{henriques2013simulations}
Henriques B.~M.,  White S.~D.,  Thomas P.~A.,  Angulo R.~E.,  Guo Q.,  Lemson
  G.,    Springel V.,  2013, Monthly Notices of the Royal Astronomical Society,
  431, 3373

\bibitem[\protect\citeauthoryear{Hilton, Hasselfield, Sif{\'o}n, Baker,
  Barrientos, Battaglia, Bond, Crichton, Das, Devlin et~al.,}{Hilton
  et~al.}{2013}]{Hilton13}
Hilton M.,  Hasselfield M.,  Sif{\'o}n C.,  Baker A.~J.,  Barrientos L.~F.,
  Battaglia N.,  Bond J.~R.,  Crichton D.,  Das S.,  Devlin M.~J.,    et~al.,
  2013, \mnras, p. stt1535

\bibitem[\protect\citeauthoryear{Hudson, Gillis, Coupon, Hildebrandt, Erben,
  Heymans, Hoekstra, Kitching, Mellier, Miller et~al.,}{Hudson
  et~al.}{2015}]{hudson2015cfhtlens}
Hudson M.~J.,  Gillis B.~R.,  Coupon J.,  Hildebrandt H.,  Erben T.,  Heymans
  C.,  Hoekstra H.,  Kitching T.~D.,  Mellier Y.,  Miller L.,    et~al., 2015,
  Monthly Notices of the Royal Astronomical Society, 447, 298

\bibitem[\protect\citeauthoryear{Ilbert, McCracken, Le~F{\`e}vre, Capak,
  Dunlop, Karim, Renzini, Caputi, Boissier, Arnouts et~al.,}{Ilbert
  et~al.}{2013}]{Ilbert13}
Ilbert O.,  McCracken H.,  Le~F{\`e}vre O.,  Capak P.,  Dunlop J.,  Karim A.,
  Renzini M.,  Caputi K.,  Boissier S.,  Arnouts S.,    et~al., 2013, \aap,
  556, A55

\bibitem[\protect\citeauthoryear{Ilbert, Salvato, Le~Floc'h, Aussel, Capak,
  McCracken, Mobasher, Kartaltepe, Scoville, Sanders et~al.,}{Ilbert
  et~al.}{2010b}]{ilbert2010galaxy}
Ilbert O.,  Salvato M.,  Le~Floc'h E.,  Aussel H.,  Capak P.,  McCracken H.,
  Mobasher B.,  Kartaltepe J.,  Scoville N.,  Sanders D.,    et~al., 2010b, The
  Astrophysical Journal, 709, 644

\bibitem[\protect\citeauthoryear{Ilbert, Salvato, Le~Floc'h, Aussel, Capak,
  McCracken, Mobasher, Kartaltepe, Scoville, Sanders et~al.,}{Ilbert
  et~al.}{2010a}]{Ilbert10}
Ilbert O.,  Salvato M.,  Le~Floc'h E.,  Aussel H.,  Capak P.,  McCracken H.,
  Mobasher B.,  Kartaltepe J.,  Scoville N.,  Sanders D.,    et~al., 2010a,
  \apj, 709, 644

\bibitem[\protect\citeauthoryear{Kettula, Giodini, Van~Uitert, Hoekstra,
  Finoguenov, Lerchster, Erben, Heymans, Hildebrandt, Kitching et~al.,}{Kettula
  et~al.}{2015}]{Kettula15}
Kettula K.,  Giodini S.,  Van~Uitert E.,  Hoekstra H.,  Finoguenov A.,
  Lerchster M.,  Erben T.,  Heymans C.,  Hildebrandt H.,  Kitching T.,
  et~al., 2015, \mnras, 451, 5978

\bibitem[\protect\citeauthoryear{Khosroshahi, Ponman \& Jones}{Khosroshahi
  et~al.}{2007}]{khosroshahi2007scaling}
Khosroshahi H.~G.,  Ponman T.~J.,    Jones L.~R.,  2007, Monthly Notices of the
  Royal Astronomical Society, 377, 595

\bibitem[\protect\citeauthoryear{Kravtsov, Berlind, Wechsler, Klypin,
  Gottl{\"o}ber, Allgood \& Primack}{Kravtsov et~al.}{2004}]{kravtsov2004dark}
Kravtsov A.~V.,  Berlind A.~A.,  Wechsler R.~H.,  Klypin A.~A.,  Gottl{\"o}ber
  S.,  Allgood B.,    Primack J.~R.,  2004, The Astrophysical Journal, 609, 35

\bibitem[\protect\citeauthoryear{Laigle, Capak \& Scoville}{Laigle
  et~al.}{2016}]{laigle2016cosmos2015}
Laigle C.,  Capak P.,    Scoville N.,  2016, Astrophysical Journal Supplement
  Series, 224, Art

\bibitem[\protect\citeauthoryear{Leauthaud, Finoguenov, Kneib, Taylor, Massey,
  Rhodes, Ilbert, Bundy, Tinker, George et~al.,}{Leauthaud
  et~al.}{2010}]{leauthaud2009weak}
Leauthaud A.,  Finoguenov A.,  Kneib J.-P.,  Taylor J.~E.,  Massey R.,  Rhodes
  J.,  Ilbert O.,  Bundy K.,  Tinker J.,  George M.~R.,    et~al., 2010, The
  Astrophysical Journal, 709, 97

\bibitem[\protect\citeauthoryear{Leauthaud, George, Behroozi, Bundy, Tinker,
  Wechsler, Conroy, Finoguenov \& Tanaka}{Leauthaud et~al.}{2012}]{Leauthaud12}
Leauthaud A.,  George M.~R.,  Behroozi P.~S.,  Bundy K.,  Tinker J.,  Wechsler
  R.~H.,  Conroy C.,  Finoguenov A.,    Tanaka M.,  2012, \apj, 746, 95

\bibitem[\protect\citeauthoryear{Lewis, Balogh, De~Propris, Couch, Bower,
  Offer, Bland-Hawthorn, Baldry, Baugh, Bridges et~al.,}{Lewis
  et~al.}{2002}]{lewis20022df}
Lewis I.,  Balogh M.,  De~Propris R.,  Couch W.,  Bower R.,  Offer A.,
  Bland-Hawthorn J.,  Baldry I.~K.,  Baugh C.,  Bridges T.,    et~al., 2002,
  Monthly Notices of the Royal Astronomical Society, 334, 673

\bibitem[\protect\citeauthoryear{{Lidman}, {Suherli}, {Muzzin}, {Wilson},
  {Demarco}, {Brough}, {Rettura}, {Cox}, {DeGroot}, {Yee}, {Gilbank},
  {Hoekstra}, {Balogh}, {Ellingson}, {Hicks}, {Nantais}, {Noble}, {Lacy},
  {Surace} \& {Webb}}{{Lidman} et~al.}{2012}]{Lidman2012}
{Lidman} C.,  {Suherli} J.,  {Muzzin} A.,  {Wilson} G.,  {Demarco} R.,
  {Brough} S.,  {Rettura} A.,  {Cox} J.,  {DeGroot} A.,  {Yee} H.~K.~C.,
  {Gilbank} D.,  {Hoekstra} H.,  {Balogh} M.,  {Ellingson} E.,  {Hicks} A.,
  {Nantais} J.,  {Noble} A.,  {Lacy} M.,  {Surace} J.,    {Webb} T.,  2012,
  \mnras, 427, 550

\bibitem[\protect\citeauthoryear{Lin, Mohr \& Stanford}{Lin
  et~al.}{2003}]{Lin03}
Lin Y.-T.,  Mohr J.~J.,    Stanford S.~A.,  2003, \apj, 591, 749

\bibitem[\protect\citeauthoryear{McCarthy, Bower \& Balogh}{McCarthy
  et~al.}{2007}]{McCarthy07}
McCarthy I.~G.,  Bower R.~G.,    Balogh M.~L.,  2007, \mnras, 377, 1457

\bibitem[\protect\citeauthoryear{McCarthy, Schaye, Ponman, Bower, Booth,
  Dalla~Vecchia, Crain, Springel, Theuns \& Wiersma}{McCarthy
  et~al.}{2010}]{mccarthy2010case}
McCarthy I.~G.,  Schaye J.,  Ponman T.~J.,  Bower R.~G.,  Booth C.~M.,
  Dalla~Vecchia C.,  Crain R.~A.,  Springel V.,  Theuns T.,    Wiersma R.~P.,
  2010, Monthly Notices of the Royal Astronomical Society, 406, 822

\bibitem[\protect\citeauthoryear{McCracken, Milvang-Jensen, Dunlop, Franx,
  Fynbo, Le~F{\`e}vre, Holt, Caputi, Goranova, Buitrago et~al.,}{McCracken
  et~al.}{2012}]{mccracken2012ultravista}
McCracken H.,  Milvang-Jensen B.,  Dunlop J.,  Franx M.,  Fynbo J.,
  Le~F{\`e}vre O.,  Holt J.,  Caputi K.,  Goranova Y.,  Buitrago F.,    et~al.,
  2012, \aap, 544, A156

\bibitem[\protect\citeauthoryear{McDonald, Stalder, Bayliss, Allen, Applegate,
  Ashby, Bautz, Benson, Bleem, Brodwin et~al.,}{McDonald
  et~al.}{2016}]{mcdonald2016star}
McDonald M.,  Stalder B.,  Bayliss M.,  Allen S.,  Applegate D.,  Ashby M.,
  Bautz M.,  Benson B.,  Bleem L.,  Brodwin M.,    et~al., 2016, The
  Astrophysical Journal, 817, 86

\bibitem[\protect\citeauthoryear{McGaugh, Schombert, De~Blok \&
  Zagursky}{McGaugh et~al.}{2009}]{McGaugh09}
McGaugh S.~S.,  Schombert J.~M.,  De~Blok W.,    Zagursky M.~J.,  2009, \apj,
  708, L14

\bibitem[\protect\citeauthoryear{Maraston}{Maraston}{2005}]{maraston2005evolutionary}
Maraston C.,  2005, Monthly Notices of the Royal Astronomical Society, 362, 799

\bibitem[\protect\citeauthoryear{Mathews, Faltenbacher, Brighenti \&
  Buote}{Mathews et~al.}{2005}]{mathews2005baryonically}
Mathews W.~G.,  Faltenbacher A.,  Brighenti F.,    Buote D.~A.,  2005, The
  Astrophysical Journal Letters, 634, L137

\bibitem[\protect\citeauthoryear{{McCarthy}, {Schaye}, {Bower}, {Ponman},
  {Booth}, {Dalla Vecchia} \& {Springel}}{{McCarthy}
  et~al.}{2011}]{mccarthy2011}
{McCarthy} I.~G.,  {Schaye} J.,  {Bower} R.~G.,  {Ponman} T.~J.,  {Booth}
  C.~M.,  {Dalla Vecchia} C.,    {Springel} V.,  2011, \mnras, 412, 1965

\bibitem[\protect\citeauthoryear{Mo \& White}{Mo \&
  White}{2002}]{mo2002abundance}
Mo H.,  White S.,  2002, Monthly Notices of the Royal Astronomical Society,
  336, 112

\bibitem[\protect\citeauthoryear{Mohr, Mathiesen \& Evrard}{Mohr
  et~al.}{1999}]{Mohr99}
Mohr J.~J.,  Mathiesen B.,    Evrard A.~E.,  1999, \apj, 517, 627

\bibitem[\protect\citeauthoryear{{Moster}, {Naab} \& {White}}{{Moster}
  et~al.}{2013}]{Moster13}
{Moster} B.~P.,  {Naab} T.,    {White} S.~D.~M.,  2013, \mnras, 428, 3121

\bibitem[\protect\citeauthoryear{Moster, Somerville, Maulbetsch, Van~den Bosch,
  Macci{\`o}, Naab \& Oser}{Moster et~al.}{2010}]{Moster10}
Moster B.~P.,  Somerville R.~S.,  Maulbetsch C.,  Van~den Bosch F.~C.,
  Macci{\`o} A.~V.,  Naab T.,    Oser L.,  2010, \apj, 710, 903

\bibitem[\protect\citeauthoryear{Rosenblatt}{Rosenblatt}{1956}]{rosenblatt1956}
Rosenblatt M.,  1956, Ann. Math. Statist., 27, 832

\bibitem[\protect\citeauthoryear{Roussel, Sadat \& Blanchard}{Roussel
  et~al.}{2000}]{Roussel00}
Roussel H.,  Sadat R.,    Blanchard A.,  2000, Arxiv preprint astro-ph/0007168

\bibitem[\protect\citeauthoryear{Simionescu, Allen, Mantz, Werner, Takei,
  Morris, Fabian, Sanders, Nulsen, George et~al.,}{Simionescu
  et~al.}{2011}]{Simionescu11}
Simionescu A.,  Allen S.~W.,  Mantz A.,  Werner N.,  Takei Y.,  Morris R.~G.,
  Fabian A.~C.,  Sanders J.~S.,  Nulsen P.~E.,  George M.~R.,    et~al., 2011,
  Science, 331, 1576

\bibitem[\protect\citeauthoryear{{Springel}, {White}, {Jenkins}, {Frenk},
  {Yoshida}, {Gao}, {Navarro}, {Thacker}, {Croton}, {Helly}, {Peacock}, {Cole},
  {Thomas}, {Couchman}, {Evrard}, {Colberg} \& {Pearce}}{{Springel}
  et~al.}{2005}]{springle2005cosmological}
{Springel} V.,  {White} S.~D.~M.,  {Jenkins} A.,  {Frenk} C.~S.,  {Yoshida} N.,
   {Gao} L.,  {Navarro} J.,  {Thacker} R.,  {Croton} D.,  {Helly} J.,
  {Peacock} J.~A.,  {Cole} S.,  {Thomas} P.,  {Couchman} H.,  {Evrard} A.,
  {Colberg} J.,    {Pearce} F.,  2005, \nat, 435, 629

\bibitem[\protect\citeauthoryear{Stott, Collins, Sahl{\'e}n, Hilton,
  Lloyd-Davies, Capozzi, Hosmer, Liddle, Mehrtens, Miller et~al.,}{Stott
  et~al.}{2010}]{Stott10}
Stott J.,  Collins C.,  Sahl{\'e}n M.,  Hilton M.,  Lloyd-Davies E.,  Capozzi
  D.,  Hosmer M.,  Liddle A.,  Mehrtens N.,  Miller C.,    et~al., 2010, \apj,
  718, 23

\bibitem[\protect\citeauthoryear{van~der Burg, Muzzin, Hoekstra, Wilson, Lidman
  \& Yee}{van~der Burg et~al.}{2014}]{Burg14}
van~der Burg R.~F.,  Muzzin A.,  Hoekstra H.,  Wilson G.,  Lidman C.,    Yee
  H.,  2014, \aap, 561, A79

\bibitem[\protect\citeauthoryear{{van der Burg}, {Muzzin}, {Hoekstra},
  {Wilson}, {Lidman} \& {Yee}}{{van der Burg} et~al.}{2014}]{vanderburg2014}
{van der Burg} R.~F.~J.,  {Muzzin} A.,  {Hoekstra} H.,  {Wilson} G.,  {Lidman}
  C.,    {Yee} H.~K.~C.,  2014, \aap, 561, A79

\bibitem[\protect\citeauthoryear{Vikhlinin, Kravtsov, Forman, Jones,
  Markevitch, Murray \& Van~Speybroeck}{Vikhlinin
  et~al.}{2006}]{vikhlinin2006chandra}
Vikhlinin A.,  Kravtsov A.,  Forman W.,  Jones C.,  Markevitch M.,  Murray S.,
    Van~Speybroeck L.,  2006, The Astrophysical Journal, 640, 691

\bibitem[\protect\citeauthoryear{Wake, Whitaker, Labb{\'e}, Van~Dokkum, Franx,
  Quadri, Brammer, Kriek, Lundgren, Marchesini et~al.,}{Wake
  et~al.}{2011}]{Wake11}
Wake D.~A.,  Whitaker K.~E.,  Labb{\'e} I.,  Van~Dokkum P.~G.,  Franx M.,
  Quadri R.,  Brammer G.,  Kriek M.,  Lundgren B.~F.,  Marchesini D.,
  et~al., 2011, \apj, 728, 46

\bibitem[\protect\citeauthoryear{Wetzel, Tinker \& Conroy}{Wetzel
  et~al.}{2012}]{wetzel2012galaxy}
Wetzel A.~R.,  Tinker J.~L.,    Conroy C.,  2012, Monthly Notices of the Royal
  Astronomical Society, 424, 232

\bibitem[\protect\citeauthoryear{White \& Frenk}{White \&
  Frenk}{1991}]{white1991galaxy}
White S.~D.,  Frenk C.~S.,  1991, The Astrophysical Journal, 379, 52

\bibitem[\protect\citeauthoryear{White, Navarro, Evrard \& Frenk}{White
  et~al.}{1993}]{White93}
White S.~D.,  Navarro J.~F.,  Evrard A.~E.,    Frenk C.~S.,  1993, Nature, 366,
  429

\bibitem[\protect\citeauthoryear{Wuyts, Schreiber, van~der Wel, Magnelli, Guo,
  Genzel, Lutz, Aussel, Barro, Berta et~al.,}{Wuyts
  et~al.}{2011}]{wuyts2011galaxy}
Wuyts S.,  Schreiber N. M.~F.,  van~der Wel A.,  Magnelli B.,  Guo Y.,  Genzel
  R.,  Lutz D.,  Aussel H.,  Barro G.,  Berta S.,    et~al., 2011, The
  Astrophysical Journal, 742, 96

\bibitem[\protect\citeauthoryear{Yang, Mo \& Van Den~Bosch}{Yang
  et~al.}{2003}]{yang2003constraining}
Yang X.,  Mo H.,    Van Den~Bosch F.~C.,  2003, Monthly Notices of the Royal
  Astronomical Society, 339, 1057

\bibitem[\protect\citeauthoryear{Yang, Mo \& Van~den Bosch}{Yang
  et~al.}{2009}]{Yang09}
Yang X.,  Mo H.,    Van~den Bosch F.~C.,  2009, \apj, 695, 900

\bibitem[\protect\citeauthoryear{Yang, Mo, Van~den Bosch, Pasquali, Li \&
  Barden}{Yang et~al.}{2007}]{yang2007galaxy}
Yang X.,  Mo H.,  Van~den Bosch F.~C.,  Pasquali A.,  Li C.,    Barden M.,
  2007, The Astrophysical Journal, 671, 153

\bibitem[\protect\citeauthoryear{Yang, Mo, van~den Bosch, Zhang \& Han}{Yang
  et~al.}{2012}]{Yang12}
Yang X.,  Mo H.,  van~den Bosch F.~C.,  Zhang Y.,    Han J.,  2012, \apj, 752,
  41

\bibitem[\protect\citeauthoryear{Zheng, Coil \& Zehavi}{Zheng
  et~al.}{2007}]{Zheng07}
Zheng Z.,  Coil A.~L.,    Zehavi I.,  2007, \apj, 667, 760

\end{thebibliography}


@article{Allen04,
  title={Constraints on dark energy from Chandra observations of the largest relaxed galaxy clusters},
  author={Allen, SW and Schmidt, RW and Ebeling, H and Fabian, AC and Van Speybroeck, L},
  journal={\mnras},
  volume={353},
  number={2},
  pages={457--467},
  year={2004},
  publisher={Oxford University Press}
}
@article{Allen08,
	title={Improved constraints on dark energy from Chandra X-ray observations of the largest relaxed galaxy clusters},
	author={Allen, SW and Rapetti, DA and Schmidt, RW and Ebeling, H and Morris, RG and Fabian, AC},
	journal={\mnras},
	volume={383},
	number={3},
	pages={879--896},
	year={2008},
	publisher={Oxford University Press}
}
@article{Andreon10,
	title={The stellar mass fraction and baryon content of galaxy clusters and groups},
	author={Andreon, S},
	journal={\mnras},
	volume={407},
	number={1},
	pages={263--276},
	year={2010},
	publisher={Oxford University Press}
}
@article{Allen11,
	title={Cosmological parameters from observations of galaxy clusters},
	author={Allen, Steven W and Evrard, August E and Mantz, Adam B},
	journal={arXiv preprint arXiv:1103.4829},
	year={2011}
}
@article{Balogh14,
	title={The GEEC2 spectroscopic survey of Galaxy groups at 0.8< z< 1},
	author={Balogh, Michael L and McGee, Sean L and Mok, Angus and Wilman, David J and Finoguenov, Alexis and Bower, Richard G and Mulchaey, John S and Parker, Laura C and Tanaka, Masayuki},
	journal={\mnras},
	volume={443},
	number={3},
	pages={2679--2694},
	year={2014},
	publisher={Oxford University Press}
}
@article{Behroozi13,
	title={The average star formation histories of galaxies in dark matter halos from z= 0-8},
	author={Behroozi, Peter S and Wechsler, Risa H and Conroy, Charlie},
	journal={\apj},
	volume={770},
	number={1},
	pages={57},
	year={2013},
	publisher={IOP Publishing}
}
@article{Behroozi10,
	title={A comprehensive analysis of uncertainties affecting the stellar mass-halo mass relation for 0< z< 4},
	author={Behroozi, Peter S and Conroy, Charlie and Wechsler, Risa H},
	journal={\apj},
	volume={717},
	number={1},
	pages={379},
	year={2010},
	publisher={IOP Publishing}
}
@article{Bell2003,
	title={The optical and near-infrared properties of galaxies. I. Luminosity and stellar mass functions},
	author={Bell, Eric F and McIntosh, Daniel H and Katz, Neal and Weinberg, Martin D},
	journal={\apjs},
	volume={149},
	number={2},
	pages={289},
	year={2003},
	publisher={IOP Publishing}
}
@article{bernardi07,
	title={The luminosities, sizes, and velocity dispersions of brightest cluster galaxies: implications for formation history},
	author={Bernardi, Mariangela and Hyde, Joseph B and Sheth, Ravi K and Miller, Chris J and Nichol, Robert C},
	journal={\aj},
	volume={133},
	number={4},
	pages={1741},
	year={2007},
	publisher={IOP Publishing}
}
@article{bernstein01,
	title={Models for the magnitude-distribution of brightest cluster galaxies},
	author={Bernstein, JP and Bhavsar, Suketu P},
	journal={\mnras},
	volume={322},
	number={3},
	pages={625--630},
	year={2001},
	publisher={Oxford University Press}
}
@article{Bower06,
	title={Breaking the hierarchy of galaxy formation},
	author={Bower, RG and Benson, AJ and Malbon, R and Helly, JC and Frenk, CS and Baugh, CM and Cole, Shaun and Lacey, Cedric G},
	journal={\mnras},
	volume={370},
	number={2},
	pages={645--655},
	year={2006},
	publisher={Oxford University Press}
}
@article{bhavsar85,
	title={First ranked galaxies in groups and clusters},
	author={Bhavsar, Suketu P and Barrow, John D},
	journal={\mnras},
	volume={213},
	number={4},
	pages={857--869},
	year={1985},
	publisher={Oxford University Press}
}
@article{Bulbul16,
	title={Probing the Outskirts of the Early-Stage Galaxy Cluster Merger A1750},
	author={Bulbul, Esra and Randall, Scott W and Bayliss, Matthew and Miller, Eric and Andrade-Santos, Felipe and Johnson, Ryan and Bautz, Mark and Blanton, Elizabeth L and Forman, William R and Jones, Christine and others},
	journal={\apj},
	volume={818},
	number={2},
	pages={131},
	year={2016},
	publisher={IOP Publishing}
}
@article{Chiu16,
	title={Baryon content of massive galaxy clusters at 0.57< z< 1.33},
	author={Chiu, I and Mohr, J and McDonald, M and Bocquet, S and Ashby, MLN and Bayliss, M and Benson, BA and Bleem, LE and Brodwin, M and Desai, S and others},
	journal={\mnras},
	volume={455},
	number={1},
	pages={258--275},
	year={2016},
	publisher={Oxford University Press}
}
@article{Coupon15,
	title={The galaxy--halo connection from a joint lensing, clustering and abundance analysis in the CFHTLenS/VIPERS field},
	author={Coupon, J and Arnouts, S and van Waerbeke, L and Moutard, T and Ilbert, O and van Uitert, E and Erben, T and Garilli, B and Guzzo, L and Heymans, C and others},
	journal={\mnras},
	volume={449},
	number={2},
	pages={1352--1379},
	year={2015},
	publisher={Oxford University Press}
}
@article{David95,
	title={Cosmological implications of ROSAT observations of groups and clusters of galaxies},
	author={David, Laurence P and Jones, Christine and Forman, William},
	journal={\apj},
	volume={445},
	pages={578--590},
	year={1995}
}
@article{deLucia07,
	title={The hierarchical formation of the brightest cluster galaxies},
	author={De Lucia, Gabriella and Blaizot, J{\'e}r{\'e}my},
	journal={\mnras},
	volume={375},
	number={1},
	pages={2--14},
	year={2007},
	publisher={Oxford University Press}
}
@article{Dvorkin15,
	title={Evolution of the gas mass fraction in galaxy clusters},
	author={Dvorkin, Irina and Rephaeli, Yoel},
	journal={\mnras},
	volume={450},
	number={1},
	pages={896--904},
	year={2015},
	publisher={Oxford University Press}
}
@article{erfanianfar13,
	title={X-ray groups of galaxies in the AEGIS deep and wide fields},
	author={Erfanianfar, G and Finoguenov, A and Tanaka, M and Lerchster, M and Nandra, K and Laird, E and Connelly, JL and Bielby, R and Mirkazemi, M and Faber, SM and others},
	journal={\apj},
	volume={765},
	number={2},
	pages={117},
	year={2013},
	publisher={IOP Publishing}
}
@article{Evrard97,
	title={The Intracluster gas fraction in X-ray clusters: Constraints on the clustered mass density},
	author={Evrard, August E},
	journal={\mnras},
	volume={292},
	number={2},
	pages={289--297},
	year={1997},
	publisher={Oxford University Press}
}
@article{Ettori09,
	title={The cluster gas mass fraction as a cosmological probe: a revised study},
	author={Ettori, S and Morandi, Andrea and Tozzi, P and Balestra, I and Borgani, S and Rosati, P and Lovisari, L and Terenziani, F},
	journal={\aap},
	volume={501},
	number={1},
	pages={61--73},
	year={2009},
	publisher={EDP Sciences}
}
@article{Ettori06,
	title={The baryon fraction in hydrodynamical simulations of galaxy clusters},
	author={Ettori, S and Dolag, K and Borgani, S and Murante, G},
	journal={\mnras},
	volume={365},
	number={3},
	pages={1021--1030},
	year={2006},
	publisher={Oxford University Press}
}
@article{Finoguenov07,
	title={The XMM-Newton Wide-Field Survey in the COSMOS Field: Statistical Properties of Clusters of Galaxies},
	author={Finoguenov, A and Guzzo, L and Hasinger, G and Scoville, NZ and Aussel, H and B{\"o}hringer, H and Brusa, M and Capak, P and Cappelluti, N and Comastri, A and others},
	journal={\apjs},
	volume={172},
	number={1},
	pages={182},
	year={2007},
	publisher={IOP Publishing}
}
@article{George11,
	title={Galaxies in X-Ray Groups. I. Robust Membership Assignment and the Impact of Group Environments on Quenching},
	author={George, Matthew R and Leauthaud, Alexie and Bundy, Kevin and Finoguenov, Alexis and Tinker, Jeremy and Lin, Yen-Ting and Mei, Simona and Kneib, Jean-Paul and Aussel, Herv{\'e} and Behroozi, Peter S and others},
	journal={\apj},
	volume={742},
	number={2},
	pages={125},
	year={2011},
	publisher={IOP Publishing}
}
@article{Giodini09,
	title={STELLAR AND TOTAL BARYON MASS FRACTIONS IN GROUPS AND CLUSTERS SINCE REDSHIFT 1Based on observations obtained with XMM-Newton, an ESA science mission with instruments and contributions directly funded by ESA Member States and NASA; also based on data collected at: the NASA/ESA Hubble Space Telescope, obtained at the Space Telescope Science Institute, which is operated by AURA Inc., under NASA contract NAS 5-26555, the Subaru Telescope, which is operated by the National Astronomical Observatory of Japan, the European Southern Observatory, Chile, under Large Program 175. A-0839, and the Canada-France-Hawaii Telescope operated by the National Research Council of Canada, the Centre National de la Recherche Scientifique de France and the University of Hawaii.},
	author={Giodini, S and Pierini, D and Finoguenov, A and Pratt, GW and Boehringer, H and Leauthaud, A and Guzzo, L and Aussel, H and Bolzonella, M and Capak, P and others},
	journal={\apj},
	volume={703},
	number={1},
	pages={982},
	year={2009},
	publisher={IOP Publishing}
}
@article{Gonzalez13,
	title={GALAXY CLUSTER BARYON FRACTIONS REVISITEDBased on observations obtained with XMM-Newton, an ESA science mission with instruments and contributions directly funded by ESA Member States and NASA.},
	author={Gonzalez, Anthony H and Sivanandam, Suresh and Zabludoff, Ann I and Zaritsky, Dennis},
	journal={\apj},
	volume={778},
	number={1},
	pages={14},
	year={2013},
	publisher={IOP Publishing}
}
@article{Gonzalez07,
	title={A census of baryons in galaxy clusters and groups},
	author={Gonzalez, Anthony H and Zaritsky, Dennis and Zabludoff, Ann I},
	journal={\apj},
	volume={666},
	number={1},
	pages={147},
	year={2007},
	publisher={IOP Publishing}
}
@article{Gozaliasl14,
	title={Mining the gap: evolution of the magnitude gap in X-ray galaxy groups from the 3-square-degree XMM coverage of CFHTLS},
	author={Gozaliasl, G and Finoguenov, A and Khosroshahi, HG and Mirkazemi, M and Salvato, M and Jassur, DMZ and Erfanianfar, G and Popesso, P and Tanaka, M and Lerchster, M and others},
	journal={\aap},
	volume={566},
	pages={A140},
	year={2014},
	publisher={EDP Sciences}
}
@article{Gozaliasl14A,
	title={Evolution of the galaxy luminosity function in progenitors of fossil groups},
	author={Gozaliasl, G and Khosroshahi, HG and Dariush, AA and Finoguenov, A and Jassur, DMZ and Molaeinezhad, A},
	journal={\aap},
	volume={571},
	pages={A49},
	year={2014},
	publisher={EDP Sciences}
}
@article{Guo11,
	title={From dwarf spheroidals to cD galaxies: simulating the galaxy population in a $\Lambda$CDM cosmology},
	author={Guo, Qi and White, Simon and Boylan-Kolchin, Michael and De Lucia, Gabriella and Kauffmann, Guinevere and Lemson, Gerard and Li, Cheng and Springel, Volker and Weinmann, Simone},
	journal={\mnras},
	volume={413},
	number={1},
	pages={101--131},
	year={2011},
	publisher={Oxford University Press}
}
@article{Gunn75,
	title={Spectrophotometry of faint cluster galaxies and the Hubble diagram-an approach to cosmology},
	author={Gunn, JE and Oke, JB},
	journal={\apj},
	volume={195},
	pages={255--268},
	year={1975}
}
@article{Henriques15,
	title={Galaxy formation in the Planck cosmology--I. Matching the observed evolution of star formation rates, colours and stellar masses},
	author={Henriques, Bruno MB and White, Simon DM and Thomas, Peter A and Angulo, Raul and Guo, Qi and Lemson, Gerard and Springel, Volker and Overzier, Roderik},
	journal={\mnras},
	volume={451},
	number={3},
	pages={2663--2680},
	year={2015},
	publisher={Oxford University Press}
}
@article{Hilton13,
	title={The Atacama Cosmology Telescope: the stellar content of galaxy clusters selected using the Sunyaev--Zel'dovich effect},
	author={Hilton, Matt and Hasselfield, Matthew and Sif{\'o}n, Crist{\'o}bal and Baker, Andrew J and Barrientos, L Felipe and Battaglia, Nicholas and Bond, J Richard and Crichton, Devin and Das, Sudeep and Devlin, Mark J and others},
	journal={\mnras},
	pages={stt1535},
	year={2013},
	publisher={Oxford University Press}
}
@article{Hoessel85,
	title={CCD observations of Abell clusters. IV-Surface photometry of 175 brightest cluster galaxies},
	author={Hoessel, JG and Schneider, DP},
	journal={\aj},
	volume={90},
	pages={1648--1664},
	year={1985}
}
@article{Ilbert10,
	title={GALAXY STELLAR MASS ASSEMBLY BETWEEN 0.2< z< 2 FROM THE S-COSMOS SURVEYBased on observations with the NASA/ESA Hubble Space Telescope, obtained at the Space Telescope Science Institute, which is operated by AURA Inc., under NASA contract NAS 5-26555. Also based on observations made with the Spitzer Space Telescope, which is operated by the Jet Propulsion Laboratory, California Institute of Technology, under NASA contract 1407. Also based on data collected at: the Subaru Telescope, which is operated by the National Astronomical Observatory of Japan; the XMM-Newton, an ESA science mission with instruments and contributions directly funded by ESA Member States and NASA; the European Southern Observatory under Large Program 175. A-0839, Chile; Kitt Peak National Observatory, Cerro Tololo Inter-American Observatory, and the National Optical Astronomy Observatory, which are operated by the Association of Universities for Research in Astronomy, Inc.(AURA) under cooperative agreement with the National Science Foundation; and the Canada-France-Hawaii Telescope with MegaPrime/MegaCam operated as a joint project by the CFHT Corporation, CEA/DAPNIA, the NRC and CADC of Canada, the CNRS of France, TERAPIX, and the University of Hawaii.},
	author={Ilbert, O and Salvato, M and Le Floc'h, E and Aussel, H and Capak, P and McCracken, HJ and Mobasher, B and Kartaltepe, J and Scoville, N and Sanders, DB and others},
	journal={\apj},
	volume={709},
	number={2},
	pages={644},
	year={2010},
	publisher={IOP Publishing}
}
@article{Ilbert13,
	title={Mass assembly in quiescent and star-forming galaxies since z≃ 4 from UltraVISTA},
	author={Ilbert, O and McCracken, HJ and Le F{\`e}vre, O and Capak, P and Dunlop, J and Karim, A and Renzini, MA and Caputi, K and Boissier, S and Arnouts, S and others},
	journal={\aap},
	volume={556},
	pages={A55},
	year={2013},
	publisher={EDP Sciences}
}
@article{Kettula15,
	title={CFHTLenS: weak lensing calibrated scaling relations for low-mass clusters of galaxies},
	author={Kettula, K and Giodini, S and Van Uitert, E and Hoekstra, H and Finoguenov, A and Lerchster, M and Erben, T and Heymans, C and Hildebrandt, H and Kitching, TD and others},
	journal={\mnras},
	volume={451},
	number={2},
	pages={5978--5999},
	year={2015},
	publisher={Oxford University Press}
}
@article{Khosroshahi14,
	title={Optically selected fossil groups; X-ray observations and galaxy properties},
	author={Khosroshahi, Habib G and Gozaliasl, Ghassem and Rasmussen, Jesper and Molaeinezhad, Alireza and Ponman, Trevor and Dariush, Ali A and Sanderson, Alastair JR},
	journal={\mnras},
	volume={443},
	number={1},
	pages={318--327},
	year={2014},
	publisher={Oxford University Press}
}
@article{Leauthaud12,
	title={The integrated stellar content of dark matter halos},
	author={Leauthaud, Alexie and George, Matthew R and Behroozi, Peter S and Bundy, Kevin and Tinker, Jeremy and Wechsler, Risa H and Conroy, Charlie and Finoguenov, Alexis and Tanaka, Masayuki},
	journal={\apj},
	volume={746},
	number={1},
	pages={95},
	year={2012},
	publisher={IOP Publishing}
}
@article{Liu09,
	title={Major dry mergers in early-type brightest cluster galaxies},
	author={Liu, FS and Mao, Shude and Deng, ZG and Xia, XY and Wen, ZL},
	journal={\mnras},
	volume={396},
	number={4},
	pages={2003--2010},
	year={2009},
	publisher={Oxford University Press}
}
@article{Lin03,
	title={Near-infrared properties of galaxy clusters: luminosity as a binding mass predictor and the state of cluster baryons},
	author={Lin, Yen-Ting and Mohr, Joseph J and Stanford, S Adam},
	journal={\apj},
	volume={591},
	number={2},
	pages={749},
	year={2003},
	publisher={IOP Publishing}
}
@article{McCarthy07,
	title={Revisiting the baryon fractions of galaxy clusters: a comparison with WMAP 3-yr results},
	author={McCarthy, Ian G and Bower, Richard G and Balogh, Michael L},
	journal={\mnras},
	volume={377},
	number={4},
	pages={1457--1463},
	year={2007},
	publisher={Oxford University Press}
}
@article{McGaugh09,
	title={The baryon content of cosmic structures},
	author={McGaugh, Stacy S and Schombert, James M and De Blok, WJG and Zagursky, Matthew J},
	journal={\apj},
	volume={708},
	number={1},
	pages={L14},
	year={2009},
	publisher={IOP Publishing}
}
@article{Mohr99,
	title={Properties of the intracluster medium in an ensemble of nearby galaxy clusters},
	author={Mohr, Joseph J and Mathiesen, Benjamin and Evrard, August E},
	journal={\apj},
	volume={517},
	number={2},
	pages={627},
	year={1999},
	publisher={IOP Publishing}
}
@ARTICLE{Moster13,
	author = {{Moster}, B.~P. and {Naab}, T. and {White}, S.~D.~M.},
	title = "{Galactic star formation and accretion histories from matching galaxies to dark matter haloes}",
	journal = {\mnras},
	archivePrefix = "arXiv",
	eprint = {1205.5807},
	keywords = {galaxies: evolution, galaxies: high-redshift, galaxies: statistics, galaxies: stellar content, cosmology: theory, dark matter},
	year = 2013,
	month = feb,
	volume = 428,
	pages = {3121-3138},
	doi = {10.1093/mnras/sts261},
	adsurl = {http://adsabs.harvard.edu/abs/2013MNRAS.428.3121M},
	adsnote = {Provided by the SAO/NASA Astrophysics Data System}
}

@article{Moster10,
	title={Constraints on the relationship between stellar mass and halo mass at low and high redshift},
	author={Moster, Benjamin P and Somerville, Rachel S and Maulbetsch, Christian and Van den Bosch, Frank C and Macci{\`o}, Andrea V and Naab, Thorsten and Oser, Ludwig},
	journal={\apj},
	volume={710},
	number={2},
	pages={903},
	year={2010},
	publisher={IOP Publishing}
}
@article{Panter07,
	title={The star formation histories of galaxies in the Sloan Digital Sky Survey},
	author={Panter, Benjamin and Jimenez, Raul and Heavens, Alan F and Charlot, Stephane},
	journal={\mnras},
	volume={378},
	number={4},
	pages={1550--1564},
	year={2007},
	publisher={Oxford University Press}
}
@article{Planck15,
	title={Planck 2015 results. XV. Gravitational lensing},
	author={Ade, PAR and Aghanim, N and Arnaud, M and Ashdown, M and Aumont, J and Baccigalupi, C and Banday, AJ and Barreiro, RB and Bartlett, JG and Bartolo, N and others},
	journal={arXiv preprint arXiv:1502.01591},
	year={2015}
}
@article{postman95,
	title={Brightest cluster galaxies as standard candles},
	author={Postman, Marc and Lauer, Tod R},
	journal={\apj},
	volume={440},
	pages={28--47},
	year={1995}
}
@article{Roussel00,
	title={The baryon content of groups and clusters of galaxies},
	author={Roussel, H and Sadat, R and Blanchard, A},
	journal={Arxiv preprint astro-ph/0007168},
	year={2000}
}
@article{oegerle91,
	title={Fundamental parameters of brightest cluster galaxies},
	author={Oegerle, William R and Hoessel, John G},
	journal={\apj},
	volume={375},
	pages={15--24},
	year={1991}
}
@article{sandage76,
	title={The absolute magnitude of first-ranked cluster galaxies as a function of cluster richness},
	author={Sandage, Allan},
	journal={\apj},
	volume={205},
	pages={6--12},
	year={1976}
}
@article{sandage72,
	title={The redshift-distance relation. I. Angular diameter of first ranked cluster galaxies as a function of redshift: the aperture correction to magnitudes},
	author={Sandage, Allan},
	journal={\apj},
	volume={173},
	pages={485},
	year={1972}
}
@article{Simionescu11,
	title={Baryons at the Edge of the X-ray--Brightest Galaxy Cluster},
	author={Simionescu, Aurora and Allen, Steven W and Mantz, Adam and Werner, Norbert and Takei, Yoh and Morris, R Glenn and Fabian, Andrew C and Sanders, Jeremy S and Nulsen, Paul EJ and George, Matthew R and others},
	journal={Science},
	volume={331},
	number={6024},
	pages={1576--1579},
	year={2011},
	publisher={American Association for the Advancement of Science}
}
@article{Stott10,
	title={The XMM Cluster Survey: The Build-up of Stellar Mass in Brightest Cluster Galaxies at High Redshift},
	author={Stott, JP and Collins, CA and Sahl{\'e}n, Martin and Hilton, M and Lloyd-Davies, E and Capozzi, D and Hosmer, M and Liddle, AR and Mehrtens, N and Miller, CJ and others},
	journal={\apj},
	volume={718},
	number={1},
	pages={23},
	year={2010},
	publisher={IOP Publishing}
}
@article{Burg14,
	title={A census of stellar mass in ten massive haloes at z\~{} 1 from the GCLASS Survey},
	author={van der Burg, Remco FJ and Muzzin, Adam and Hoekstra, Henk and Wilson, Gillian and Lidman, Chris and Yee, HKC},
	journal={\aap},
	volume={561},
	pages={A79},
	year={2014},
	publisher={EDP Sciences}
}
@article{Burg07,
	title={How special are brightest group and cluster galaxies?},
	author={Von Der Linden, Anja and Best, Philip N and Kauffmann, Guinevere and White, Simon DM},
	journal={\mnras},
	volume={379},
	number={3},
	pages={867--893},
	year={2007},
	publisher={Oxford University Press}
}
@article{Wake11,
	title={Galaxy clustering in the NEWFIRM Medium Band Survey: the relationship between stellar mass and dark matter halo mass at 1< z< 2},
	author={Wake, David A and Whitaker, Katherine E and Labb{\'e}, Ivo and Van Dokkum, Pieter G and Franx, Marijn and Quadri, Ryan and Brammer, Gabriel and Kriek, Mariska and Lundgren, Britt F and Marchesini, Danilo and others},
	journal={\apj},
	volume={728},
	number={1},
	pages={46},
	year={2011},
	publisher={IOP Publishing}
}
@article{White78,
	title={Core condensation in heavy halos: a two-stage theory for galaxy formation and clustering},
	author={White, Simon DM and Rees, MJ},
	journal={\mnras},
	volume={183},
	number={3},
	pages={341--358},
	year={1978},
	publisher={Oxford University Press}
}
@article{White93,
	title={The baryon content of galaxy clusters: a challenge to cosmological orthodoxy},
	author={White, Simon DM and Navarro, Julio F and Evrard, August E and Frenk, Carlos S},
	journal={Nature},
	volume={366},
	number={6454},
	pages={429--433},
	year={1993}
}
@article{Yang09,
	title={Galaxy Groups in the SDSS DR4. III. The Luminosity and Stellar Mass Functions},
	author={Yang, Xiaohu and Mo, HJ and Van den Bosch, Frank C},
	journal={\apj},
	volume={695},
	number={2},
	pages={900},
	year={2009},
	publisher={IOP Publishing}
}
@article{Yang12,
	title={Evolution of the Galaxy--Dark Matter Connection and the Assembly of Galaxies in Dark Matter Halos},
	author={Yang, Xiaohu and Mo, HJ and van den Bosch, Frank C and Zhang, Youcai and Han, Jiaxin},
	journal={\apj},
	volume={752},
	number={1},
	pages={41},
	year={2012},
	publisher={IOP Publishing}
}
@article{Zheng07,
	title={Galaxy evolution from halo occupation distribution modeling of DEEP2 and SDSS galaxy clustering},
	author={Zheng, Zheng and Coil, Alison L and Zehavi, Idit},
	journal={\apj},
	volume={667},
	number={2},
	pages={760},
	year={2007},
	publisher={IOP Publishing}
}
@article{mccracken2012ultravista,
  title={UltraVISTA: a new ultra-deep near-infrared survey in COSMOS},
  author={McCracken, HJ and Milvang-Jensen, B and Dunlop, J and Franx, M and Fynbo, JPU and Le F{\`e}vre, O and Holt, J and Caputi, KI and Goranova, Y and Buitrago, F and others},
  journal={\aap},
  volume={544},
  pages={A156},
  year={2012},
  publisher={EDP Sciences}
}
@article{wuyts2011star,
  title={ON STAR FORMATION RATES AND STAR FORMATION HISTORIES OF GALAXIES OUT TO z~ 3},
  author={Wuyts, Stijn and Schreiber, Natascha M F{\"o}rster and Lutz, Dieter and Nordon, Raanan and Berta, Stefano and Altieri, Bruno and Andreani, Paola and Aussel, Herv{\'e} and Bongiovanni, Angel and Cepa, Jordi and others},
  journal={\apj},
  volume={738},
  number={1},
  pages={106},
  year={2011},
  publisher={IOP Publishing}
}
@article{springle2005cosmological,
  title={The cosmological simulation code Gadget-2 code},
  author={Springle, V},
  journal={MNRAS},
  volume={364},
  pages={1105--1134},
  year={2005}
}
@article{bell2003estimating,
  title={Estimating star formation rates from infrared and radio luminosities: the origin of the radio-infrared correlation},
  author={Bell, Eric F},
  journal={The Astrophysical Journal},
  volume={586},
  number={2},
  pages={794},
  year={2003},
  publisher={IOP Publishing}
}
\end{document}